\pdfoutput=1

\documentclass[11pt,twoside,a4paper,cmspaper,final,collab]{cms-tdr}
\def\svnVersion{473933P}\def\cmsCernNoTag{CERN-EP-2017-307}\def\cmsCernDate{\today}\def\cmsMessage{Published in the European Physical Journal C as \href{http://dx.doi.org/10.1140/epjc/s10052-018-6146-9}{\doi{10.1140/epjc/s10052-018-6146-9.}}}

\begin{document}\cmsNoteHeader{HIG-15-007}

\hyphenation{had-ron-i-za-tion}
\hyphenation{cal-or-i-me-ter}
\hyphenation{de-vices}
\RCS$Revision: 473904 $
\RCS$HeadURL: svn+ssh://svn.cern.ch/reps/tdr2/papers/HIG-15-007/trunk/HIG-15-007.tex $
\RCS$Id: HIG-15-007.tex 473904 2018-09-04 22:17:56Z veelken $

\newlength\cmsSmallFigWidth
\ifthenelse{\boolean{cms@external}}{\setlength\cmsSmallFigWidth{0.57\columnwidth}}{\setlength\cmsSmallFigWidth{0.30\textwidth}}
\newlength\cmsFigWidth
\ifthenelse{\boolean{cms@external}}{\setlength\cmsFigWidth{0.95\columnwidth}}{\setlength\cmsFigWidth{0.44\textwidth}}
\ifthenelse{\boolean{cms@external}}{\providecommand{\cmsLeft}{top\xspace}}{\providecommand{\cmsLeft}{left\xspace}}
\ifthenelse{\boolean{cms@external}}{\providecommand{\cmsRight}{bottom\xspace}}{\providecommand{\cmsRight}{right\xspace}}

\providecommand{\tauh}{\ensuremath{\Pgt_{\mathrm{h}}}\xspace}
\providecommand{\taue}{\ensuremath{\Pgt_{\Pe}}\xspace}
\providecommand{\taum}{\ensuremath{\Pgt_{\Pgm}}\xspace}
\newcommand{\Ppizero}{\ensuremath{\Pgpz}}
\newcommand{\Pnu}{\ensuremath{\PGn}}
\newcommand{\Pnut}{\ensuremath{\PGn_{\Pgt}}}
\newcommand{\APnue}{\ensuremath{\PAGn_{\Pe}}}
\newcommand{\APnum}{\ensuremath{\PAGn_{\Pgm}}}
\newcommand{\APnut}{\ensuremath{\PAGn_{\Pgt}}}
\renewcommand{\PJgy}{\ensuremath{\cmsSymbolFace{J}/\psi}\xspace}
\newcommand{\PUpsilon}{\ensuremath{\Upsilon}\xspace}
\renewcommand{\Pgg}{\ensuremath{\PGg}\xspace}
\newcommand{\Pggx}{\ensuremath{\Pgg^{*}}\xspace}
\newcommand{\PHiggs}{\ensuremath{\textrm{H}}\xspace}
\newcommand{\APq}{\ensuremath{\bar{\Pq}}}
\newcommand{\Pbottom}{\ensuremath{\PQb}}
\newcommand{\Phadron}{\ensuremath{\mathrm{h}}}

\newcommand{\pT}{\ensuremath{p_{\mathrm{T}}}\xspace}
\newcommand{\pTvec}{\ensuremath{{\vec p}_{\mathrm{T}}}\xspace}
\newcommand{\kT}{\ensuremath{k_{\mathrm{T}}}\xspace}
\newcommand{\vecMET}{\ensuremath{{\vec p}_{\mathrm{T}}^{\ \textrm{miss}}}\xspace}
\renewcommand{\MET}{\ensuremath{p_{\mathrm{T}}^{\,\textrm{miss}}}\xspace}
\providecommand{\mT}{\ensuremath{m_{\mathrm{T}}}\xspace}
\newcommand{\mVis}{\ensuremath{m_{\textrm{vis}}}\xspace}
\newcommand{\Pzetamiss}{\ensuremath{P_{\zeta}^{\,\textrm{miss}}}\xspace}
\newcommand{\Pzetavis}{\ensuremath{P_{\zeta}^{\,\textrm{vis}}}\xspace}
\newcommand{\pTell}{\ensuremath{p_{\mathrm{T}}^{\,\ell}}\xspace}
\newcommand{\pTvece}{\ensuremath{{\vec p}_{\mathrm{T}}}^{\ \mathrm{e}}\xspace}
\newcommand{\pTvecmu}{\ensuremath{{\vec p}_{\mathrm{T}}}^{\ \mu}\xspace}

\newcommand{\sig}{\ensuremath{\textrm{sig}}}
\newcommand{\fit}{\ensuremath{\textrm{fit}}}
\newcommand{\out}{\ensuremath{\textrm{out}}}
\newcommand{\gen}{\ensuremath{\textrm{gen}}}
\newcommand{\obs}{\ensuremath{\textrm{obs}}}
\newcommand{\misid}{\ensuremath{\textrm{misid}}}
\newcommand{\fake}{\ensuremath{\textrm{fake}}}
\newcommand{\FF}{\ensuremath{F_\textrm{F}}}
\newcommand{\jet}{\ensuremath{\textrm{jet}}}

\newcommand{\cf}{{\it cf.}\xspace}
\renewcommand{\ie}{i.e.\xspace}

\newcommand{\oneProngZeroPizero}{\ensuremath{\mathrm{h}^{\pm}}\xspace}
\newcommand{\oneProngOnePizero}{\ensuremath{\mathrm{h}^{\pm}\Ppizero}\xspace}
\newcommand{\oneProngTwoPizero}{\ensuremath{\mathrm{h}^{\pm}\Ppizero\Ppizero}\xspace}
\newcommand{\oneProngPizeros}{\ensuremath{\mathrm{h}^{\pm}\Ppizero\text{s}}\xspace}
\newcommand{\threeProngZeroPizero}{\ensuremath{\mathrm{h}^{\pm}\mathrm{h}^{\mp}\mathrm{h}^{\pm}}\xspace}

\providecommand{\suppMaterial}{supplemental material [URL will be inserted by publisher]}

\cmsNoteHeader{HIG-15-007}
\title{
Measurement of the \texorpdfstring{$\PZ/\gamma^{*} \to \tau\tau$}{Z to tau-tau} cross section in pp collisions at \texorpdfstring{$\sqrt{s} = $ 13\TeV}{sqrt(s) = 13 TeV} and validation of \texorpdfstring{$\tau$}{tau} lepton analysis techniques
}
\titlerunning{Measurement of the \texorpdfstring{$\PZ \to \tau\tau$}{Z to tau-tau} cross section at \texorpdfstring{13\TeV}{13 TeV}}
\date{\today}

\abstract{
A measurement is presented of the $\PZ/\gamma^{*} \to \tau\tau$ cross section in $\textrm{pp}$ collisions at $\sqrt{s} = 13\TeV$, using data recorded by the CMS experiment at the LHC, corresponding to an integrated luminosity of 2.3\fbinv. The product of the inclusive cross section and branching fraction is measured to be $\sigma(\textrm{pp} \to \PZ/\gamma^{*}\text{+X}) \,  \mathcal{B}(\PZ/\gamma^{*} \to \tau\tau) = 1848 \pm 12\stat \pm 67\text{~(syst+lumi)}\unit{pb}$, in agreement with the standard model expectation, computed at next-to-next-to-leading order accuracy in perturbative quantum chromodynamics. The measurement is used to validate new analysis techniques relevant for future measurements of $\tau$ lepton production. The measurement also provides the reconstruction efficiency and energy scale for $\tau$ decays to hadrons$ + \nu_{\tau}$ final states, determined with respective relative uncertainties of 2.2\% and 0.9\%.
}

\hypersetup{%
pdfauthor={CMS Collaboration},%
pdftitle={Measurement of the Z to tau-tau cross section in pp collisions at sqrt(s) = 13 TeV and validation of tau lepton analysis techniques},%
pdfsubject={CMS},%
pdfkeywords={CMS, physics}}

\maketitle

\section{Introduction}
\label{sec:introduction}

Final states with $\Pgt$ leptons are important experimental signatures at the CERN LHC.
In particular, the recently reported observation of decays of standard model (SM) Higgs bosons ($\PHiggs$)~\cite{Higgs-Discovery_ATLAS,Higgs-Discovery_CMS,Higgs-Discovery_CMS_longPaper} into pairs of $\Pgt$ leptons~\cite{HIG-15-002}
suggests additional searches in the context of 
new charged~\cite{HIG-11-019,HIG-14-023,HIG-12-005,Aad:2014kga} 
and neutral~\cite{HIG-10-002,HIG-11-029,HIG-13-021,HIG-14-019,HIG-14-033,HIG-16-043,Aad:2011rv,Aad:2014vgg,Aad:2015oqa} Higgs bosons,
lepton-flavor violation~\cite{HIG-14-005,Aad:2015gha,Aad:2015pfa},
supersymmetry~\cite{SUS-12-004,SUS-13-002,SUS-13-003,Aad:2012ypy,Aad:2014mra,Aad:2014yka,Aad:2014nua,Aad:2015tin},
leptoquarks~\cite{EXO-12-002,EXO-14-008},
extra spatial dimensions~\cite{HIG-15-013,Aad:2015xja},
and massive gauge bosons~\cite{EXO-11-031,EXO-12-011,Aad:2015osa}.

With a lifetime of $2.9 \times 10^{-13}\unit{s}$,
the $\Pgt$ lepton usually decays before reaching the innermost detector.
Approximately two thirds of $\Pgt$ leptons decay into a hadronic system and a $\Pgt$ neutrino.
Constrained by the $\Pgt$ lepton mass of $1.777\GeV$,
the hadronic system is characterized by low particle multiplicities, typically consisting of either one or three charged pions or kaons, and up to two neutral pions.
The hadrons produced in $\Pgt$ decays therefore also tend to be highly collimated.
The $\Pgt$ lepton decays into an electron or muon and two neutrinos with a probability of $35\%$.
We denote the electron and muon produced in $\Pgt \to \Pe\Pnu\Pnu$ and $\Pgt \to \Pgm\Pnu\Pnu$ decays by $\taue$ and $\taum$, to distinguish them from prompt electrons and muons, respectively.
The hadronic system produced in a $\Pgt \to \mbox{hadrons} + \Pnut$ decay is denoted by the symbol $\tauh$.

The Drell--Yan (DY)~\cite{Drell:1970wh} production of $\Pgt$ lepton pairs ($\Pq\APq \to \cPZ/\Pggx \to \Pgt\Pgt$) is interesting for several reasons.
First, the process $\cPZ/\Pggx \to \Pgt\Pgt$ represents a reference signal to study the efficiency to reconstruct and identify $\tauh$,
as well as to measure the $\tauh$ energy scale.
Moreover, $\cPZ/\Pggx \to \Pgt\Pgt$ production constitutes the dominant irreducible background to analyses of SM $\PHiggs \to \Pgt\Pgt$ events,
and to searches for new resonances decaying to $\Pgt$ lepton pairs.
The cross section for DY production exceeds the one for SM $\PHiggs$ production by about two orders of magnitude.
Signals from new resonances are expected to be even more rare.
It is therefore important to control with a precision reaching the sub-percent level
the rate for $\cPZ/\Pggx \to \Pgt\Pgt$ production, as well as its distribution in kinematic observables.
In addition, the reducible backgrounds relevant for the study of $\PZ/\Pggx \to \Pgt\Pgt$
are also relevant for studies of SM $\PHiggs$ production and to searches for new resonances.

This paper reports a precision measurement of the inclusive $\Pp\Pp \to \cPZ/\Pggx\text{+X} \to \Pgt\Pgt\text{+X}$ cross section.
The measurement demonstrates that $\cPZ/\Pggx \to \Pgt\Pgt$ production is well understood,
and provides ways to validate techniques relevant in future analyses of $\Pgt$ lepton production.
Most notably, a method based on control samples in data is introduced for determining background contributions
arising from the misidentification of quark or gluon jets as $\tauh$.
Measurements of the $\tauh$ identification (ID) efficiency and of the $\tauh$ energy scale~\cite{TAU-14-001}
are obtained as byproducts of the analysis.

The cross section for DY production of $\Pgt$ lepton pairs
was previously measured by the CMS and ATLAS experiments
in proton-proton ($\Pp\Pp$) collisions at $\sqrt{s} = 7\TeV$ at the LHC~\cite{EWK-10-013,Aad:2011kt},
and in proton-antiproton collisions at $\sqrt{s} = 1.96\TeV$ by the CDF and D0 experiments at the Fermilab Tevatron~\cite{Abulencia:2007iy,Abazov:2004vd,Abazov:2008ff}.
In this study, we present the $\Pp\Pp \to \cPZ/\Pggx\text{+X} \to \Pgt\Pgt\text{+X}$ cross section measured at $\sqrt{s} = 13\TeV$,
using data recorded by the CMS experiment, corresponding to an integrated luminosity of 2.3\fbinv.
Events are selected in the $\taue\tauh$, $\taum\tauh$, $\tauh\tauh$, $\taue\taum$, and $\taum\taum$ decay channels.
The $\taue\taue$ channel is not considered in this analysis,
as it was studied previously in the context of the SM $\PHiggs \to \Pgt\Pgt$ analysis,
and found to be the least sensitive of these channels~\cite{HIG-13-004}.
The $\Pp\Pp \to \cPZ/\Pggx\text{+X} \to \Pgt\Pgt\text{+X}$ cross section
is obtained through a simultaneous fit of $\Pgt$ lepton pair mass distributions
in all decay channels.

The paper is organized as follows.
The CMS detector is described briefly in Section~\ref{sec:detector}.
Section~\ref{sec:datasamples_and_MonteCarloSimulation} describes the data and the Monte Carlo (MC) simulations used in the analysis.
The reconstruction of electrons, muons, $\tauh$, and jets, along with various kinematic quantities, is described in Section~\ref{sec:eventReconstruction}.
Section~\ref{sec:eventSelection} details the selection of events in the different decay channels,
followed in Section~\ref{sec:backgroundEstimation} by a description of the procedures used to estimate background contributions.
The systematic uncertainties relevant for the measurement of the $\Pp\Pp \to \cPZ/\Pggx\text{+X} \to \Pgt\Pgt\text{+X}$ cross section are described in Section~\ref{sec:systematicUncertainties},
and the extraction of the signal is given in Section~\ref{sec:signalExtraction}.
The results are presented in Section~\ref{sec:results},
and the paper concludes with a summary in Section~\ref{sec:summary}.

\section{The CMS detector}
\label{sec:detector}

The central feature of the CMS apparatus is a superconducting solenoid of 6\unit{m} internal diameter, providing a magnetic field of 3.8\unit{T}.
A silicon pixel and strip tracker,
a lead tungstate crystal electromagnetic calorimeter (ECAL), and a brass and scintillator hadron calorimeter (HCAL),
each composed of a barrel and two endcap sections,
are positioned within the solenoid volume.
The silicon tracker measures charged particles within the pseudorapidity range $\abs{\eta}< 2.5$.
Trajectories of isolated muons with $\pT = 100\GeV$, emitted at $\abs{\eta} < 1.4$, are reconstructed with an efficiency close to 100\%
and resolutions of 2.8\% in $\pT$, and with uncertainties of 10 and 30\mum in their respective transverse and longitudinal impact parameters relative to their points of origin~\cite{TRK-11-001}.
The ECAL is a fine-grained hermetic calorimeter with quasi-projective geometry,
segmented in the barrel region of $\abs{\eta} < 1.48$, as well as in the two endcaps that extend up to $\abs{\eta} < 3.0$.
Similarly, the HCAL barrel and endcaps cover the region $\abs{\eta} < 3.0$.
Forward calorimeters extend the coverage up to $\abs{\eta} < 5.0$.
Muons are measured and identified in the range $\abs{\eta}< 2.4$ using
gas-ionization detectors embedded in the steel flux-return yoke outside the solenoid.
A two-level trigger system is used to reduce the rate of recorded
events to a level suitable for data acquisition and storage.
The first level (L1) of the CMS trigger system, composed of specialized hardware processors,
uses information from the calorimeters and muon detectors to select the most interesting events in a fixed time interval of less than 4\mus.
The high-level trigger processor farm decreases the event rate from around 100\unit{kHz} to less than 1\unit{kHz} before storage and subsequent analysis.
Details of the CMS detector and its performance, together with a definition of the coordinate system and kinematic variables, can be found in Ref.~\cite{Chatrchyan:2008zzk}.

\section{Data and Monte Carlo simulation}
\label{sec:datasamples_and_MonteCarloSimulation}

The data were recorded in $\Pp\Pp$ collisions at 25\unit{ns} bunch spacing
and are required to satisfy standard data quality criteria.
The analysed data correspond to an integrated luminosity of 2.3\fbinv.

{\tolerance=600
The $\cPZ/\Pggx \to \Pgt\Pgt$ signal and the $\cPZ/\Pggx \to \Pe\Pe$,
$\cPZ/\Pggx \to \Pgm\Pgm$, $\PW$+jets, $\cPqt\cPaqt$, single top
quark, and diboson ($\PW\PW$, $\PW\cPZ$, and $\cPZ\cPZ$) background processes are modelled through samples of MC simulated events.
Background contributions arising from multijet production via quantum chromodynamic interactions are determined from data.
The $\cPZ/\Pggx \to \ell\ell$ (where $\ell$ refers to $\Pe$, $\Pgm$, or $\Pgt$) and
$\PW$+jets events are generated using leading-order (LO) matrix elements (ME) in quantum chromodynamics, implemented in the program
\MGvATNLO $2.2.2$~\cite{MadGraph5_aMCatNLO},
and $\cPqt\cPaqt$ and single top quark events are generated using the next-to-leading order (NLO) program
{\sc powheg} v$2$~\cite{POWHEG1,POWHEG2,POWHEG3,POWHEG_ttbar,POWHEG_singletop}.
The diboson events are modelled using the NLO ME program implemented in \MGvATNLO.
The background events are complemented with SM $\PHiggs \to \Pgt\Pgt$ events,
generated for an $\PHiggs$ mass of $m_{\PHiggs} = 125\GeV$,
using the implementation of the gluon-gluon and vector boson fusion processes in {\sc powheg}~\cite{POWHEG_ggH,POWHEG_qqH}.
All events are generated using the {\sc NNPDF3.0}~\cite{NNPDF1,NNPDF2,NNPDF3} set of parton distribution functions (PDF).
Parton showers and parton hadronization are modelled using \PYTHIA $8.212$~\cite{pythia8} and the CUETP8M1 underlying-event tune~\cite{PYTHIA_CUETP8M1tune_CMS},
which is based on the Monash tune~\cite{PYTHIA_MonashTune}.
The decays of $\Pgt$ leptons, including polarization effects, are modelled through \PYTHIA.
The $\cPZ/\Pggx \to \ell\ell$, $\PW$+jets, and $\cPqt\cPaqt$ events are normalized
to cross sections computed at next-to-next-to-leading order (NNLO)
accuracy~\cite{FEWZ3,TTbarXsectionNNLO}.
A reweighting is applied to MC-generated $\cPqt\cPaqt$ and $\cPZ/\Pggx \to \ell\ell$ events to improve the respective
modelling of the $\pT$ spectrum of the top quarks~\cite{Chatrchyan:2012saa,TOP-12-028} and the dilepton mass and $\pT$ spectra relative to data.
The weights applied to simulated $\cPZ/\Pggx \to \ell\ell$ events are obtained
from studies of the distributions in dilepton mass and $\pT$ in $\cPZ/\Pggx \to \Pgm\Pgm$ events.
The cross sections for single top quark~\cite{singleTopXsectionNNLO1,singleTopXsectionNNLO2,singleTopXsectionNNLO3} and diboson~\cite{MCFMdiBosonXsection} production are computed at NLO accuracy.
\par}

Minimum bias events generated with \PYTHIA are overlaid on all simulated events
to account for the presence of additional inelastic $\Pp\Pp$ interactions, referred to as pileup (PU), which take place in the same, previous, or subsequent bunch crossings as the hard-scattering interaction.
The pileup distribution in simulated events matches that in data,
amounting to, on average, ${\approx}12$ inelastic $\Pp\Pp$ interactions per bunch crossing.
All generated events are passed through a detailed simulation of the CMS apparatus, based on \GEANTfour~\cite{geant4},
and reconstructed using the same version of the CMS reconstruction software as used for data.

\section{Event reconstruction}
\label{sec:eventReconstruction}

The information provided by all CMS subdetectors is employed in a
particle-flow (PF)
algorithm~\cite{PRF-14-001}
to identify and reconstruct individual particles in the event, namely muons, electrons, photons, charged and neutral hadrons.
These particles are then used to reconstruct jets, $\tauh$ candidates and the vector imbalance in missing transverse momentum in the event, referred to as $\vecMET$,
as well as to quantify the isolation of leptons.

Electrons are reconstructed using an algorithm~\cite{EGM-13-001} that matches trajectories in the silicon tracker to energy depositions in the ECAL.
Trajectories of electron candidates are reconstructed using a dedicated algorithm that accounts for the emission of bremsstrahlung photons.
The energy loss due to bremsstrahlung is determined by searching for energy depositions in the ECAL emitted tangentially to the track.
A multivariate (MVA) approach based on boosted decision trees (BDT)~\cite{TMVA}
is employed to distinguish electrons from hadrons that mimic electron signatures.
Observables that quantify the quality of the electron track,
the compactness of the electron cluster in directions transverse and longitudinal relative to the electron motion,
and the matching of the track momentum and direction to the sum and positions of energy depositions in the ECAL are used as inputs to the BDT.
The BDT is trained on samples of genuine and false electrons,
produced in MC simulation.
Additional requirements are applied to remove electrons originating from photon conversions.

The identification of muons is based on
linking track segments reconstructed in the silicon tracking detector and in the muon system~\cite{Chatrchyan:2012xi}.
The matching is done both by starting from a track in the muon system and
starting from a track in the inner detector.
When a link is established, the track parameters are refitted using the combination of hits in the inner and outer detectors,
and the reconstructed trajectory is referred to as a global muon track.
Quality criteria are applied on the multiplicity of hits, the number of matched segments, and the quality of the fit to a global muon track, the latter being quantified through a $\chi^{2}$ criterion.

Electrons and muons in signal events are expected to be isolated,
while leptons from heavy flavour (charm and bottom quark) decays,
as well as from in-flight decays of pions and kaons,
are often reconstructed within jets.
Isolated leptons are distinguished from leptons in jets through a sum, denoted by the symbol $I_{\ell}$, of the scalar $\pT$ values of additional charged particles, neutral hadrons,
and photons reconstructed using the PF algorithm within a cone in $\eta$ and azimuth $\phi$ (in radians)
of size $\Delta R = \sqrt{\smash[b]{(\Delta\eta)^{2} + (\Delta\phi)^{2}}} = 0.3$, centred around the lepton direction.
Neutral hadrons and photons within the innermost region of the cone, $\Delta R < 0.01$, are excluded from the isolation sum for muons
to prevent the footprint of the muon in ECAL and HCAL from causing the muon to fail isolation criteria.
When computing the isolation of electrons reconstructed in the ECAL endcap region,
we exclude photons within $\Delta R < 0.08$ and charged particles within $\Delta R < 0.015$ of the direction of the electron,
to avoid counting photons emitted in bremsstrahlung and tracks originating from the conversion of such photons.
As the amount of material that electrons traverse in the barrel region before reaching the ECAL is smaller, 
the resulting probability for bremsstrahlung emission and photon conversion is sufficiently reduced 
so as not to require exclusion of particles in the innermost cone from the isolation sum.
Efficiency loss due to pileup is kept minimal
by considering only charged particles originating from the lepton production vertex (``charged from PV'').
The contribution from the neutral component of pileup to the isolation of the lepton is taken into account by means of $\Delta\beta$ corrections~\cite{PRF-14-001},
which enter the computation of the isolation $I_{\ell}$, as follows:
\begin{linenomath}
\begin{equation}
I_{\ell} = \sum_{\substack{\textrm{charged} \\ \textrm{from PV}}} \pT + \textrm{max} \left\{ 0, \sum_{\textrm{neutrals}} \pT - \Delta\beta \right\},
\label{eq:lepIsolationDeltaBeta}
\end{equation}
\end{linenomath}
where $\ell$ corresponds to either $\Pe$ or $\Pgm$, and the sums extend over, respectively,
the charged particles that originate from the lepton production vertex and the neutral particles.
The ``$\textrm{max}$'' function represents taking the largest of the two values within the brackets.
The $\Delta\beta$ corrections are computed by summing the scalar $\pT$ of charged particles
that are within a cone of size $\Delta R = 0.3$ around the lepton direction, but do not originate from the lepton production vertex,
(``charged from PU'') and scaling that sum by a factor of one-half:
\begin{linenomath}
\begin{equation}
\Delta\beta = 0.5 \, \sum_{\substack{\textrm{charged} \\ \textrm{from PU}}} \pT.
\end{equation}
\end{linenomath}
The factor of $0.5$ approximates the phenomenological ratio of neutral-to-charged hadron production in the hadronization of inelastic $\Pp\Pp$ collisions.

Collision vertices are reconstructed using a deterministic annealing algorithm~\cite{Chabanat:2005zz,Fruhwirth:2007hz},
with the reconstructed vertex position required to be compatible with the location of the LHC beam in the $x$--$y$ plane.
The primary collision vertex (PV) is taken to be the vertex that has the maximum $\sum \pT^{2}$ of tracks associated to it.
Electrons, muons, and $\tauh$ candidates are required to be compatible with originating from the PV.

{\tolerance=4800
Hadronic $\Pgt$ decays are reconstructed using the ``hadrons+strips'' (HPS) algorithm~\cite{TAU-14-001}, which is used to separate the
individual decay modes of the $\Pgt$ into
$\Pgt^{-} \to \Phadron^{-}\Pnut$, $\Pgt^{-} \to \Phadron^{-}\Ppizero\Pnut$, $\Pgt^{-} \to \Phadron^{-}\Ppizero\Ppizero\Pnut$, and $\Pgt^{-} \to \Phadron^{-}\Phadron^{+}\Phadron^{-}\Pnut$,
where $\Phadron^{\pm}$ denotes either a charged pion or kaon (the decay modes of $\Pgt^{+}$ are assumed to be identical to their partner $\Pgt^{-}$ modes through charge conjugation invariance).
The $\tauh$ candidates are constructed by combining the charged PF hadrons with neutral pions.
The neutral pions are reconstructed by clustering the PF photons within rectangular strips,
narrow in the $\eta$, but wide in the $\phi$ directions,
to account for the non-negligible probability for photons produced in $\Ppizero \to \Pgg\Pgg$ decays
to convert into electron-positron pairs when traversing the all-silicon tracking detector of CMS
and the broadening of energy depositions in the ECAL that occurs when this happens.
For the same reason, electrons and positrons reconstructed through the PF algorithm are considered in the reconstruction of the neutral pions
besides photons.
The momentum of the $\tauh$ candidate is taken as the vector sum over the momenta of the charged hadrons and neutral pions used in reconstructing the $\tauh$ decay mode, assuming the pion-mass hypotheses.
We do not use the strips of $0.20 \times 0.05$ size in the $\eta$--$\phi$ plane, used in previous analyses~\cite{HIG-11-019,HIG-14-023,HIG-12-005,HIG-10-002,HIG-11-029,HIG-13-021,HIG-14-019,HIG-14-033,HIG-14-005,SUS-12-004,SUS-13-002,SUS-13-003,EXO-12-002,EXO-14-008,HIG-15-013,EXO-11-031,EXO-12-011,EWK-10-013,HIG-13-004},
but an improved version of the strip reconstruction developed during the $\sqrt{s} = 13\TeV$ run.
In the improved version, the size of the strip is adjusted as function of $\pT$,
taking into consideration the bending of charged particles in the magnetic field increasing inversely with $\pT$.
More details on strip reconstruction and validation of the algorithm with data are given in Ref.~\cite{TAU-16-002}.
The main handle for distinguishing $\tauh$ from the large background of quark and gluon jets relies on the use of tight isolation requirements.
The sums of scalar $\pT$ values from photons and from charged particles originating from the PV
within a cone of $\Delta R = 0.5$ centred around the $\tauh$ direction, are used as input to an MVA-based $\tauh$ ID discriminant.
The set of input variables is complemented with the scalar $\pT$ sum of charged particles not originating from the PV,
by the $\tauh$ decay mode, and by observables that are sensitive to the lifetime of the $\Pgt$.
The transverse impact parameter of the ``leading'' (highest $\pT$) track of each $\tauh$ candidate relative to the PV
is used for $\tauh$ candidates reconstructed in any decay mode.
For $\tauh$ candidates reconstructed in the $\Pgt^{-} \to \Phadron^{-}\Phadron^{+}\Phadron^{-}\Pnut$ decay mode,
a fit of the three tracks to a common secondary vertex (SV) is attempted, and the distance between SV and PV is used as additional input to the MVA.
The MVA is trained on genuine $\tauh$ and jets generated in MC simulation.
Four working points (WP), referred to as barely, minimally, moderately, and tightly constrained,
are defined through changes made in the selections on the MVA output.
The thresholds are adjusted as functions of the $\pT$ of the $\tauh$ candidate,
such that the $\tauh$ identification efficiency for each WP is independent of $\pT$.
The moderate and tight WP used to select events in different channels
provide efficiencies of $55$ and $45\%$, and misidentification rates for jets of typically $1$ and $0.5\%$, depending on the $\pT$ of the jet~\cite{TAU-16-002}.
Additional discriminants are employed to separate $\tauh$ from electrons and muons.
The separation of $\tauh$ from electrons is performed via another MVA-based discriminant~\cite{TAU-16-002} that utilizes input observables that quantify the matching between the sum of energy depositions in the ECAL and the momentum of the leading track of the $\tauh$ candidate,
as well as variables that distinguish electromagnetic from hadronic showers.
The cutoff-based discriminant described in Ref.~\cite{TAU-14-001} is used to separate $\tauh$ from muons.
It is based on matching the leading track of the $\tauh$ candidate with energy depositions in the ECAL and HCAL, as well as with track segments in the muon detectors.
\par}

Jets within the range $\vert \eta \vert < 4.7$ are reconstructed
using the anti-$\textrm{k}_{\text{t}}$ algorithm~\cite{Cacciari:2008gp,Cacciari:2011ma} with a distance parameter $R = 0.4$.
Reconstructed jets are required not to overlap with identified electrons, muons, or $\tauh$ candidates within $\Delta R < 0.5$,
and to pass a set of minimal identification criteria that aim to reject jets arising from calorimeter noise~\cite{JME-10-003}.
The energy of reconstructed jets is calibrated as function of jet $\pT$ and $\eta$~\cite{JME-13-004}.
Average energy density corrections calculated using the {\sc FastJet} algorithm~\cite{Cacciari:2008gn, Cacciari:2007fd}
are applied to compensate pileup effects.
Jets originating from the hadronization of $\Pbottom$ quarks
are identified using the ``combined secondary vertex'' (CSV) algorithm~\cite{Chatrchyan:2012jua},
which exploits observables related to the long lifetime of $\Pbottom$ hadrons and the higher particle multiplicity and mass
of $\Pbottom$ jets compared to light-quark and gluon jets.

{\tolerance=600
The vector $\vecMET$, with its magnitude referred to as $\MET$, is reconstructed using an MVA regression algorithm~\cite{JME-13-003}.
To reduce the impact of pileup on the resolution in $\MET$,
the algorithm utilizes the fact that pileup produces jets predominantly of low $\pT$,
while leptons and high-$\pT$ jets are almost exclusively produced through hard scattering processes.
\par}

The $\cPZ/\Pggx \to \Pgt\Pgt$ signal is distinguished from backgrounds by means of the mass of the $\Pgt$ lepton pair.
The mass, denoted by the symbol $m_{\Pgt\Pgt}$, is reconstructed using the {\sc SVfit} algorithm~\cite{SVfit}.
The algorithm is based on a likelihood approach and uses as inputs the measured momenta of the visible decay products of both $\Pgt$ leptons,
the reconstructed $\MET$, and an event-by-event estimate of the $\MET$ resolution.
The latter is computed as described in Refs.~\cite{JME-10-009,JME-13-003}.
The inputs are combined with a probabilistic model for leptonic and hadronic $\Pgt$ decays
to estimate the momenta of the neutrinos produced in these decays.
The algorithm achieves a resolution in $m_{\Pgt\Pgt}$ of ${\approx}15\%$ relative to the mass of the $\Pgt$ lepton pairs at the generator level.

\section{Event selection}
\label{sec:eventSelection}

The events selected in the $\taue\tauh$, $\taum\tauh$, $\tauh\tauh$, $\taue\taum$, and $\taum\taum$ channels
are recorded by combining single-electron and single-muon triggers,
triggers that are based on the presence of two $\tauh$ candidates in the event,
and triggers based on the presence of both an electron and a muon.

The $\taue\tauh$ and $\taum\tauh$ channels utilize single-electron and -muon triggers
with $\pT$ thresholds of $23$ and $18\GeV$, respectively.
Selected events are required to contain
an electron of $\pT > 24\GeV$ or a muon of $\pT > 19\GeV$, both with $\lvert \eta \rvert < 2.1$,
and a $\tauh$ candidate with $\pT > 20\GeV$ and $\lvert \eta \rvert < 2.3$.
The electron or muon is required to pass an isolation requirement of $I_{\ell} < 0.10 \, \pTell$, computed according to Eq.~(\ref{eq:lepIsolationDeltaBeta}).
The $\tauh$ candidate is required to pass the moderate WP of the MVA-based $\tauh$ ID discriminant,
and to have a charge opposite to that of the electron or muon.
The $\tauh$ candidate is further required to pass a tight or minimal requirement on the discriminant that separates hadronic $\Pgt$ decays from electrons,
and a minimal or tight selection on the discriminant that separates $\tauh$ from muons.
Background arising from $\PW$+jets and $\cPqt\cPaqt$ production is reduced
by requiring the transverse mass of electron or muon and $\vecMET$ to satisfy $\mT < 40\GeV$.
The transverse mass is defined by:
\begin{linenomath}
\begin{equation}
\mT = \sqrt{2 \, \pTell \, \MET \, \left( 1 - \cos \Delta\phi \right)} \, ,
\label{eq:defMt}
\end{equation}
\end{linenomath}
where the symbol $\ell$ refers to the electron or muon,
and $\Delta\phi$ denotes the angle in the transverse plane between the lepton momentum and the $\vecMET$ vector.
Events containing additional electrons with $\pT > 10\GeV$ and $\lvert \eta \rvert < 2.5$, or muons with $\pT > 10\GeV$ and $\lvert \eta \rvert < 2.4$,
passing minimal identification and isolation criteria,
are rejected to reduce backgrounds from $\cPZ/\Pggx \to \Pe\Pe$ and $\Pgm\Pgm$ events, and from diboson production.

A trigger based on the presence of two $\tauh$ candidates is used to record events in the $\tauh\tauh$ channel.
The trigger selects events containing two isolated calorimeter energy deposits at the L1 trigger stage,
which are subsequently required to pass a simplified version of the PF-based offline $\tauh$ reconstruction at the high-level trigger stage.
The latter applies additional isolation criteria.
The $\pT$ threshold for both $\tauh$ candidates is $35\GeV$.
The trigger efficiency increases with $\pT$ of the $\tauh$,
because different algorithms are used to reconstruct the $\pT$ at the L1 trigger stage and in the offline reconstruction.
The trigger reaches an efficiency plateau of ${\approx}80\%$ for events in which both $\tauh$ candidates have $\pT > 60\GeV$.
Selected events are required to contain two $\tauh$ candidates with $\pT > 40\GeV$ and $\lvert \eta \rvert < 2.1$
that have opposite charge and satisfy
the tight WP of the MVA-based $\tauh$ ID discriminant,
as well as the minimal criteria on the discriminants used to separate $\tauh$ from electrons and muons.
Events containing electrons with $\pT > 10\GeV$ and $\lvert \eta \rvert < 2.5$ or muons with $\pT > 10\GeV$ and $\lvert \eta \rvert < 2.4$,
passing minimal identification and isolation criteria,
are rejected to avoid overlap with the $\taue\tauh$ and $\taum\tauh$ channels.

\begin{figure*}[ht!]
\centering
\includegraphics[width=\cmsFigWidth]{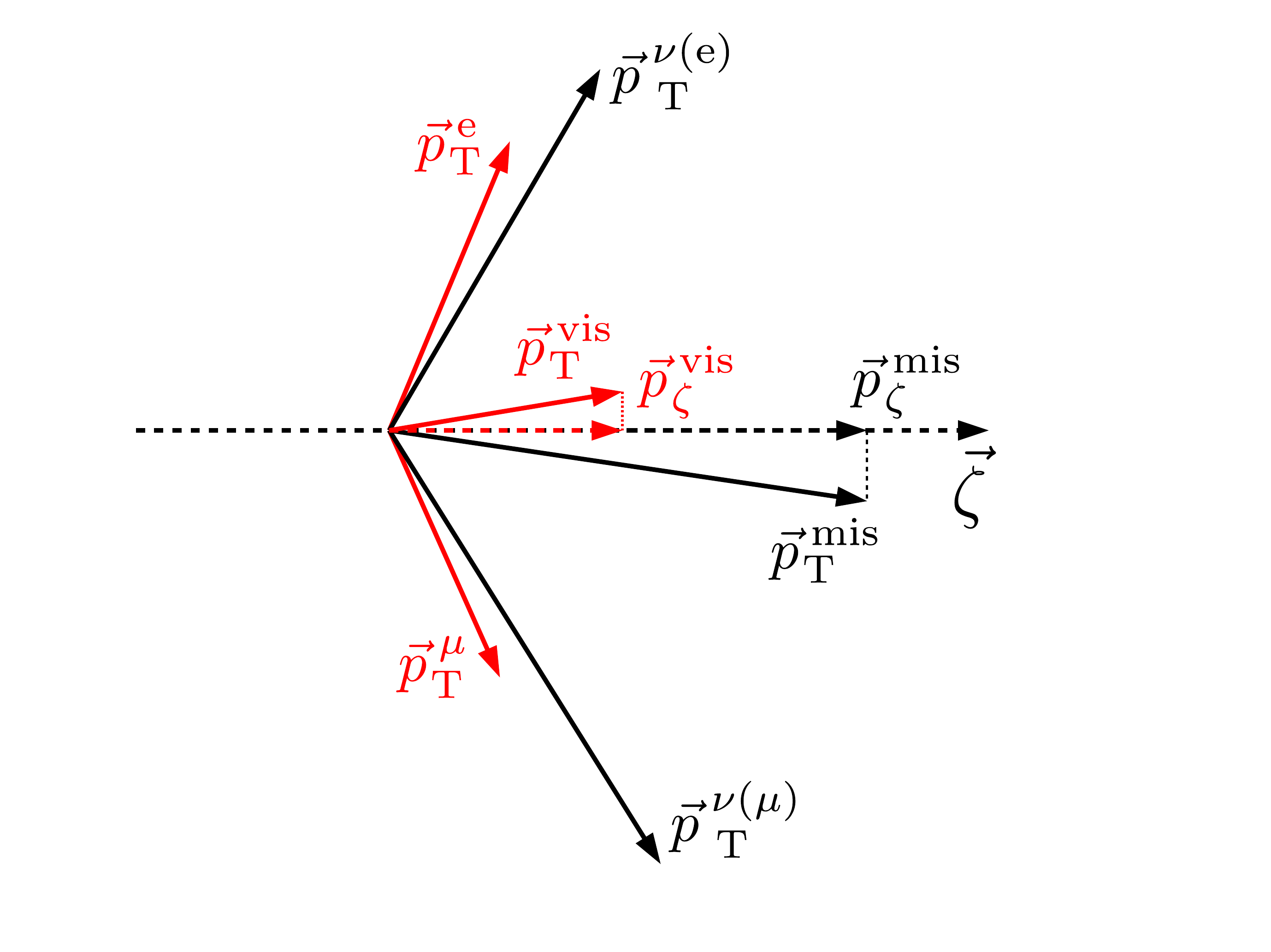}  \hfil
\includegraphics[width=\cmsFigWidth]{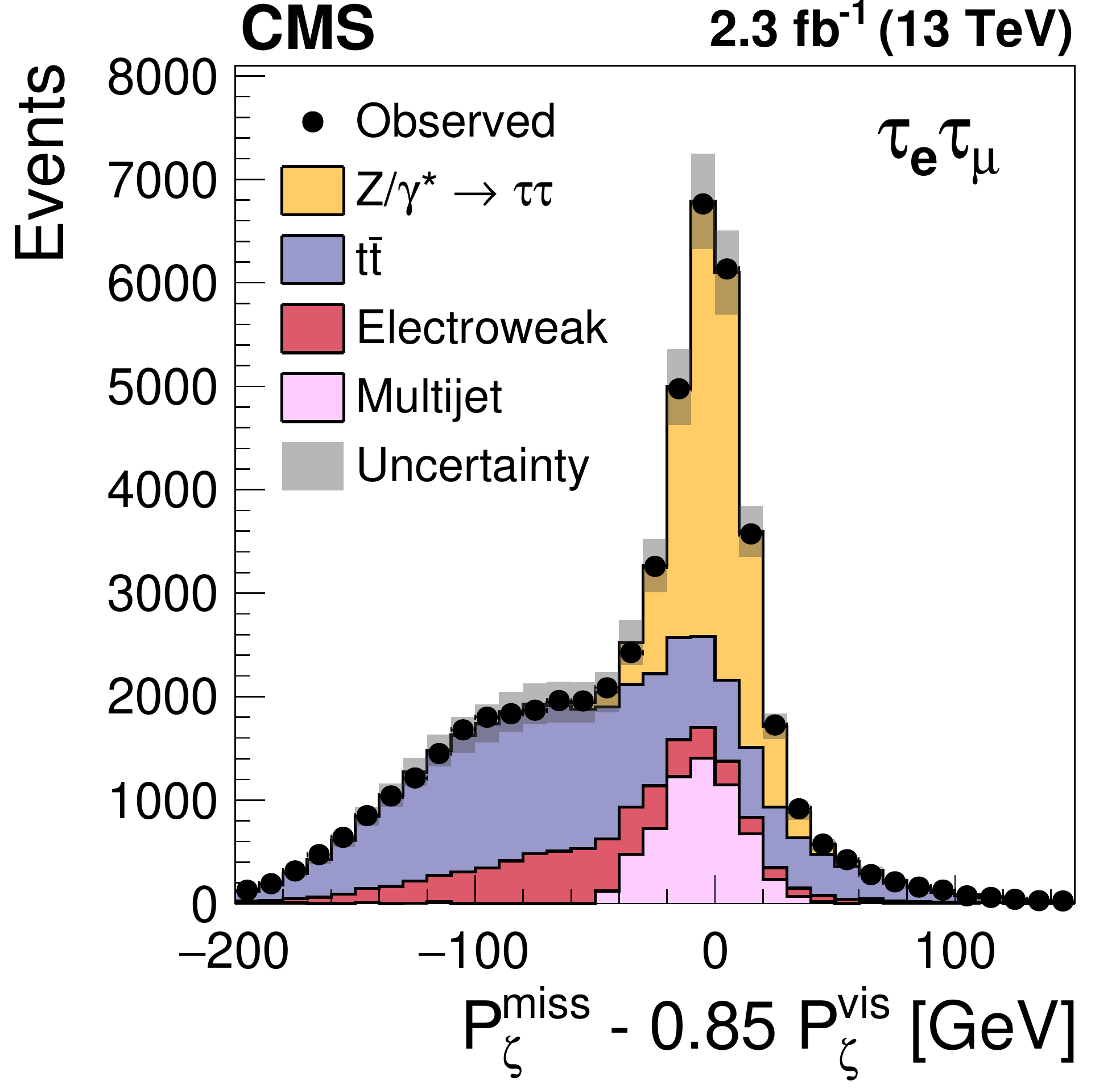}

\caption{
(Left) Construction of the projections $\Pzetamiss$ and $\Pzetavis$,
and (right) the distribution in the observable $\Pzetamiss - 0.85 \, \Pzetavis$
for events selected in the $\taue\taum$ channel, before imposing the condition $\Pzetamiss - 0.85 \, \Pzetavis > -20\GeV$.
Also indicated is the separation of the background into its main components.
The sum of background contributions from $\PW$+jets, single top quark, and diboson production is referred to as ``electroweak'' background.
The symbols $\pTvec^{\ \Pnu(\Pe)}$ and $\pTvec^{\ \Pnu(\Pgm)}$ refer to the vectorial sum of transverse momenta
of the two neutrinos produced in the respective $\Pgt \to \Pe\Pnu\Pnu$ and $\Pgt \to \Pgm\Pnu\Pnu$ decays.
}
\label{fig:Pzeta}
\end{figure*}

Events in the $\taue\taum$ channel are recorded with the triggers based on the presence of an electron and a muon.
The acceptance for the $\cPZ/\Pggx \to \Pgt\Pgt$ signal is increased by using two complementary triggers.
The first trigger selects events that contain an electron with $\pT > 12\GeV$ and a muon with $\pT > 17\GeV$,
while events containing an electron with $\pT > 17\GeV$ and a muon with $\pT > 8\GeV$ are recorded through the second trigger.
The offline event selection demands the presence of an electron with $\pT > 13\GeV$ and $\vert \eta \vert < 2.5$,
in conjunction with a muon of $\pT > 10\GeV$ and $\lvert \eta \rvert < 2.4$.
Either the electron or the muon is required to pass a threshold of $\pT > 18\GeV$,
to ensure that at least one of the two triggers is fully efficient.
Electrons and muons are further required to satisfy isolation criteria of $I_{\ell} < 0.15 \, \pTell$, and to have opposite charge.
Background from $\cPqt\cPaqt$ production is reduced through a cutoff on a topological discriminant~\cite{CDFrefPzeta} based on the projections:
\begin{linenomath}
\begin{equation}
\Pzetamiss = \vecMET \cdot \hat{\zeta}
\qquad \mbox{and} \qquad
\Pzetavis = \left( \pTvece + \pTvecmu \right) \cdot \hat{\zeta} \, ,
\label{eq:defPzeta}
\end{equation}
\end{linenomath}
where the symbol $\hat{\zeta}$ denotes a unit vector in the direction of the bisector of the electron and muon $\pTvec$ vectors.
The discriminator takes advantage of the fact that the angle between the neutrinos and the visible $\Pgt$ lepton decay products is typically small,
causing the $\vecMET$ vector in signal events to point in the direction of the visible $\Pgt$ decay products,
which is often not true for $\cPqt\cPaqt$ background.
Selected events are required to satisfy the condition $\Pzetamiss - 0.85 \, \Pzetavis > -20\GeV$.
The reconstruction of the projections $\Pzetamiss$ and $\Pzetavis$ is illustrated in Fig.~\ref{fig:Pzeta}.
The figure also shows the distribution in the observable $\Pzetamiss - 0.85 \, \Pzetavis$ for events selected in the $\taue\taum$ channel
before that condition is applied.

The events selected in the $\taum\taum$ channel are recorded using a single-muon trigger with a $\pT$ threshold of $18\GeV$.
The two muons are required to be within the acceptance of $\vert \eta \vert < 2.4$, and to have opposite charge.
The muons of higher and lower $\pT$ are required to satisfy the conditions of $\pT > 20$ and $> 10\GeV$, respectively.
Both muons are required to pass an isolation criterion of $I_{\Pgm} < 0.15 \, \pT^{\,\Pgm}$.
The large background arising from DY production of $\Pgm$ pairs is reduced by requiring the mass of the two muons to satisfy $m_{\Pgm\Pgm} < 80\GeV$,
and through the application of a cutoff on the output of a BDT trained to separate the $\cPZ/\Pggx \to \Pgt\Pgt$ signal from the $\cPZ/\Pggx \to \Pgm\Pgm$ background.
The following observables are used as BDT inputs:
the ratio of the $\pT$ of the dimuon system to the scalar $\pT$ sum of the two muons ($\pT^{\,\Pgm\Pgm} / \sum \pT^{\,\Pgm}$),
the pseudorapidity of the dimuon system ($\eta_{\Pgm\Pgm}$),
the $\MET$,
the topological discriminant $P_{\zeta}$, computed according to Eq.~(\ref{eq:defPzeta}),
and the azimuthal angle between the muon of positive charge and the $\vecMET$ vector, denoted by the symbol $\Delta\phi(\Pgm^{+}, \vecMET)$.
The angle between the muon of negative charge and the $\vecMET$ vector, $\Delta\phi(\Pgm^{-}, \vecMET)$,
is not used as BDT input, as it is strongly anticorrelated with $\Delta\phi(\Pgm^{+}, \vecMET)$.

We refer to the events passing the selection criteria detailed in this Section
as belonging to the ``signal region'' (SR) of the analysis.

\section{Background estimation}
\label{sec:backgroundEstimation}

{\tolerance=600
The accuracy of the background estimate is improved by determining from data
the contributions from the main backgrounds, as well as from backgrounds that are difficult to model through MC simulation.
In particular, the background from multijet production falls into the latter category.
In the $\taue\tauh$, $\taum\tauh$, and $\tauh\tauh$ channels, the dominant background
is from events in which a quark or gluon jet is misidentified as $\tauh$.
The estimation of background from these ``false'' $\tauh$ sources is discussed in Section~\ref{sec:backgroundEstimation_Fakes_etau(mu)tau_tautau}.
It predominantly arises from multijet production in the $\tauh\tauh$ channel
and from $\PW$+jets events, as well as from multijet production in the $\taue\tauh$ and $\taum\tauh$ channels.
A small additional background contribution in the $\taue\tauh$, $\taum\tauh$, and $\tauh\tauh$ channels
arises from $\cPqt\cPaqt$ events with quark or gluon jets misidentified as $\tauh$.
The multijet background is also relevant in the $\taue\taum$ and $\taum\taum$ channels.
The estimation of the multijet background in these channels is described in Section~\ref{sec:backgroundEstimation_QCD_emu_mumu}.
The contribution to the SR from the $\taue\taum$ and $\taum\taum$ channels
arising from backgrounds with misidentified leptons other than multijet production is small and
not distinguished from background contributions with genuine leptons.
Significant background contributions arise from $\cPqt\cPaqt$ production in the $\taue\taum$ channel
and from the DY production of muon pairs in the $\taum\taum$ channel.
The normalization of the $\cPqt\cPaqt$ background in the $\taue\taum$ and $\taum\taum$ channels is determined from data,
using a control region that contains events with one electron, one muon, and one or more $\Pbottom$-tagged jets.
Details of the procedure are given in Section~\ref{sec:backgroundEstimation_TTbar}.
The $\cPqt\cPaqt$ normalization factor obtained from this control region is also applied to the $\cPqt\cPaqt$ background events
selected in the $\taue\tauh$, $\taum\tauh$, and $\tauh\tauh$ channels,
in which the reconstructed $\tauh$ is either due to a genuine $\tauh$ or due to the misidentification of an electron or muon.
The background rate from $\cPZ/\Pggx \to \Pe\Pe$ and $\cPZ/\Pggx \to \Pgm\Pgm$ production is determined from the data
through a maximum-likelihood (ML) fit of the $m_{\Pgt\Pgt}$ distributions in the SR,
described in Section~\ref{sec:signalExtraction}.
The contributions of minor backgrounds from single top quark and diboson production,
as well as a small contribution from $\PW$+jets background in the $\taue\taum$ and $\taum\taum$ channels,
are obtained from MC simulation.
The sum of these minor backgrounds is referred to as ``electroweak'' background.
A Higgs boson with a mass of $m_{\PHiggs} = 125\GeV$, produced at the rate and with branching fractions predicted in the SM, is considered as background.
Nevertheless, this contribution is found to be negligible.
\par}

The background estimates are summarized in Table~\ref{tab:eventYieldsPrefit}.
The quoted uncertainties represent the quadratic sum of statistical and systematic sources.

\begin{table*}[ht!]
\topcaption{
Expected number of background events in the $\taue\tauh$, $\taum\tauh$, $\tauh\tauh$, $\taue\taum$, and $\taum\taum$ channels
in data, corresponding to an integrated luminosity of 2.3\fbinv.
The uncertainties are rounded to two significant digits,
except when they are $< 10$, in which case they are rounded to one significant digit,
and the event yields are rounded to match the precision in the uncertainties.
}
\label{tab:eventYieldsPrefit}
\centering
\begin{tabular}{l|r@{$ \,\,\pm\,\, $}lr@{$ \,\,\pm\,\, $}lr@{$ \,\,\pm\,\, $}l}
Process & \multicolumn{2}{c}{$\taue\tauh$} & \multicolumn{2}{c}{$\taum\tauh$} & \multicolumn{2}{c}{$\tauh\tauh$} \\
\hline
Jets misidentified as $\tauh$                    & $5\,400$ & $880$ & $10\,200$ & $1\,300$ & $680$ & $210$ \\
$\cPqt\cPaqt$                                    & $365$ & $35$ & $651$ & $60$ & $19$ & $3$ \\
$\cPZ/\Pggx \to \Pe\Pe$, $\Pgm\Pgm$ ($\Pe$ or $\Pgm$ misidentified as $\tauh$) & $940$ & $250$ & $780$ & $210$ & \multicolumn{2}{c}{---} \\
Electroweak                                      & $96$ & $15$ & $185$ & $29$ & $43$ & $8$ \\
SM $\PHiggs$                                     & $48$ & $10$ & $100$ & $21$ & $13$ & $3$ \\
\hline
Total expected background                        & $6\,850$ & $910\phantom{\,0}$ & $11\,900$ & $1\,300\phantom{0}$ & $750$ & $210$ \\
\end{tabular}

\vspace*{0.3 cm}

\begin{tabular}{l|r@{$ \,\,\pm\,\, $}lr@{$ \,\,\pm\,\, $}l}
Process & \multicolumn{2}{c}{$\taue\taum$} & \multicolumn{2}{c}{$\taum\taum$} \\
\hline
Multijet                                         & $4\,530$ & $670\phantom{\,0}$ & $740$ & $140\phantom{\,0}$ \\
$\cPZ/\Pggx \to \Pgm\Pgm$                        & \multicolumn{2}{c}{---} & $7\,650$ & $300$ \\
$\cPqt\cPaqt$                                    & $3\,650$ & $310$ & $1\,370$ & $110$ \\
Electroweak                                      & $1\,180$ & $120$ & $312$ & $34$ \\
SM $\PHiggs$                                     & $57$ & $12$ & $18$ & $4$ \\
\hline
Total expected background                        & $9\,400$ & $760$ & $10\,100$ & $390$ \\
\end{tabular}

\end{table*}

In preparation for future analyses of $\Pgt$ lepton production,
the validity of the background-estimation procedures described in this section
is further tested in event categories that are relevant to the SM $\PHiggs \to \Pgt\Pgt$ analysis, as well as in searches for new physical phenomena.
Event categories based on jet multiplicity,
$\pT$ of the $\Pgt$ lepton pair, and on the multiplicity of $\Pbottom$ jets in the event
are used in $\PHiggs \to \Pgt\Pgt$ analyses performed by CMS in the context of the SM~\cite{HIG-13-004}
and of its minimal supersymmetric extension~\cite{HIG-10-002,HIG-11-029,HIG-13-021},
as well as in the context of searches for Higgs boson pair production~\cite{HIG-14-034}.
The validation of the background-estimation procedures in these event categories is detailed in the Appendix.

\subsection{Estimation of \texorpdfstring{false-$\tauh$}{false-tau(h)} background in \texorpdfstring{$\taue\tauh$}{tau(e)-tau(h)}, \texorpdfstring{$\taum\tauh$}{tau(mu)-tau-h}, and \texorpdfstring{$\tauh\tauh$}{tau(h)-tau(h)} channels}
\label{sec:backgroundEstimation_Fakes_etau(mu)tau_tautau}

The background arising from events in which a quark or gluon jet is misidentified as $\tauh$
in the $\taue\tauh$, $\taum\tauh$, and $\tauh\tauh$ channels
is estimated via the ``fake factor'' ($\FF$) method.
The method is based on selecting events that pass altered $\tauh$ ID criteria,
and weighting the events through suitably chosen extrapolation factors (the $\FF$).
The events passing the altered $\tauh$ ID criteria are referred to as belonging to the ``application region'' (AR) of the $\FF$ method.
Except for modifying the $\tauh$ ID criteria, the same selections are applied to events in the AR and in the SR.
The $\FF$ are measured in dedicated control regions in data.
These are referred to as ``determination regions'' (DR) of the $\FF$ method,
and are chosen such that they neither overlap with the SR nor with the AR.

The $\FF$ are determined in bins of decay mode and $\pT$ of the $\tauh$ candidate, and as a function of jet multiplicity.
In each such bin, the $\FF$ is given by the ratio:
\begin{linenomath}
\begin{equation}
\FF = \frac{N_{\textrm{nominal}}}{N_{\textrm{altered}}} \, ,
\label{eq:jetToTauFakeRate}
\end{equation}
\end{linenomath}
where $N_{\textrm{nominal}}$ corresponds to the number of events with $\tauh$ candidates that pass the nominal WP of the MVA-based $\tauh$ ID discriminant in a given channel,
and $N_{\textrm{altered}}$ is the number of events with $\tauh$ candidates that satisfy the altered $\tauh$ ID criteria.
To satisfy the altered $\tauh$ ID criteria, $\tauh$ candidates must satisfy the barely constrained WP, but fail the nominal WP.
The multiplicity of jets that is used to parametrize the $\FF$ is denoted by $N_{\jet}$, and is defined by the jets that satisfy the conditions $\pT > 20\GeV$ and $\vert \eta \vert < 4.7$,
and do not overlap with $\tauh$ candidates passing the barely constrained WP of the MVA-based $\tauh$ ID discriminant,
nor with electrons or muons within $\Delta R < 0.5$.
In each bin, the contribution from processes with genuine $\tauh$, and with electrons or muons misidentified as $\tauh$,
are estimated through MC simulation, and subtracted from the numerator as well as from the denominator in Eq.~(\ref{eq:jetToTauFakeRate}).

\begin{figure*}[h]
\centering
\includegraphics[height=120mm]{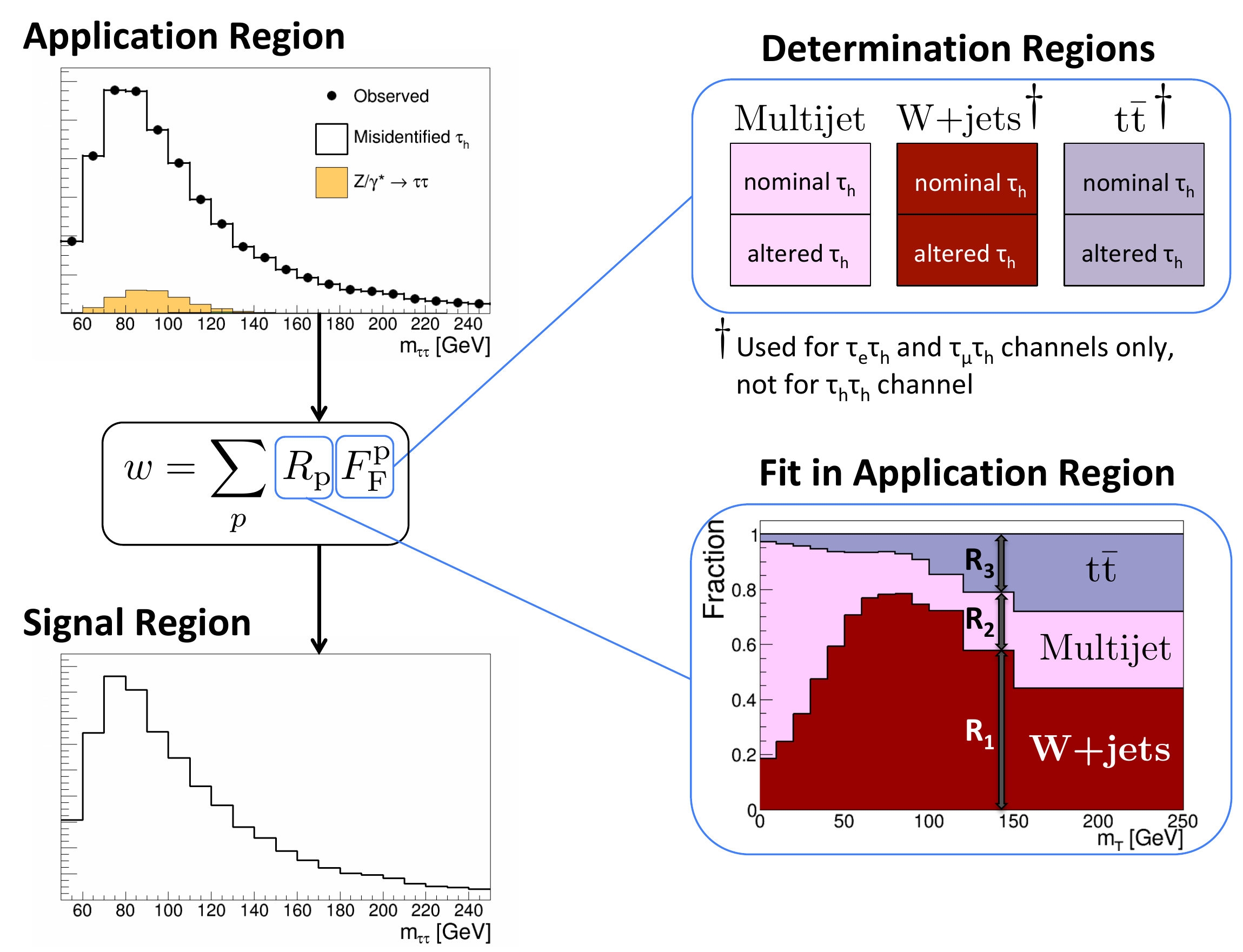}

\caption{
Schematic illustration of the $\FF$ method,
used to estimate the false-$\tauh$ background in the $\taue\tauh$, $\taum\tauh$, and $\tauh\tauh$ channels.
An event sample enriched in multijet, $\PW$+jets, and $\cPqt\cPaqt$ backgrounds is selected in the AR (top left).
The weights $w$, given by the product of the $\FF$ measured in the DR (top right) and the relative fractions $R_{\textrm{p}}$
of different background processes $\textrm{p}$ in the AR, are applied to the events selected in the AR to yield the estimate of the false-$\tauh$ background in the SR (bottom left).
The superscript $\textrm{p}$ on the symbol $\FF^{\textrm{p}}$ indicates that the $\FF$ depend on the background process $\textrm{p}$,
where $\textrm{p}$ refers to either multijet, $\PW$+jets, or $\cPqt\cPaqt$ background.
The contribution of the $\PZ/\gamma^{*} \to \tau\tau$ signal in the AR is subtracted, based on MC simulation.
The fractions $R_{\textrm{p}}$ are determined by a fit of the $\mT$ distribution in the AR (bottom right),
described in more detail in Section~\ref{sec:FF_Rp}.
The fraction $R_{1}$ includes a small contribution from DY events in which the reconstructed $\tauh$
is due to the misidentification of a quark or a gluon jet.
}
\label{fig:FFmethod_illustration}
\end{figure*}

As the probabilities for jets to be misidentified as $\tauh$ depend on the $\tauh$ ID criteria, and the latter differ in different channels,
the $\FF$ are measured separately in each one of them.
Moreover, the misidentification rates differ for multijet, $\PW$+jets, and $\cPqt\cPaqt$ events,
necessitating a measurement of the $\FF$ in the DR enriched in contributions from multijet, $\PW$+jets, and $\cPqt\cPaqt$ backgrounds.
The relative fractions of multijet, $\PW$+jets, and $\cPqt\cPaqt$ background processes in the AR, denoted by $R_{\textrm{p}}$,
are determined through a fit to the distribution in $\mT$,
and are used to weight the $\FF$ determined in the DR when computing the estimate of the false-$\tauh$ background in the SR.
The procedure is illustrated in Fig.~\ref{fig:FFmethod_illustration}.

The $\tauh$ ID criteria applied in the AR are identical to the $\tauh$ ID criteria applied in the denominator of Eq.~(\ref{eq:jetToTauFakeRate}).
More specifically, the criteria on $\pT$ and $\eta$, as well as the requirements on the discriminators that distinguish $\tauh$ from electrons and muons, are the same as in the SR.
The $\tauh$ candidates selected in the $\taue\tauh$ and $\taum\tauh$ channels are required to pass the barely constrained,
but fail the moderately constrained WP of the MVA-based $\tauh$ ID discriminant.
In the $\tauh\tauh$ channel, one of the two $\tauh$ candidates must pass the tight WP,
while the other $\tauh$ candidate is required to pass the barely constrained, but fail the tight WP,
precluding overlap of the AR with the SR.

\begin{figure*}[ht]
\centering
\includegraphics[width=\cmsFigWidth]{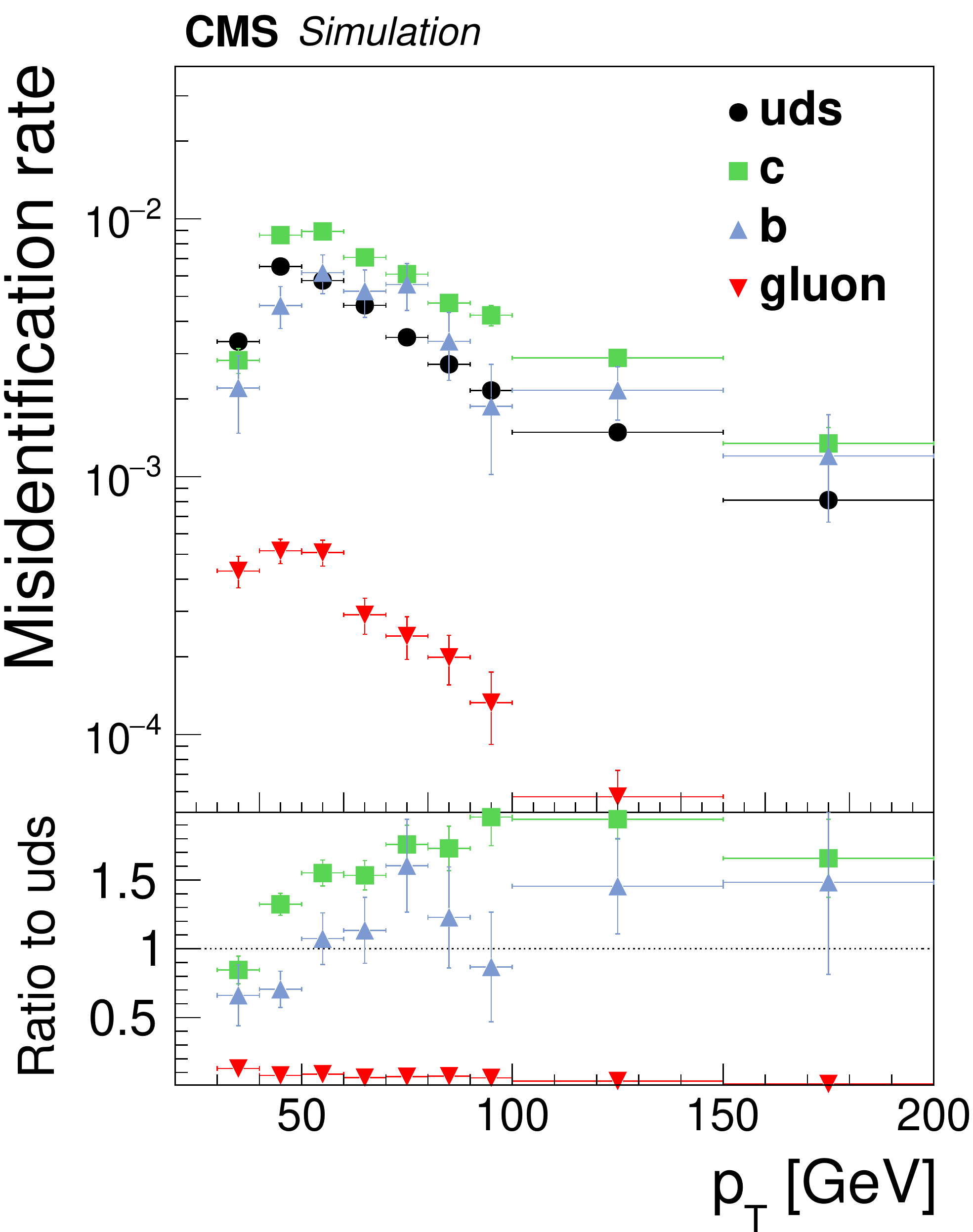} \hfil
\includegraphics[width=\cmsFigWidth]{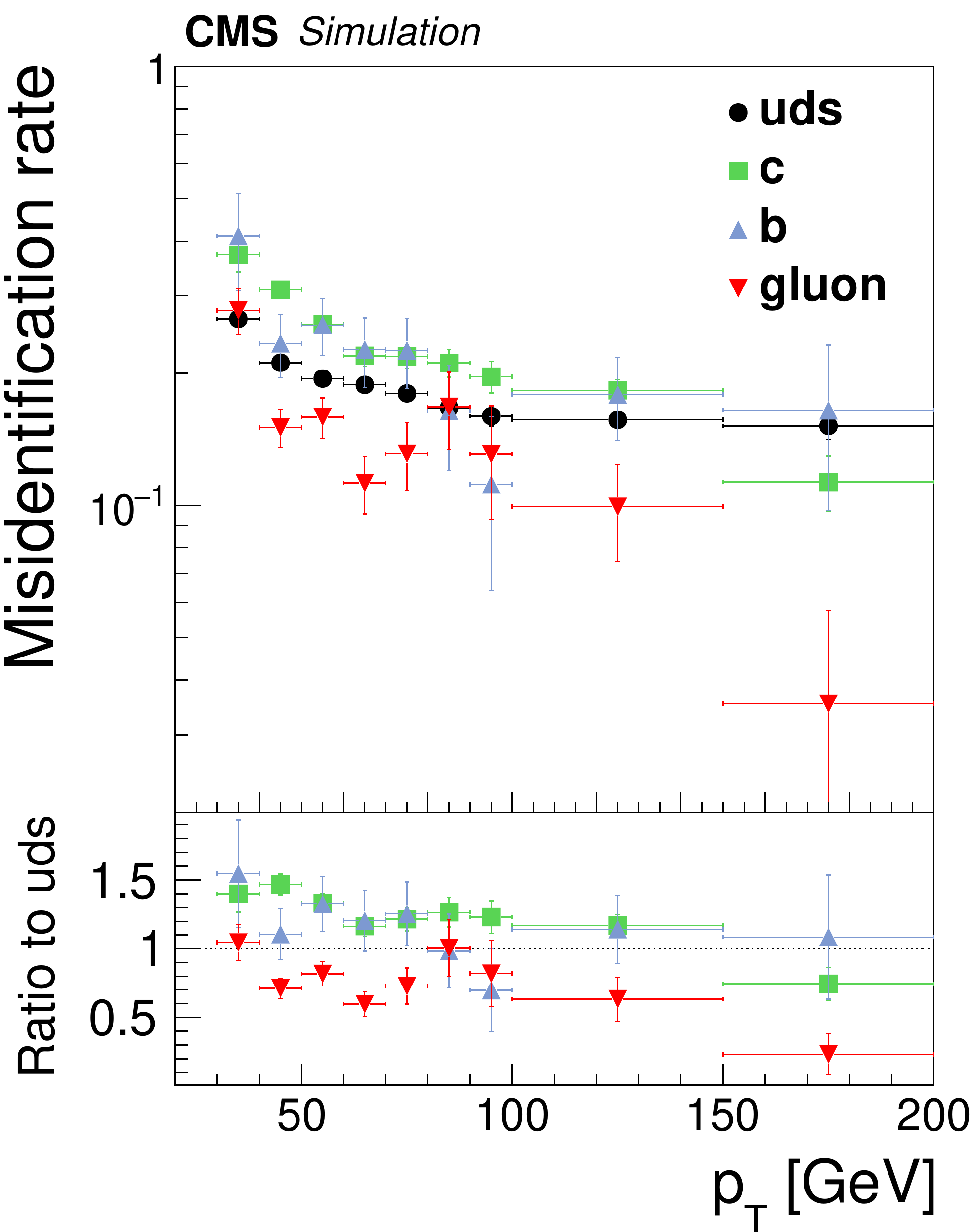}

\caption{
Probabilities for gluon and quark jets, of different flavour in simulated multijet events,
to pass the moderate WP of the MVA-based $\tauh$ ID discriminant, as a function of jet $\pT$,
for jets passing $\pT > 20\GeV$ and $\vert \eta \vert < 2.3$ (left), and for jets passing in addition the
barely constrained WP of the MVA-based $\tauh$ ID discriminant (right).
}
\label{fig:tauFakeRatesMC_vs_flavor}
\end{figure*}

{\tolerance=2400
The DR enriched in contributions from multijet, $\PW$+jets, and $\cPqt\cPaqt$ backgrounds
contain specific mixtures of gluon, light-quark ($\cPqu$, $\cPqd$, $\cPqs$),
and heavy-flavour ($\cPqc$, $\cPqb$) quark jets,
with different probabilities for misidentification as $\tauh$,
as illustrated for simulated events in Fig.~\ref{fig:tauFakeRatesMC_vs_flavor}.
The misidentification rates are shown for jets passing $\pT > 20\GeV$ and $\vert \eta \vert < 2.3$,
and for jets satisfying in addition the barely constrained WP of the MVA-based $\tauh$ ID discriminant.
In general, the misidentification rates are higher in quark jets compared to gluon jets,
as the former typically have lower particle multiplicity and are more collimated than the latter,
thereby increasing their probability to be misidentified as $\tauh$.
As it can be seen in the figure, the requirement for jets to pass minimal $\tauh$ selection criteria significantly reduce the flavour dependence of the misidentification rates.
This in turn lowers
the systematic uncertainty that arises from the limited knowledge of the flavour composition in the AR.
Residual flavour dependence of the $\FF$ is taken into account
by measuring separate sets of $\FF$ in each DR,
and determining the relative fraction $R_{\textrm{p}}$ of multijet, $\PW$+jets, and $\cPqt\cPaqt$ backgrounds in the AR of the respective channel.
Given the $\FF$ and the fractions $R_{\textrm{p}}$,
the estimate of the background from misidentified $\tauh$ in the SR is obtained
by applying the weights
\begin{linenomath}
\begin{equation}
w = \sum_{\textrm{p}} \, R_{\textrm{p}} \, \FF^{\textrm{p}}
\label{eq:jetToTauFakeRate_weight}
\end{equation}
\end{linenomath}
to events selected in the AR,
where the sum extends over the above three background processes $\textrm{p}$.
The $\FF$ refer, as usual, to Eq.~(\ref{eq:jetToTauFakeRate}).
The symbol $\FF^{\textrm{p}}$ indicates that, in addition to their dependence on $\tauh$ decay mode, $\tauh$ candidate $\pT$, and jet multiplicity,
the $\FF$ depend on the background process $\textrm{p}$,
where the superscript $\textrm{p}$ refers to either multijet, $\PW$+jets, or $\cPqt\cPaqt$ background.
In the $\tauh\tauh$ channel, the $\FF^{\textrm{p}}$ is determined by the decay mode and $\pT$
of the $\tauh$ candidate that passes the barely constrained, but fails the tight WP of the MVA-based $\tauh$ ID discriminant.
The $\tauh$ candidate that passes the tight WP does not enter the computation of the weight $w$.
\par}

The underlying assumption in the $\FF$ method is that the ratio of the number of events
from background process $\textrm{p}$ in the SR to the number of events from the same background
in the AR is equal to the ratio $N_{\textrm{nominal}}/N_{\textrm{altered}}$ that is measured in the background-specific DR.

The measurement of the $\FF$ is detailed in Section~\ref{sec:FF_measurement},
while the fractions $R_{\textrm{p}}$ are discussed in Section~\ref{sec:FF_Rp}.
The estimate of the false-$\tauh$ background obtained from the $\FF$ method is validated in control regions devoid of $\cPZ/\Pggx \to \Pgt\Pgt$ signal.
The result of this validation is presented in Section~\ref{sec:FF_validation}.

\subsubsection{Measurement of  $\FF$}
\label{sec:FF_measurement}
The $\FF$ are measured in DR chosen such that one particular background process is enhanced in each DR.
The selection criteria applied in the DR are similar to those applied in the SR.
In the following, we describe only the differences relative to the SR.

In the $\taue\tauh$ and $\taum\tauh$ channels,
three different DR are used to measure the $\FF$ for multijet, $\PW$+jets, and $\cPqt\cPaqt$ backgrounds.
The DR dominated by multijet background contains events
in which the charges of the $\tauh$ candidate and of the light lepton candidates are the same,
and the electron or muon satisfies a modified isolation criterion of $0.05 < I_{\ell}/\pTell < 0.15$.
Depending on whether the $\tauh$ candidate passes or fails the moderate WP of the MVA-based $\tauh$ ID discriminant,
the event contributes either to the numerator or to the denominator of Eq.~(\ref{eq:jetToTauFakeRate}).
The DR dominated by $\PW$+jets background is defined by
modifying the requirement for the transverse mass of lepton and $\vecMET$ to $\mT > 70\GeV$.
The contamination arising from $\cPqt\cPaqt$ background is reduced by vetoing events containing jets that
pass the $\Pbottom$ tagging criteria described in Section~\ref{sec:eventReconstruction}.
A common $\cPqt\cPaqt$ DR is used for the $\taue\tauh$ and $\taum\tauh$ channels.
The events are required to contain an electron, a muon, at least one $\tauh$ candidate,
and pass triggers based on the presence of an electron or a muon.
The offline event selection demands that
the electron satisfy the conditions $\pT > 13\GeV$ and $\vert \eta \vert < 2.5$,
the muon $\pT > 10\GeV$ and $\lvert \eta \rvert < 2.4$,
and that both pass an isolation criterion of $I_{\ell} < 0.10 \, \pTell$.
The event is furthermore required to contain at least one jet that passes the $\Pbottom$ tagging criteria
described in Section~\ref{sec:eventReconstruction}.
In case events contain multiple $\tauh$ candidates, the candidate used for the $\FF$ measurement is chosen at random.

In the $\tauh\tauh$ channel, a single DR is used,
which selects a high purity sample of multijet events, the dominant background in this channel.
The multijet DR is identical to the SR of the $\tauh\tauh$ channel,
except that the two $\tauh$ candidates are required to have the same rather than opposite charge.
One of the jets is chosen to be the ``tag'' jet, and required to pass the tight WP of the MVA-based $\tauh$ ID discriminant,
while the measurement of the $\FF$ is performed on the other jet, referred to as the ``probe'' jet.
The tag jet is chosen at random.
The $\PW$+jets and $\cPqt\cPaqt$ backgrounds are small in the $\tauh\tauh$ channel,
making it difficult to define a DR that is dominated by these backgrounds,
or that provides sufficient statistical information for the $\FF$ measurement.
The $\FF$ in the multijet DR of the $\tauh\tauh$ channel are therefore used to weight all events selected in the AR of the $\tauh\tauh$ channel.
Differences in the $\FF$ between $\PW$+jets, $\cPqt\cPaqt$, and multijet events are accounted for by adding a systematic uncertainty of $30\%$
on the part of the background from misidentified $\tauh$ expected
from the contribution of $\PW$+jets and $\cPqt\cPaqt$ background processes.
This contribution is estimated using MC simulation,
and the magnitude of the systematic uncertainty is motivated by the difference found in the $\FF$ measured in multijet, $\PW$+jets, and $\cPqt\cPaqt$ DR
in the $\taue\tauh$ and $\taum\tauh$ channels.

\begin{figure*}
\centering
\includegraphics[width=\cmsFigWidth]{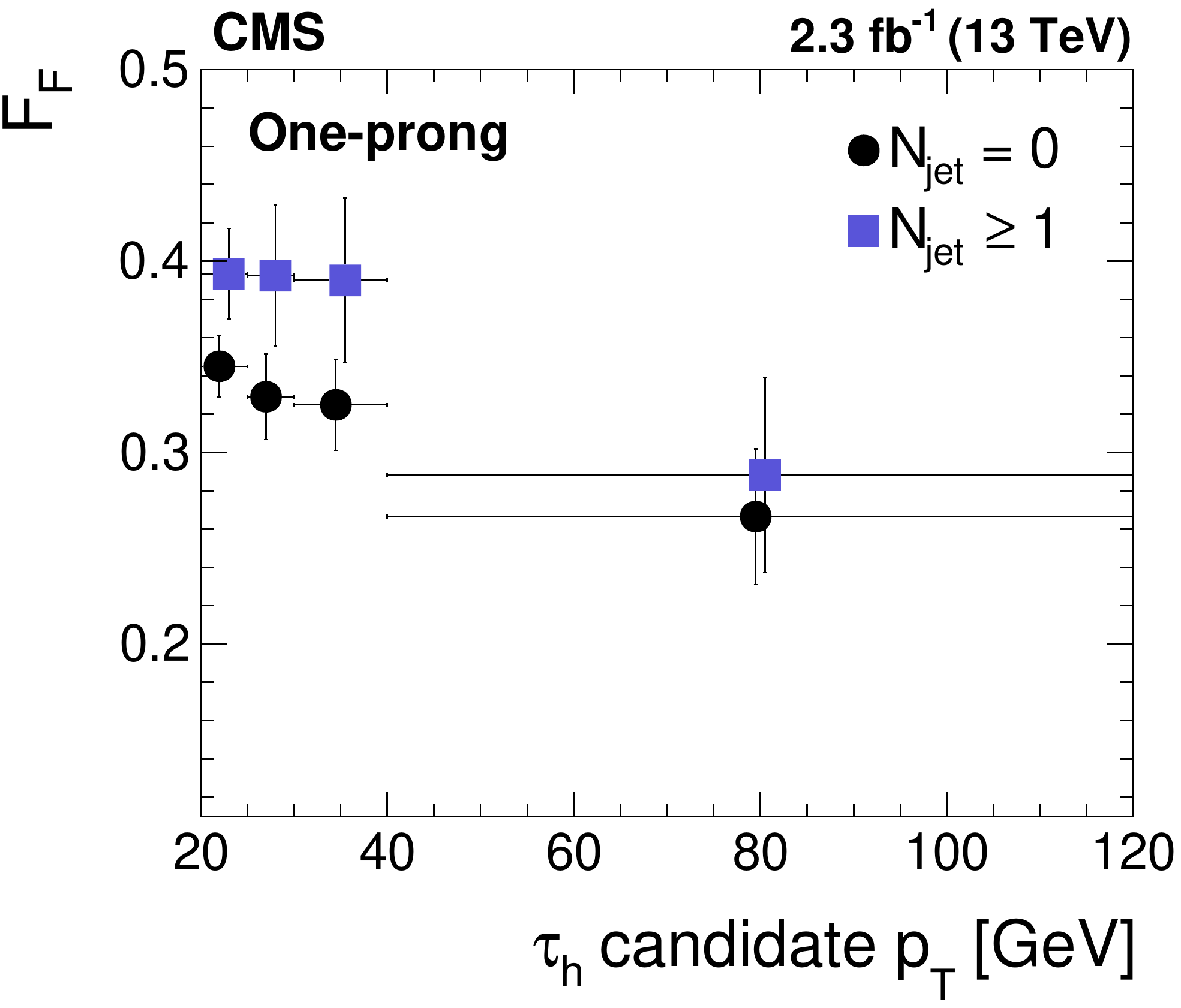} \hfil
\includegraphics[width=\cmsFigWidth]{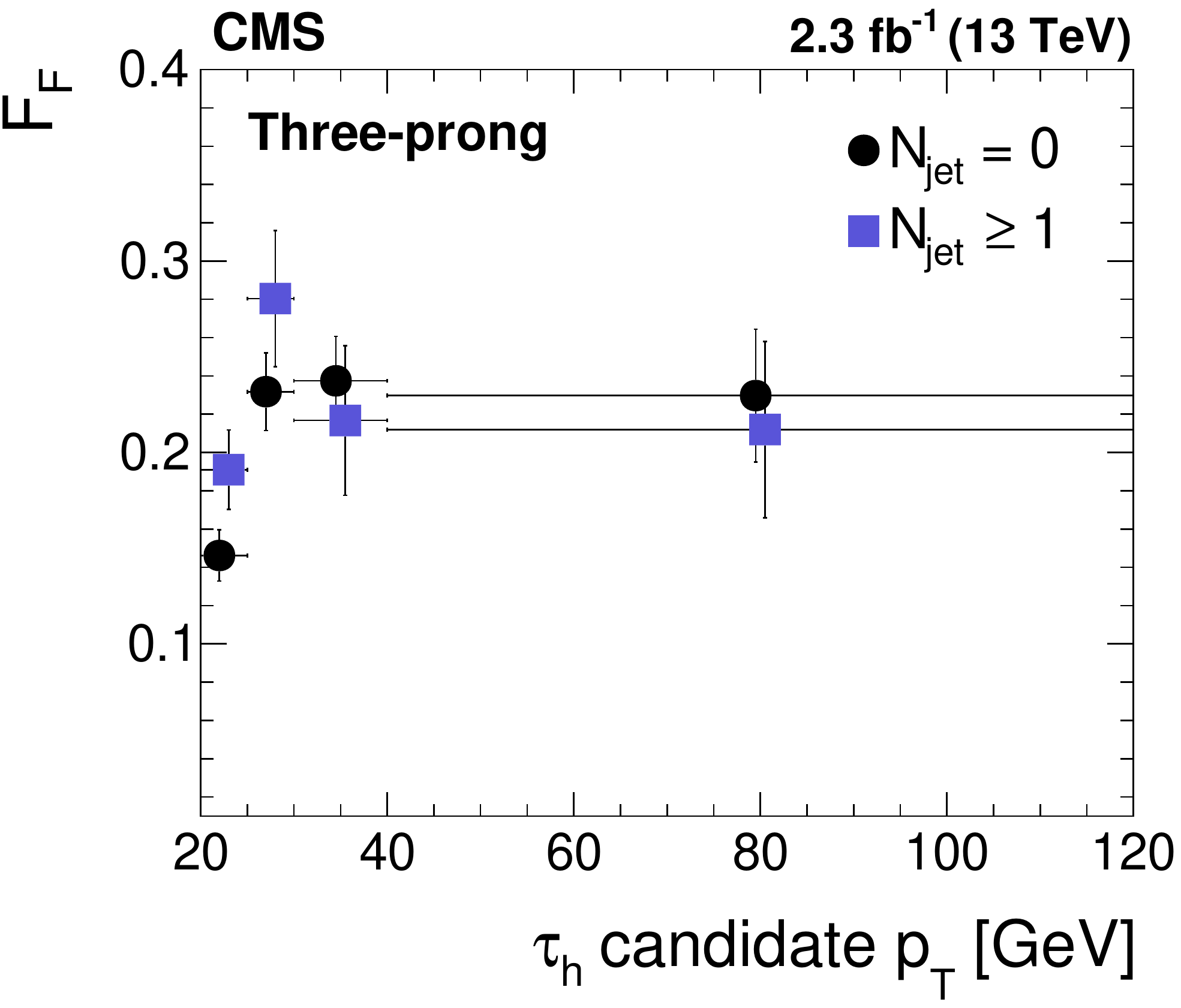} \\
\includegraphics[width=\cmsFigWidth]{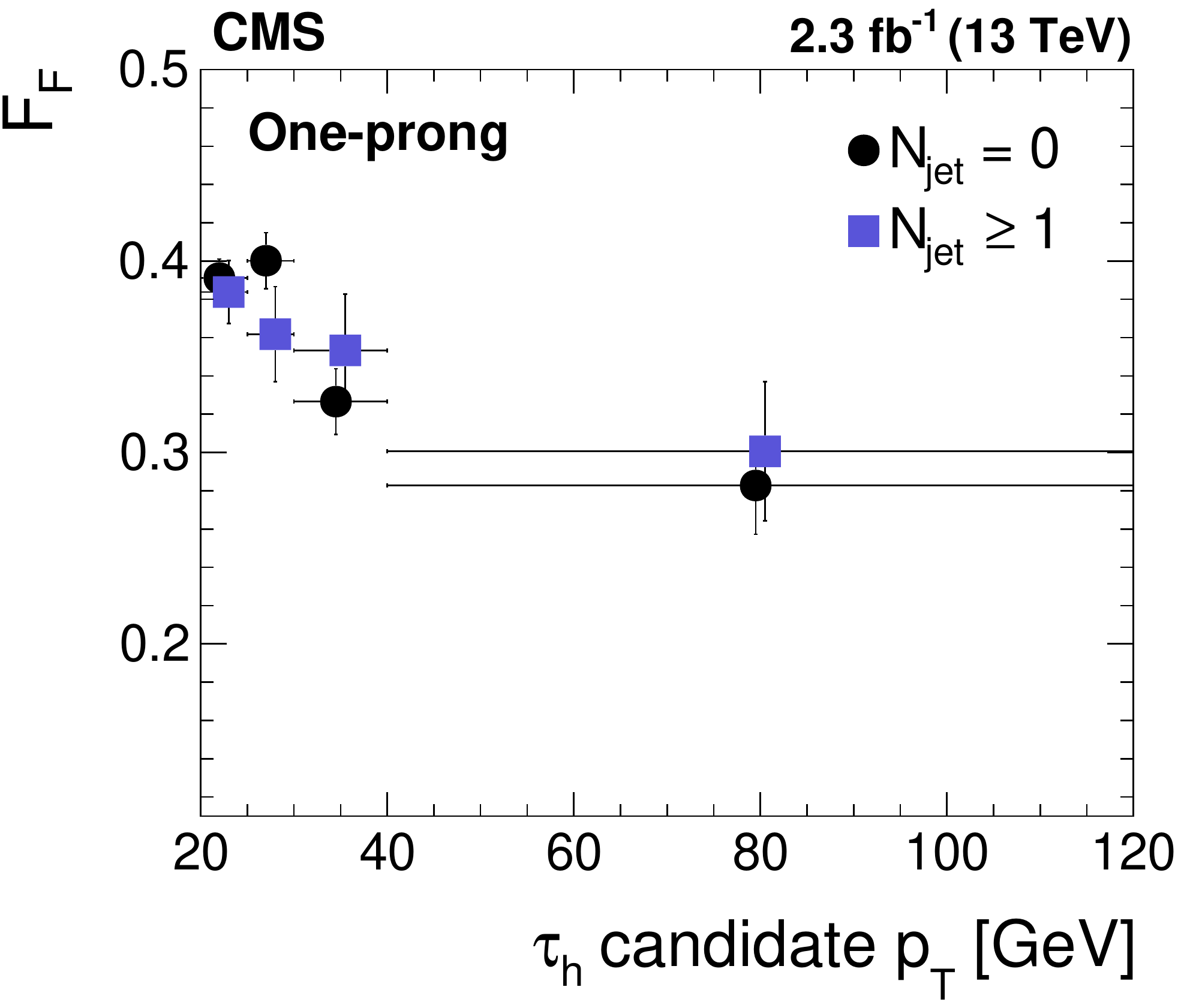} \hfil
\includegraphics[width=\cmsFigWidth]{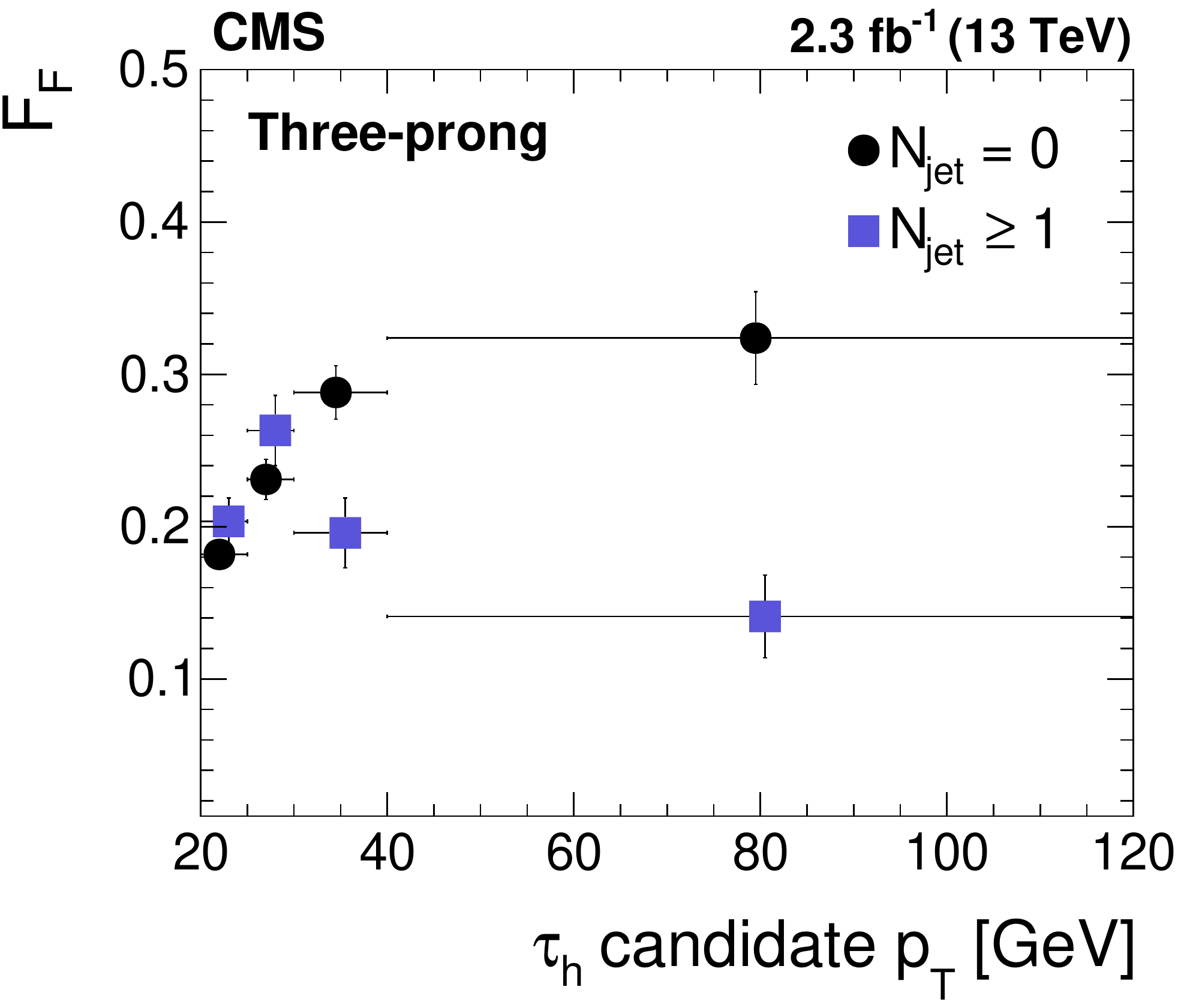} \\
\includegraphics[width=\cmsFigWidth]{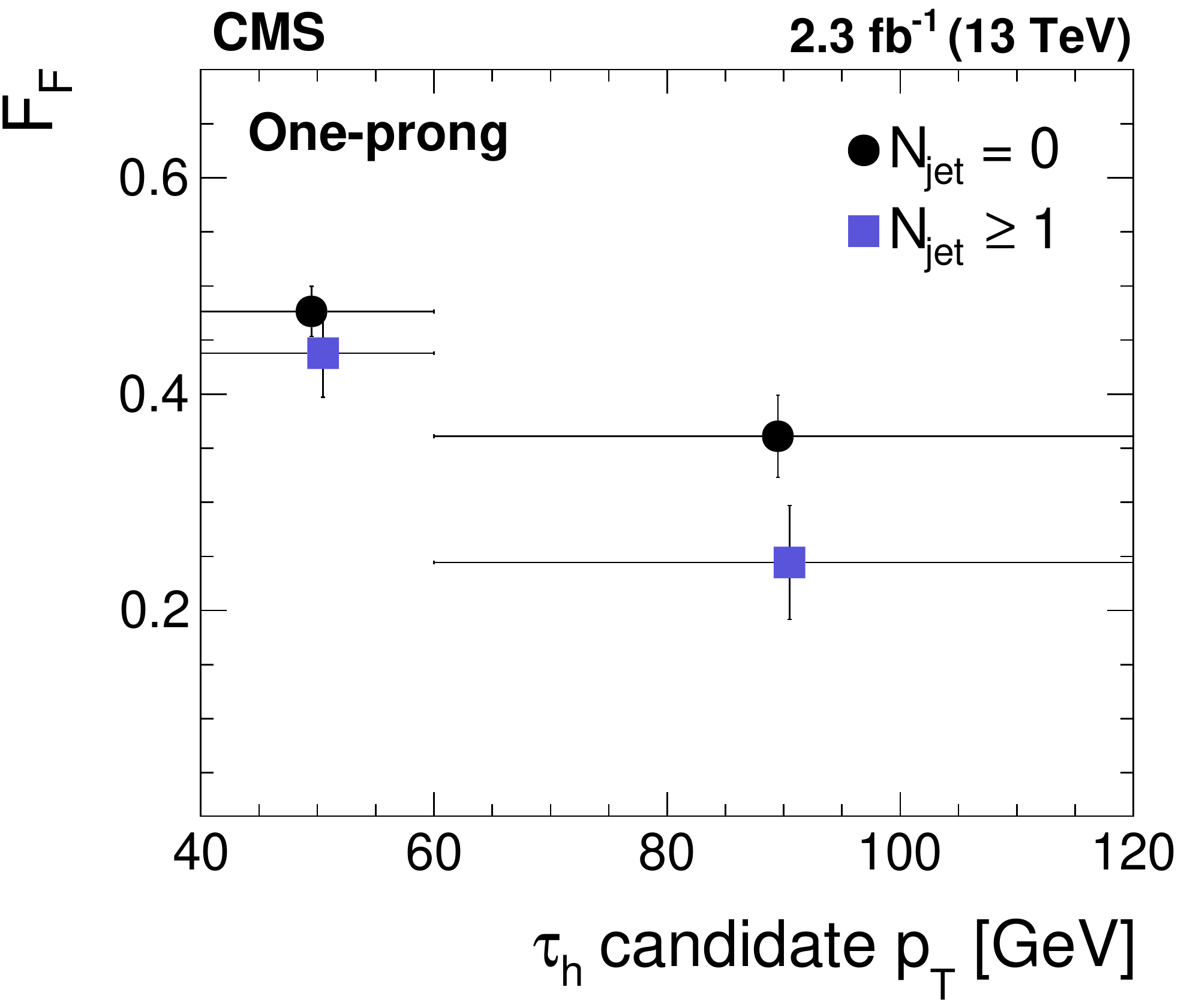} \hfil
\includegraphics[width=\cmsFigWidth]{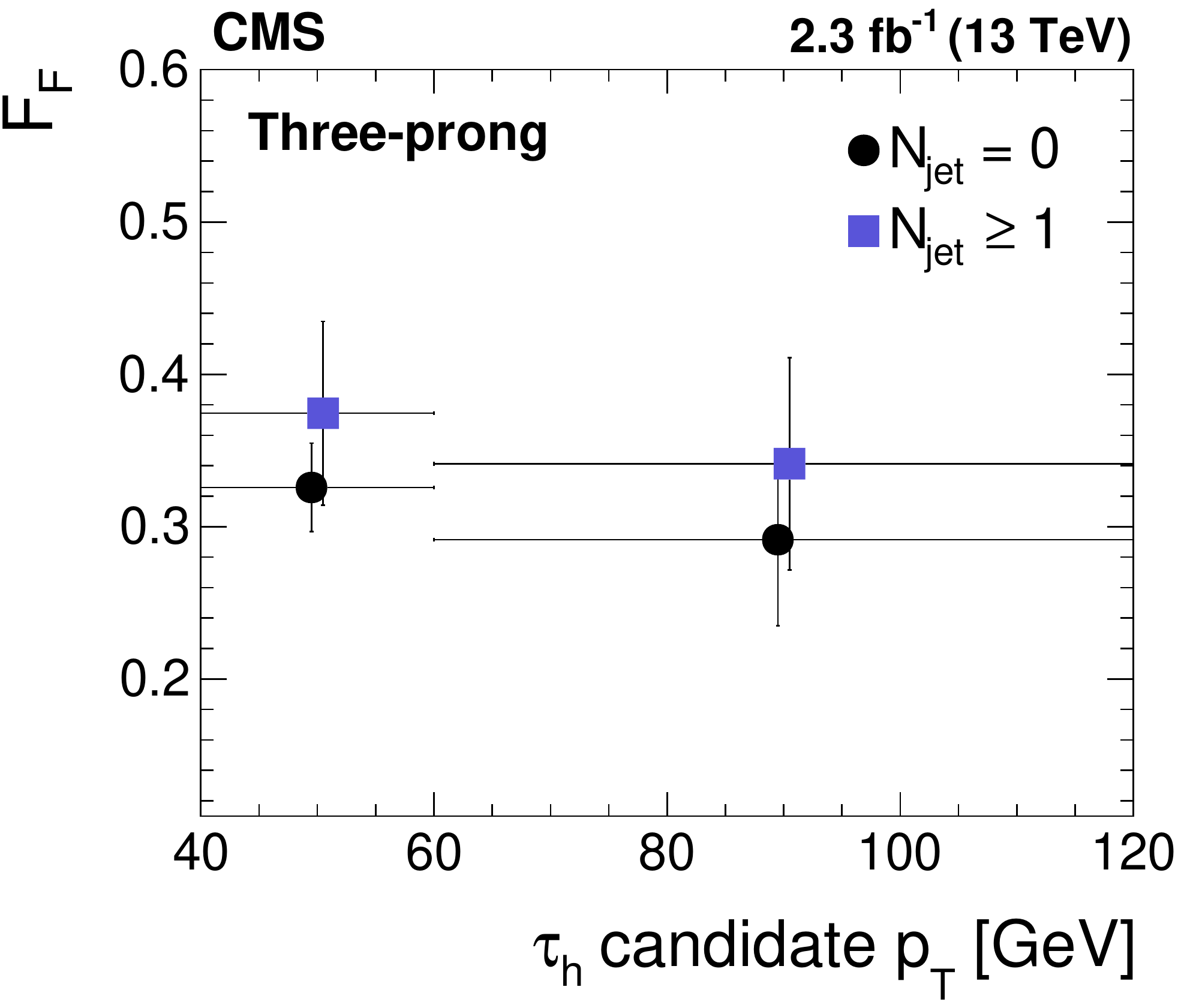}

\caption{
The $\FF$ values measured in multijet events
in the $\taue\tauh$ (upper), $\taum\tauh$ (center), and $\tauh\tauh$ (lower) channels,
presented in bins of jet multiplicity and $\tauh$ decay mode, as a function of $\tauh$ $\pT$.
The abscissae of the points are offset to distinguish the points with different jet multiplicities.
}
\label{fig:tauFakeRatesData1}
\end{figure*}

\begin{figure*}
\centering
\includegraphics[width=\cmsFigWidth]{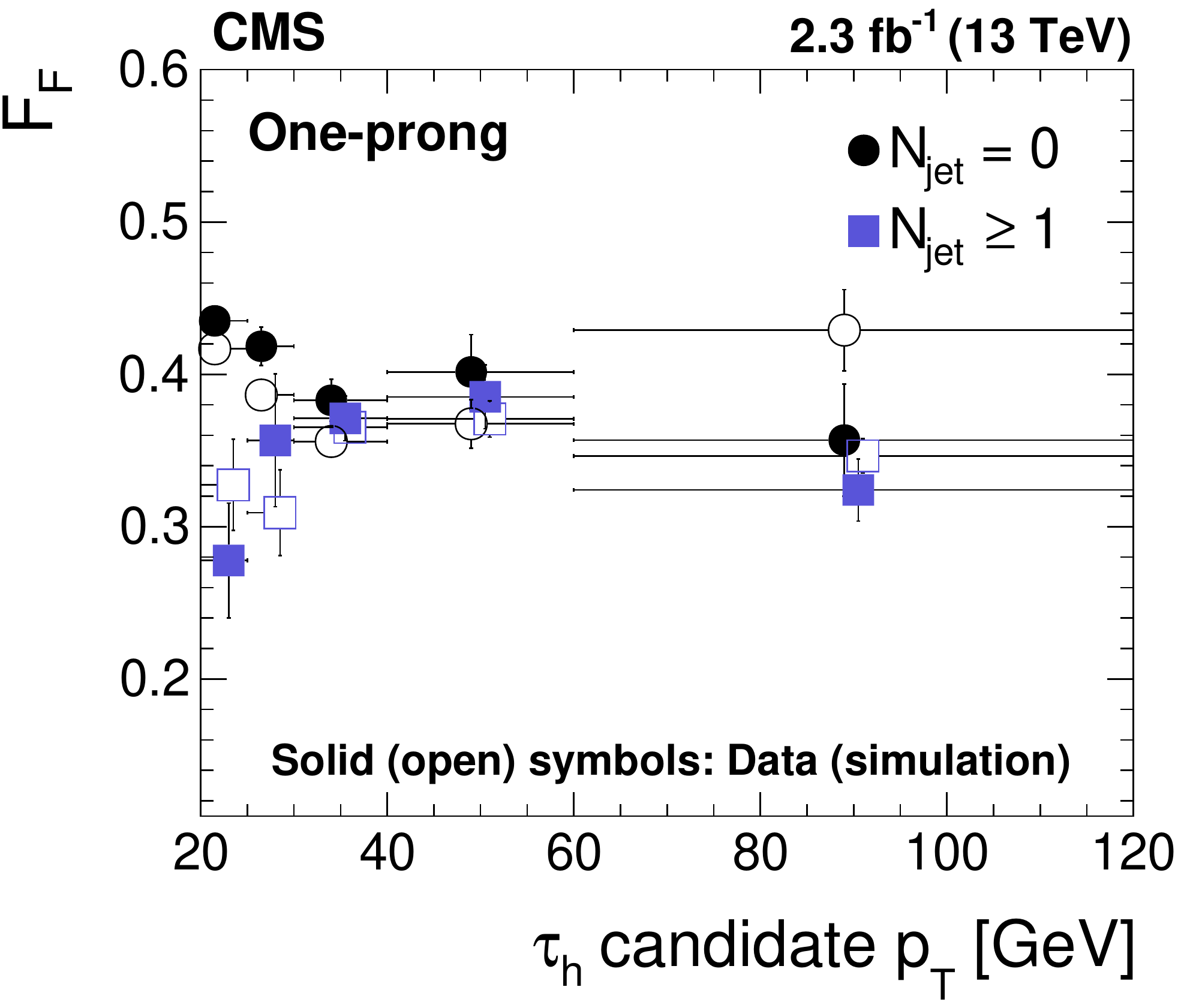} \hfil
\includegraphics[width=\cmsFigWidth]{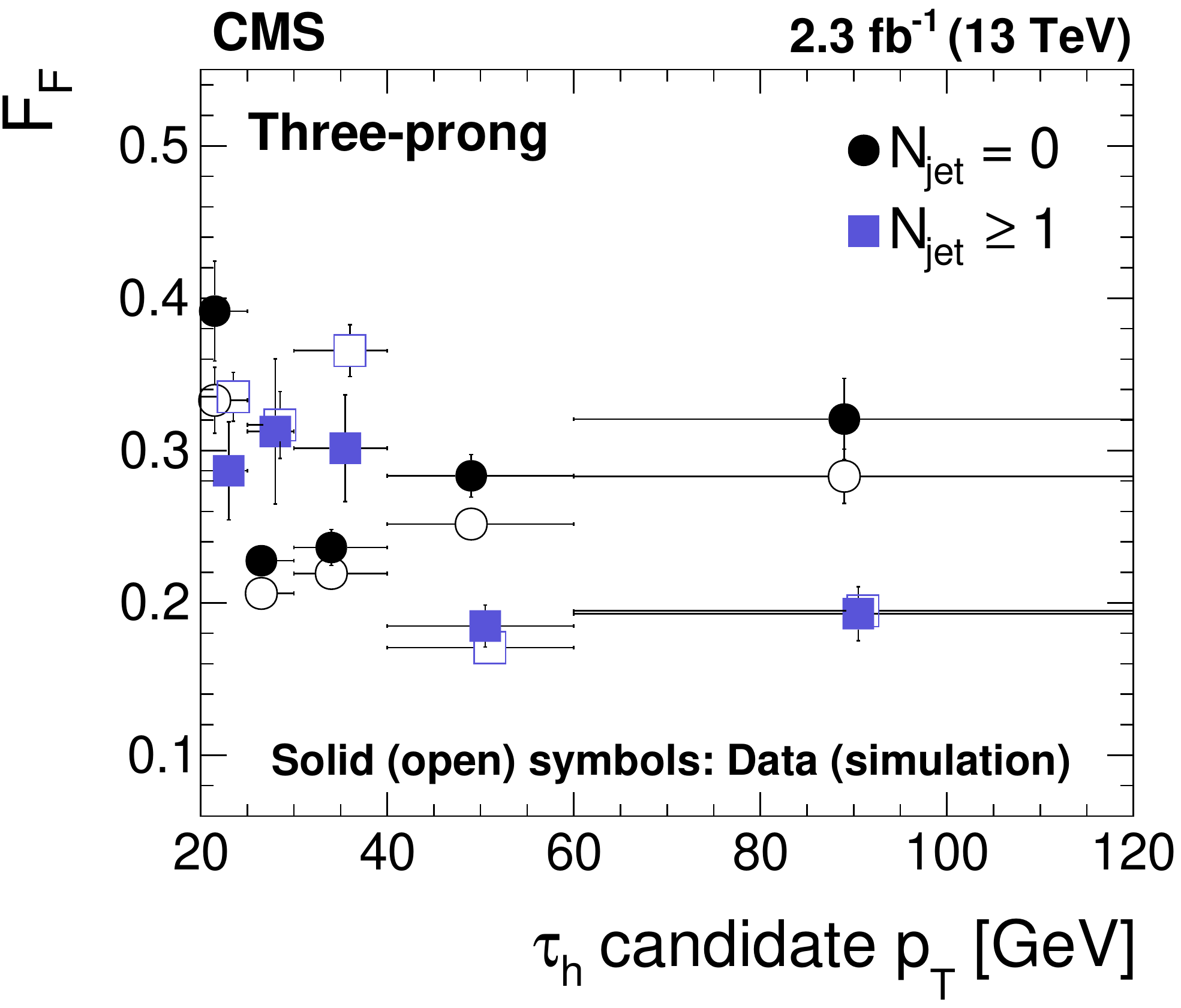} \\
\includegraphics[width=\cmsFigWidth]{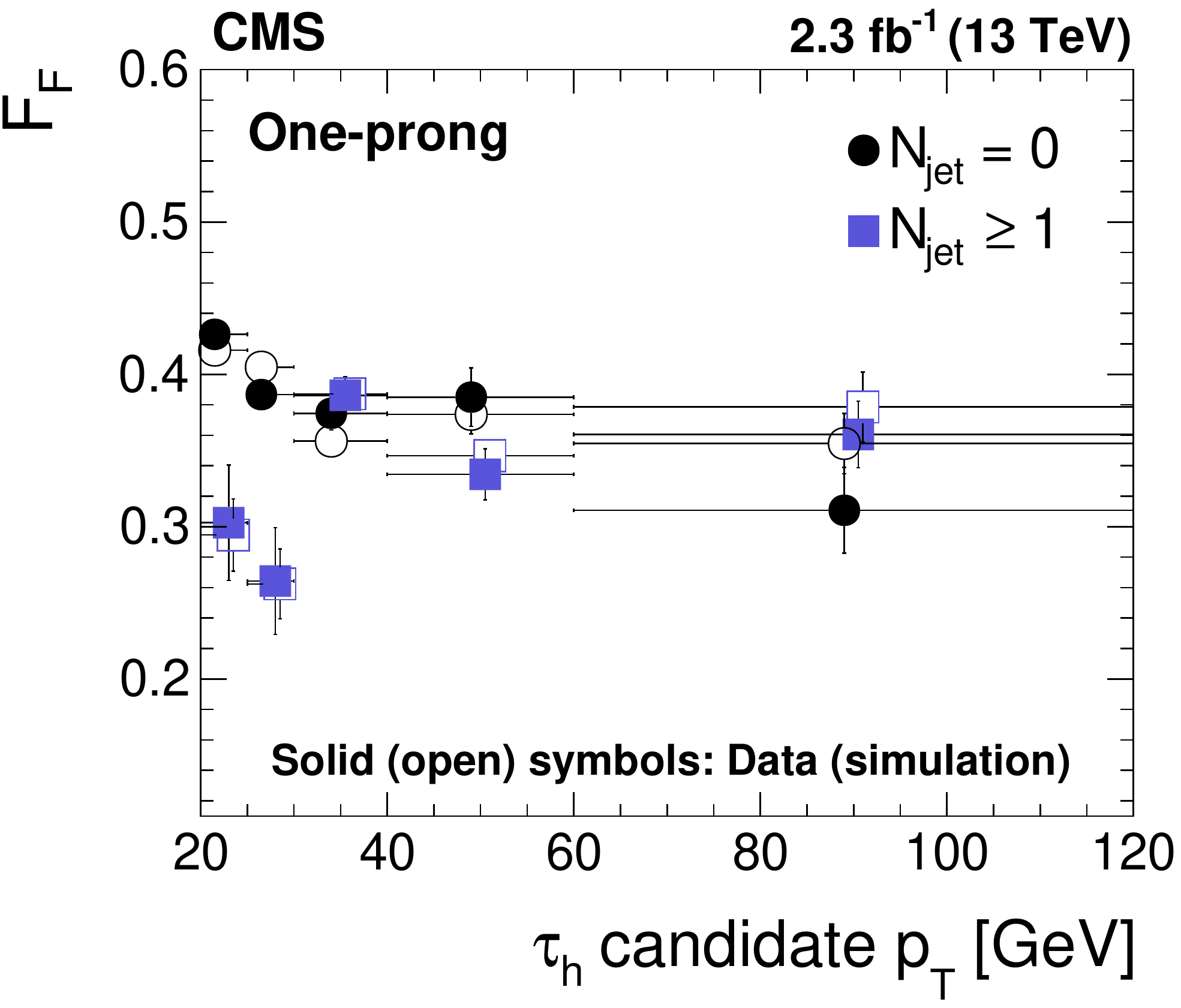} \hfil
\includegraphics[width=\cmsFigWidth]{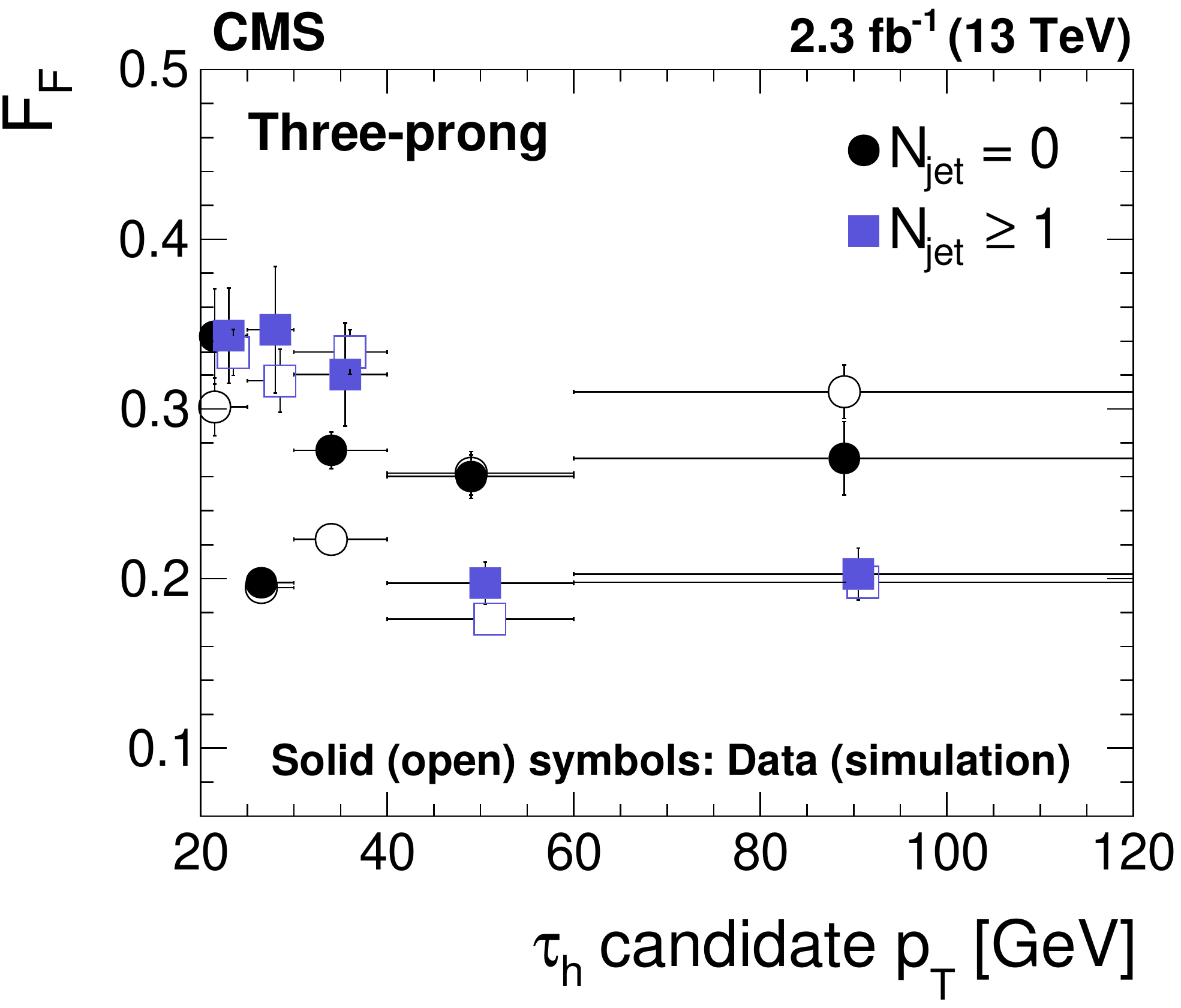} \\
\includegraphics[width=\cmsFigWidth]{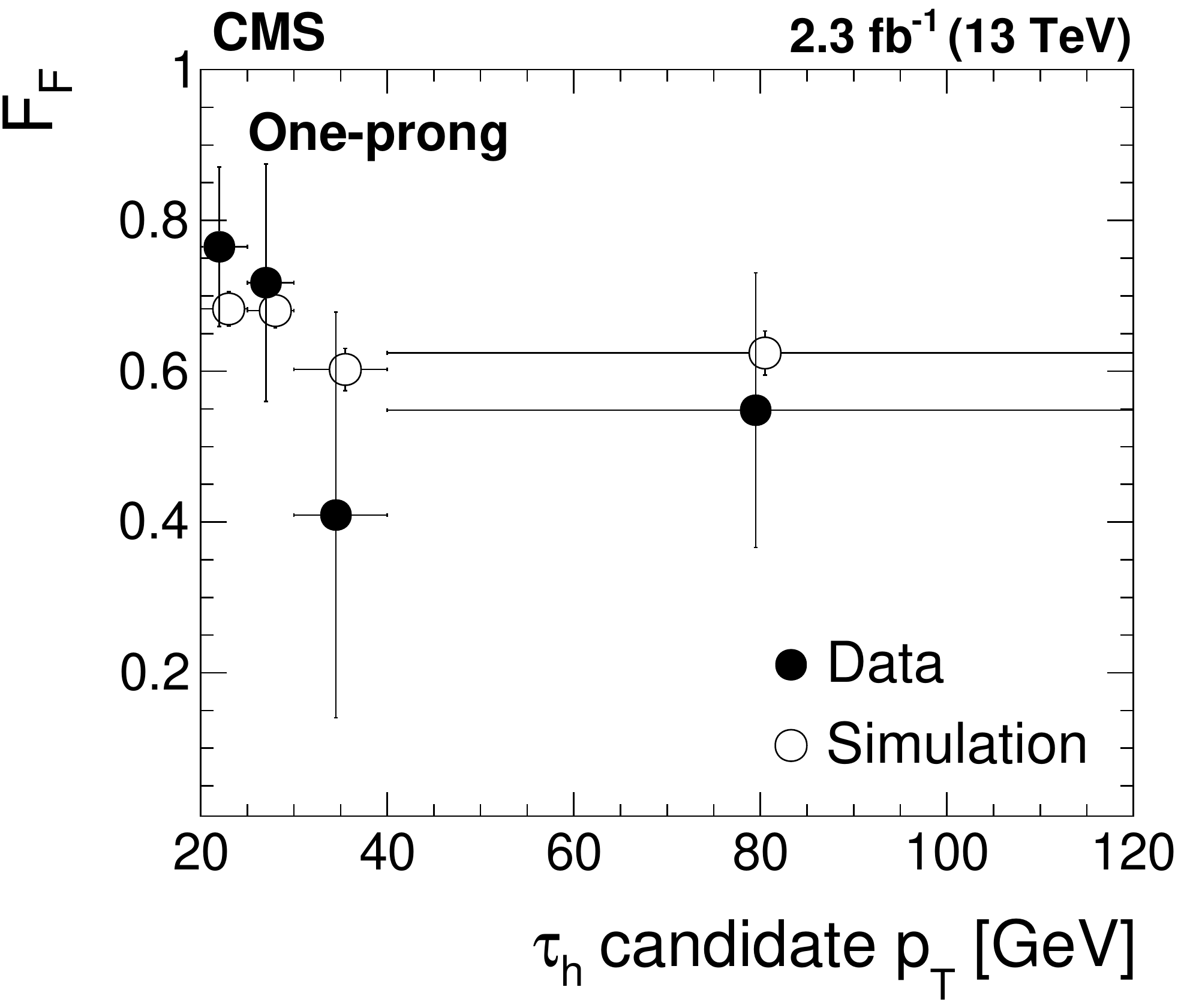} \hfil
\includegraphics[width=\cmsFigWidth]{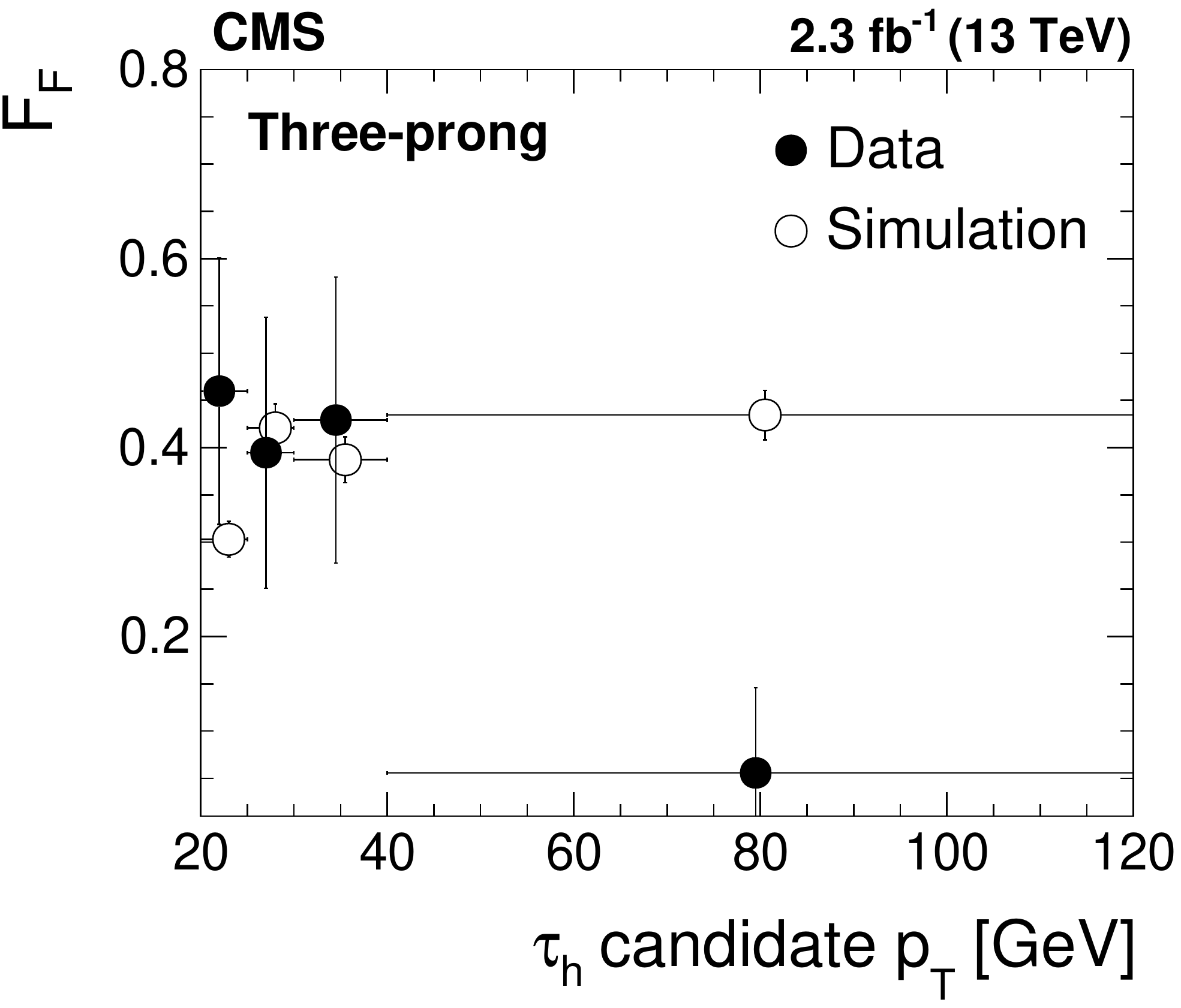}

\caption{
The $\FF$ values measured in $\PW$+jets events
in the $\taue\tauh$ (upper) and $\taum\tauh$ (center) channels and in $\cPqt\cPaqt$ events (lower),
presented in bins of jet multiplicity and $\tauh$ decay mode, as a function of $\tauh$ $\pT$.
A common $\cPqt\cPaqt$ DR is used for the $\taue\tauh$ and $\taum\tauh$ channels.
The abscissae of the points are offset to distinguish the points with different jet multiplicities.
}
\label{fig:tauFakeRatesData2}
\end{figure*}

The $\FF$ determined in the various DR are shown in Figs.~\ref{fig:tauFakeRatesData1} and~\ref{fig:tauFakeRatesData2}.
The decay modes $\Pgt^{-} \to \Phadron^{-}\Pnut$, $\Pgt^{-} \to \Phadron^{-}\Ppizero\Pnut$, and $\Pgt^{-} \to \Phadron^{-}\Ppizero\Ppizero\Pnut$
are referred to as ``one-prong'' decays and the mode $\Pgt^{-} \to \Phadron^{-}\Phadron^{+}\Phadron^{-}\Pnut$ as ``three-prong'' decays.
The measured $\FF$ are corrected
for differences in the $\tauh$ misidentification rates between DR and AR.
The magnitude of these relative corrections is ${\approx}10\%$,
as discussed below.

For the multijet DR in the $\taue\tauh$ and $\taum\tauh$ channels,
correlations between the $\FF$ and the charge of the electron or muon and the $\tauh$ candidate,
and between $\FF$ and the isolation of the electron or muon, are studied in data and taken into account as follows.
A correction for the extrapolation from events in which the charges of lepton and $\tauh$ candidate have the same sign (SS)
to events in which they have opposite sign (OS)
is obtained by comparing $\FF$ in the SS and OS events
containing electrons or muons that pass an inverted isolation criterion of $0.1 < I_{\ell}/\pTell < 0.2$.
The dependence of the $\FF$ on the isolation of the electron or muon is studied
using an event sample selected with no isolation condition applied to the lepton.
The results of this study are used to extrapolate the $\FF$ obtained in the multijet DR ($0.05 < I_{\ell}/\pTell < 0.15$)
to the SR ($I_{\ell} < 0.10 \, \pTell$).

For the DR dominated by $\PW$+jets background in the $\taue\tauh$ and $\taum\tauh$ channels,
closure tests of the $\FF$ method reveal a dependence of the $\FF$ on $\mT$,
which is not accounted for by the chosen parametrization of the $\FF$ as functions of jet multiplicity, $\tauh$ decay mode, and $\pT$.
The dependence on $\mT$ is studied using simulated $\PW$+jets events,
and used to extrapolate the $\FF$ measured in the $\PW$+jets DR ($\mT > 70\GeV$) to the SR ($\mT < 40\GeV$).

In the $\tauh\tauh$ channel,
the $\FF$ determined in the multijet DR are corrected for a dependence of the $\FF$
on the relative charge of the two $\tauh$ candidates.
This is studied in events in which the tag jet (the jet on which the FF measurement is not performed)
fails the tight WP of the MVA-based $\tauh$ ID discriminant.
The difference between the $\FF$ in OS and SS events defines this correction.

\subsubsection{Determination of $R_{\textrm{p}}$}
\label{sec:FF_Rp}
In the $\taue\tauh$ and $\taum\tauh$ channels,
the relative fractions $R_{\textrm{p}}$ of multijet, $\PW$+jets, and $\cPqt\cPaqt$ backgrounds in the AR are determined through
a fit to the distribution in $\mT$.
The distribution in $\mT$ (``template'') used to represent the multijet background in the fit
is obtained from a sample of events selected in data,
in which the $\tauh$ candidate and the electron or muon have same charge, and where at least one of the leptons satisfies
a modified isolation criterion of $0.05 < I_{\ell}/\pTell < 0.15$.
The contributions from other backgrounds to this control region are subtracted, based on MC simulation.
The distribution representing the other backgrounds in the fit are also taken from simulation.
The templates for $\cPqt\cPaqt$, diboson, and DY events are split into three components:
events in which the reconstructed $\tauh$ is due to a genuine $\tauh$,
events in which the $\tauh$ is due to the misidentification of an electron or muon,
and events in which a quark or gluon jet is misidentified as $\tauh$.
The normalization of each component is determined independently in the fit.
The relative fractions of the $\cPZ/\Pggx \to \Pgt\Pgt$ signal and all individual background processes are left unconstrained in the fit.
Finally, the fractions $R_{\textrm{p}}$ are parametrized as function of $\mT$ and are normalized
such that the contribution of all processes $\textrm{p}$ in which the reconstructed $\tauh$
is due to a misidentified jet sums to unity, $\sum_{\textrm{p}} \, R_{\textrm{p}} = 1$.

In the $\tauh\tauh$ channel,
the AR is dominated by multijet background.
The contributions from the $\cPZ/\Pggx \to \Pgt\Pgt$ signal and all background processes, except multijet production, are small
and taken from simulation.
The fraction of multijet background in the AR is determined by subtracting the sum of all processes modelled in the MC simulation
from the data in the AR, without performing a fit in this channel.

A small fraction of events in the AR of the $\taue\tauh$, $\taum\tauh$, and $\tauh\tauh$ channels
arises from DY events in which quark or gluon jets are misidentified as $\tauh$ candidates.
These events are treated as background and are included in the false-$\tauh$ estimate using the $\FF$ method.
As the analysed data do not provide a way of determining $\FF$ in DY events with sufficient statistical accuracy,
the $\FF$ measured in $\PW$+jets events are used instead
for the fraction of DY events with jets misidentified as $\tauh$ in the $\taue\tauh$ and $\taum\tauh$ channels.
The validity of this procedure is justified by studies of $\FF$ in simulated $\PW$+jets and DY events,
which indicate that the flavour composition of jets and the $\FF$ are very similar in these events.
In the $\tauh\tauh$ channel, the $\FF$ measured in multijet events are used and the systematic uncertainty
on the DY background with misidentified $\tauh$ is increased by $30\%$.

\subsubsection{Validation of the false-$\tauh$ background estimate in control regions}
\label{sec:FF_validation}
\begin{figure*}[p!]
\centering
\includegraphics[width=\cmsFigWidth]{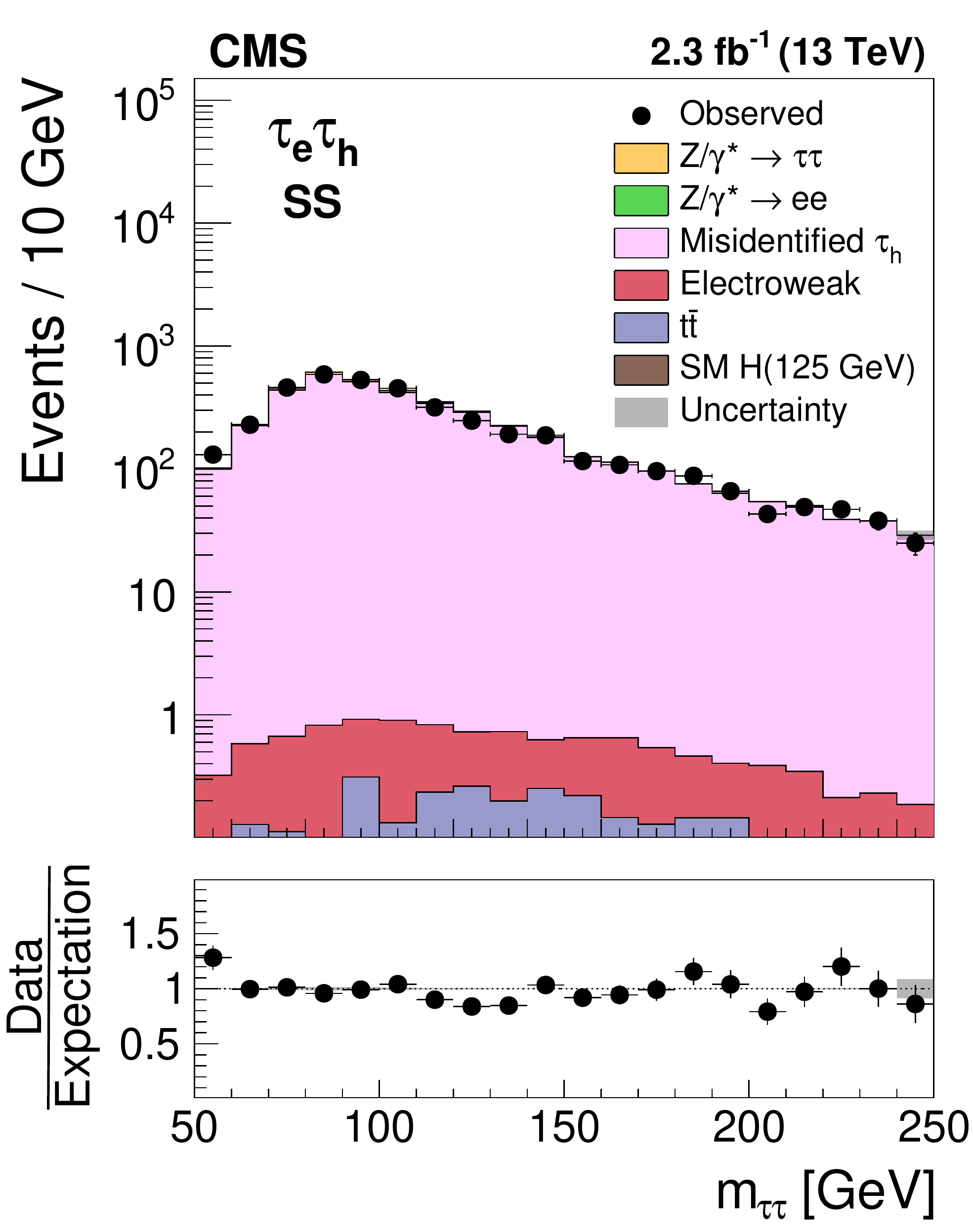} \hfil
\includegraphics[width=\cmsFigWidth]{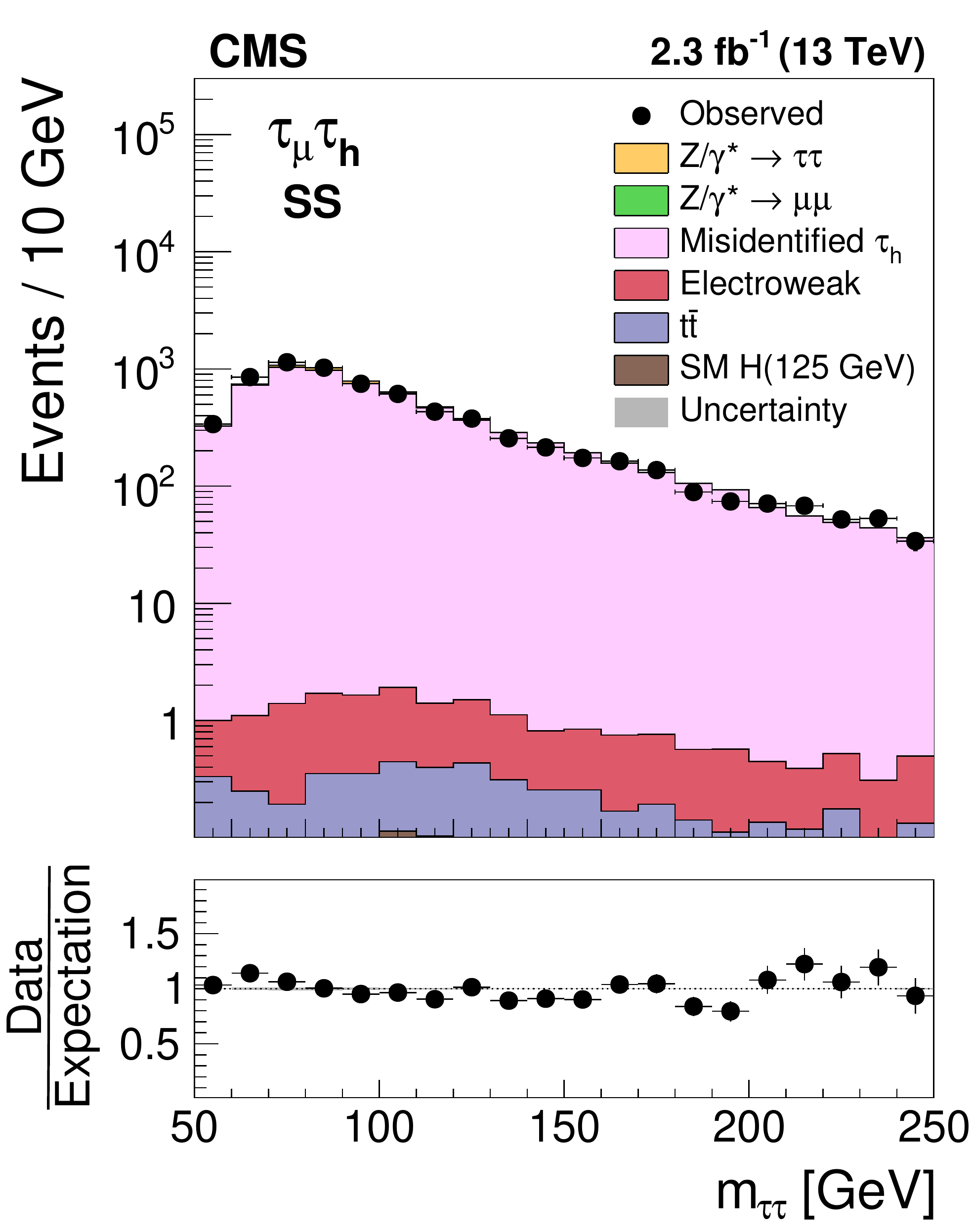} \\
\includegraphics[width=\cmsFigWidth]{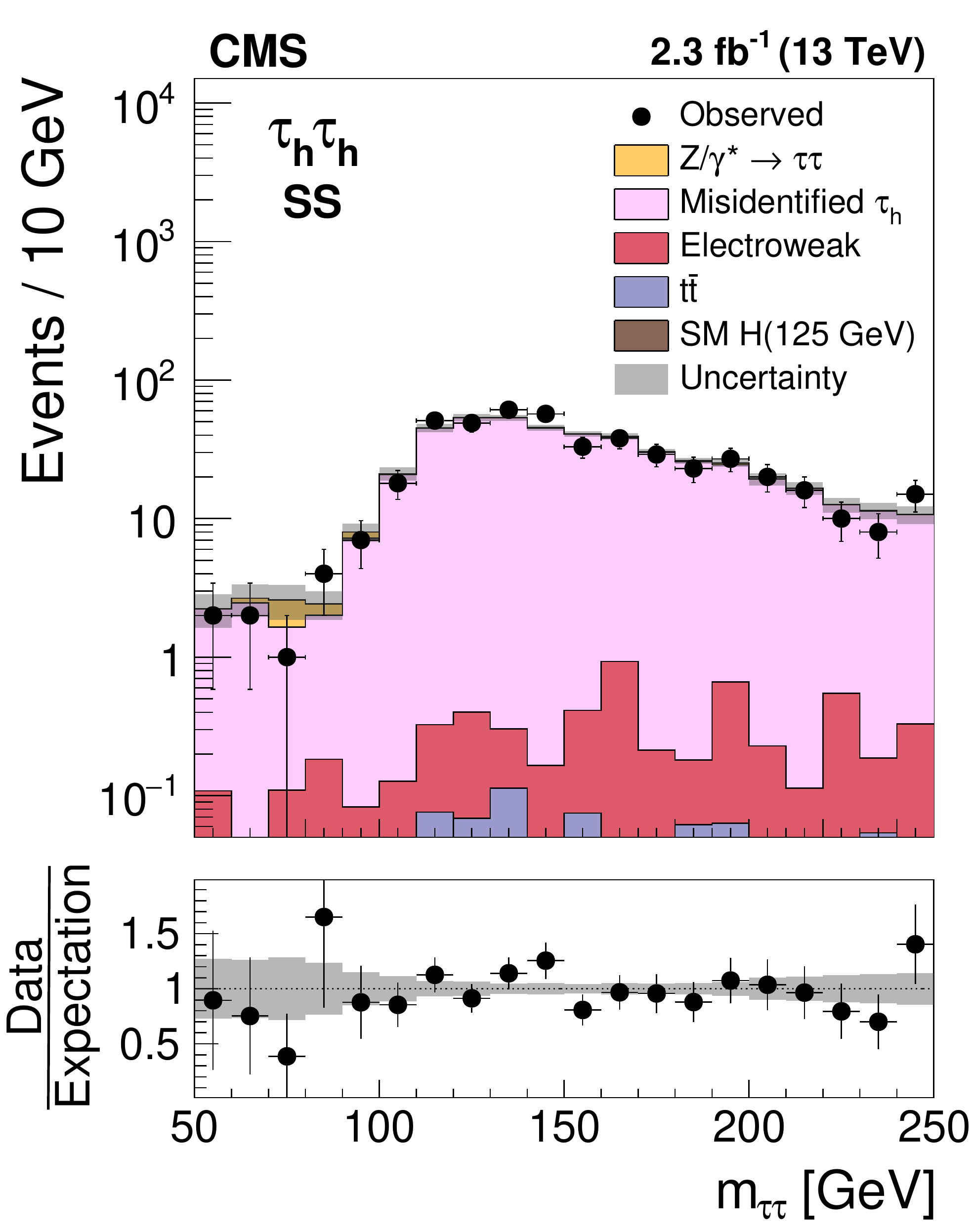}

\caption{
Distributions in $m_{\Pgt\Pgt}$
for SS events containing (upper left) $\Pe\tauh$, (upper right) $\Pgm\tauh$, and (lower) $\tauh\tauh$ pairs,
compared to expected background contributions.
}
\label{fig:mTauTauSS}
\end{figure*}

The modelling of the background from jets misidentified as $\tauh$ in the $\taue\tauh$, $\taum\tauh$, and $\tauh\tauh$ channels through the $\FF$ method
is validated by comparing the background estimates obtained in this method to the data
in control regions containing events with SS $\Pe\tauh$, $\Pgm\tauh$, and $\tauh\tauh$ pairs.
A dedicated set of $\FF$, without corrections for the extrapolation from OS to SS events, is determined for this validation.
The selection of events in the multijet DR is also altered in this validation,
to avoid overlap with the AR.
The distributions in $m_{\Pgt\Pgt}$ in events containing SS $\Pe\tauh$, $\Pgm\tauh$, and $\tauh\tauh$ pairs are shown in Fig.~\ref{fig:mTauTauSS}.
The data are compared to the sum of false-$\tauh$ background and other backgrounds.
The contribution of other backgrounds,
in which the reconstructed $\tauh$ is due either to a genuine $\tauh$ or to the misidentification of an electron or muon,
is obtained from the MC simulation.
The event yield of the $\cPZ/\Pggx \to \Pgt\Pgt$ signal in these control regions is small.
The normalization of individual backgrounds and of the $\cPZ/\Pggx \to \Pgt\Pgt$ signal is determined through a fit to the distributions in $m_{\Pgt\Pgt}$
in which the rate of each background is allowed to vary within its estimated systematic uncertainty.
The good agreement observed between the data and the background prediction in the control regions of all three channels
confirms the validity of false-$\tauh$ background estimates obtained through the $\FF$ method.

\subsection{Estimation of multijet background in \texorpdfstring{$\taue\taum$}{tau(e)-tau(mu)} and \texorpdfstring{$\taum\taum$}{tau(mu)-tau(mu)} channels}
\label{sec:backgroundEstimation_QCD_emu_mumu}

The contributions from multijet background in the SR of the $\taue\taum$ or $\taum\taum$ channels
are estimated using control regions containing events with an electron and muon or two muons of same charge, respectively.
An estimate for the contribution from multijet events in the SR is obtained
by scaling the yield of the multijet background in the SS control region by a suitably chosen extrapolation factor, defined
by the ratio of $\Pe\Pgm$ or $\Pgm\Pgm$ pairs with opposite charge
to those with same charge.
The ratio is measured in events in which at least one lepton passes an inverted isolation criterion of $I_{\ell} > 0.15 \, \pTell$.
We refer to this event sample as an isolation sideband region (SB).
The requirement $I_{\ell} > 0.15 \, \pTell$ ensures that the SB does not overlap with the SR.
A complication arises from the fact that the ratio of OS to SS pairs depends on
the lepton kinematics and the isolation criterion used in the SB.
The nominal OS/SS ratio is measured in an isolation sideband (SB1)
defined by requiring both leptons to satisfy a relaxed isolation criterion of $I_{\ell} < 0.60 \, \pTell$,
with at least one lepton passing the condition $I_{\ell} > 0.15 \, \pTell$.
The systematic uncertainty in the OS/SS ratio that arises from the choice of the upper limit on $I_{\ell}$ applied in SB1
is estimated by taking the difference between the OS/SS ratio computed in SB1 and the ratio computed in a different isolation sideband region (SB2).
The latter is defined by requiring at least one lepton to pass the condition $I_{\ell} > 0.60 \, \pTell$, without setting an
upper limit on $I_{\ell}$ in the SB2 region.
The criteria to select events in the isolation sidebands are optimized to ensure high statistical accuracy in
the measurement of the OS/SS extrapolation factor
and at the same time the minimization of differences in lepton kinematic distributions between the SR and the SB.
In both isolation sidebands,
the OS/SS ratio is measured as function of $\pT$ of the two leptons $\ell$ and $\ell'$
and of their separation $\Delta R(\ell,\ell') = \sqrt{(\eta_{\ell} - \eta_{\ell'})^{2} + (\phi_{\ell} - \phi_{\ell'})^{2}}$ in the $\eta$-$\phi$ plane.
The contributions to the SS control region, as well as to SB1 and SB2, from backgrounds other than multijet production
are subtracted, based on results from MC simulation.

\subsection{Estimation of \texorpdfstring{$\cPqt\cPaqt$}{t-tbar} background}
\label{sec:backgroundEstimation_TTbar}

While the $m_{\Pgt\Pgt}$ distribution for $\cPqt\cPaqt$ background is obtained from MC simulation,
the event yield in the $\cPqt\cPaqt$ background in the SR is determined from data, using a control region dominated by $\cPqt\cPaqt$ background.
Events in the $\cPqt\cPaqt$ control region are required to satisfy selection criteria that are similar to the requirements for the SR of the $\taue\taum$ channel,
described in Section~\ref{sec:eventSelection}.
The main differences are that the cutoff on $\Pzetamiss - 0.85 \, \Pzetavis$
is inverted to $\Pzetamiss - 0.85 \, \Pzetavis < -40\GeV$,
and a condition $\MET > 80\GeV$ is added to the event selection in the $\cPqt\cPaqt$ control region.
The $\cPqt\cPaqt$ event yield observed in the control region is a $1.01 \pm 0.07$ multiple of the expectation from the MC simulation.
The ratio of the $\cPqt\cPaqt$ event yield measured in data to the MC prediction is applied as a scale factor to simulated $\cPqt\cPaqt$ events,
to correct the $\cPqt\cPaqt$ background yield in the $\taue\taum$ and $\taum\taum$ channels,
as well as to correct the part of the $\cPqt\cPaqt$ background in the $\taue\tauh$, $\taum\tauh$, and $\tauh\tauh$ channels
that is either due to genuine $\tauh$ or due to the misidentification of an electron or muon as $\tauh$.
The latter is not included in the background estimate obtained through the $\FF$ method, but modelled in the MC simulation.

\section{Systematic uncertainties}
\label{sec:systematicUncertainties}

Imprecisely measured or imperfectly simulated effects can alter the normalization and distribution of the $m_{\Pgt\Pgt}$ mass spectrum in $\cPZ/\Pggx \to \Pgt\Pgt$ signal or background processes.
These systematic uncertainties can be categorized into theory-related and experimental sources.
The latter can be further subdivided into those associated with the reconstruction of physical objects of interest
and with estimated backgrounds.
The uncertainties related to the reconstruction of physical objects apply to the $\cPZ/\Pggx \to \Pgt\Pgt$ signal and to backgrounds
modelled in the MC simulation.
The main background contributions are determined from data, as described in Section~\ref{sec:backgroundEstimation},
and are largely unaffected by the accuracy achieved in modelling data in the MC simulation.

\begin{figure*}[h]
\centering
\includegraphics[width=\cmsSmallFigWidth]{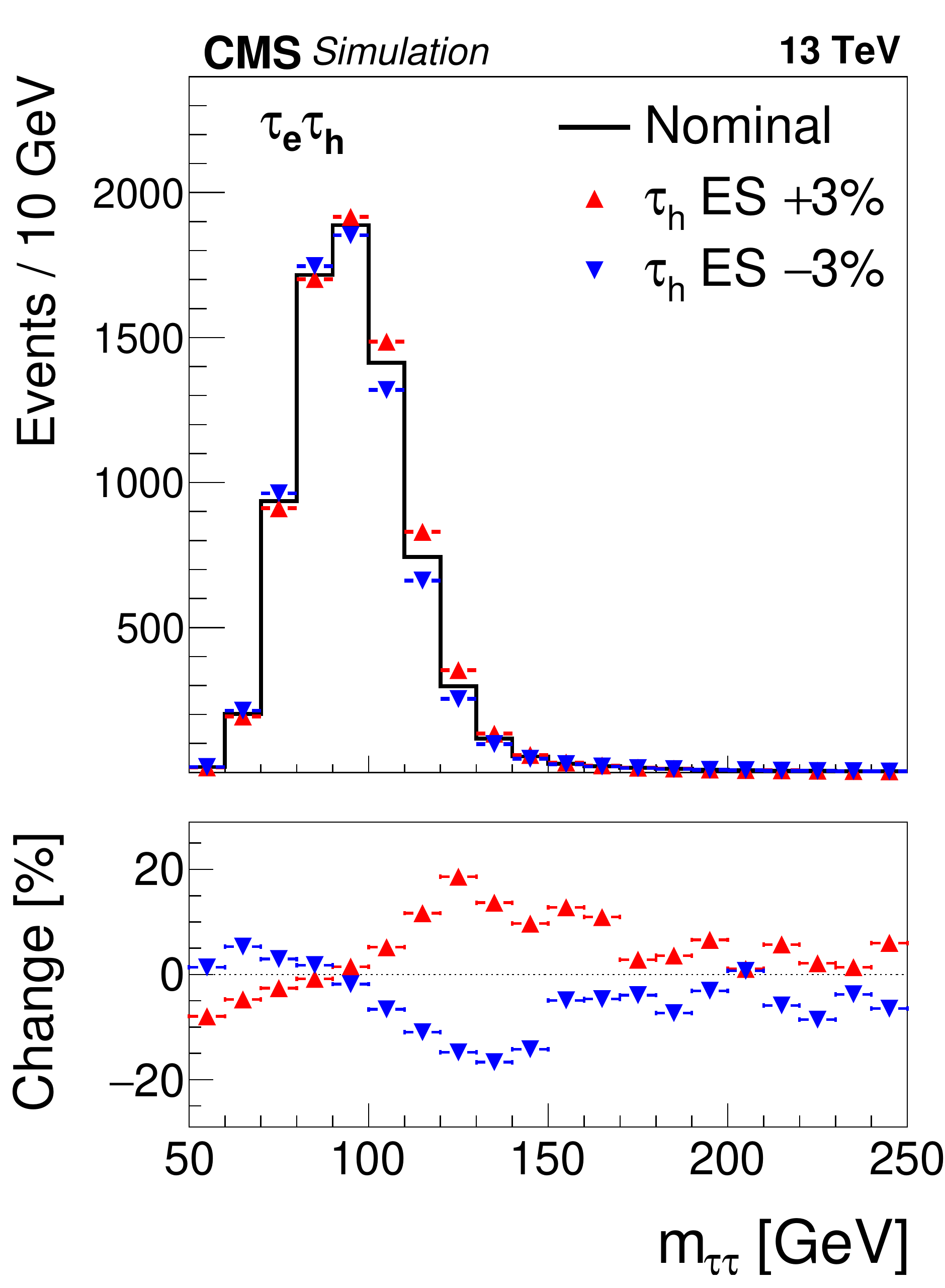} \hfil
\includegraphics[width=\cmsSmallFigWidth]{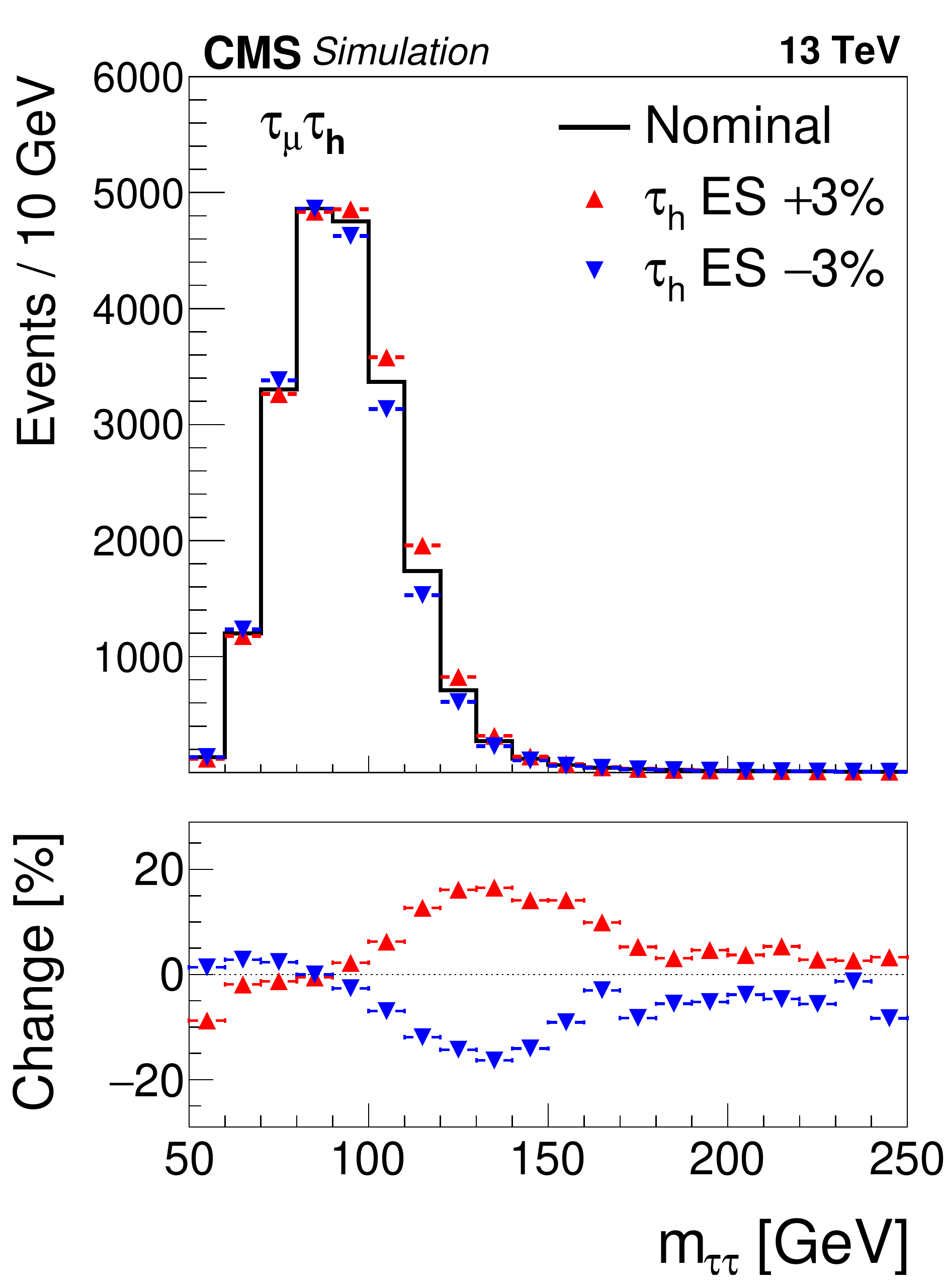} \hfil
\includegraphics[width=\cmsSmallFigWidth]{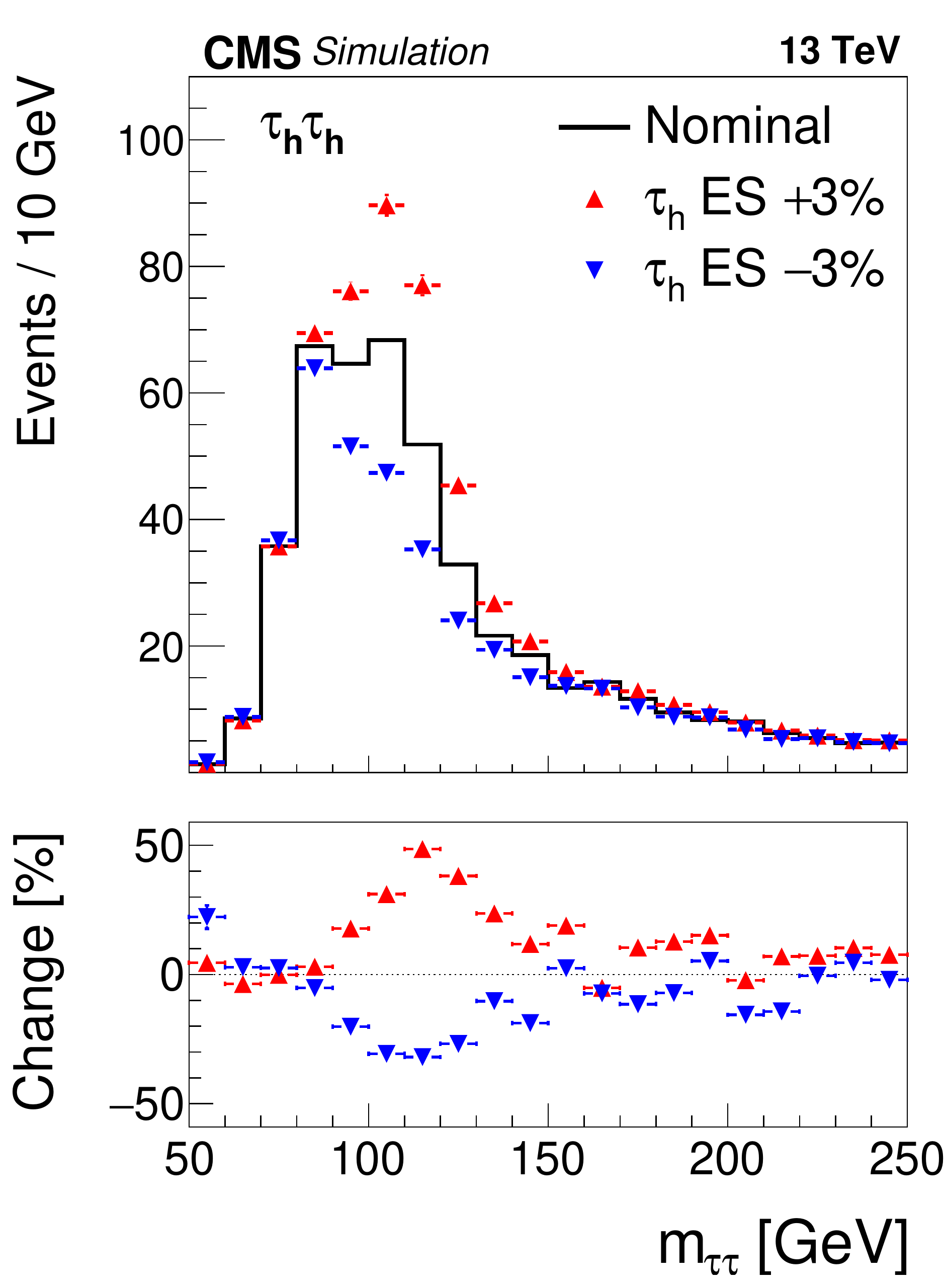}

\caption{
Distributions expected in $m_{\Pgt\Pgt}$ for $\cPZ/\Pggx \to \Pgt\Pgt$ signal events
in the (left) $\taue\tauh$, (center) $\taum\tauh$, and (right) $\tauh\tauh$ channels
for the nominal value of the $\tauh$\,ES, and after implementing $3\%$ systematic shift.
}
\label{fig:tauEnergyScaleUncertainty}
\end{figure*}

The main experimental uncertainties are related to the reconstruction and identification of electrons, muons, and $\tauh$, as follows.
The efficiency to reconstruct and identify $\tauh$ and the energy scale of $\tauh$ ($\tauh$\,ES)
is measured using $\cPZ/\Pggx \to \Pgt\Pgt \to \taum\tauh$ events.
The former is done by comparing the number of $\cPZ/\Pggx \to \Pgt\Pgt \to \taum\tauh$ events
with $\tauh$ candidates passing and failing the $\tauh$\,ID criteria,
and the latter by comparing the distributions in the $\tauh$ candidate mass,
as well as the visible mass of the muon and $\tauh$ system in data and in MC simulation~\cite{TAU-16-002},
measured with respective uncertainties of ${\approx}6$ and ${\approx}1\%$.
The events selected for the $\tauh$\,ID efficiency and $\tauh$\,ES measurements overlap with the events in the $\taum\tauh$ channel.
We account for the overlap by assigning a $3\%$ uncertainty to $\tauh$\,ES.
A $3\%$ change in the $\tauh$\,ES affects the acceptance in $\cPZ/\Pggx \to \Pgt\Pgt$ signal by $3$, $3$, and $17\%$
in the $\taue\tauh$, $\taum\tauh$, and $\tauh\tauh$ channels, respectively.
The impact on the signal acceptance and on the distribution in $m_{\Pgt\Pgt}$
is illustrated in Fig.~\ref{fig:tauEnergyScaleUncertainty}.
It has been checked that the overlap and the choice in the $\tauh$\,ES uncertainty have little impact on the final results.
The ML fit performed to measure the $\cPZ/\Pggx \to \Pgt\Pgt$ cross section, described in Section~\ref{sec:signalExtraction}, reduces the uncertainties
in the $\tauh$\,ID efficiency and in the $\tauh$\,ES to $2.2$ and $0.9\%$, respectively.
The efficiency of the $\tauh$ trigger used in the $\tauh\tauh$ channel is measured in $\cPZ/\Pggx \to \Pgt\Pgt \to \taum\tauh$ events
with an uncertainty of ${\approx}4.5\%$ per $\tauh$. The measurement is detailed in Ref.~\cite{TRG-12-001}.

{\tolerance=900
Electron and muon reconstruction, identification, isolation, and trigger efficiencies
are measured using $\cPZ/\Pggx \to \Pe\Pe$ and $\cPZ/\Pggx \to \Pgm\Pgm$ events via the ``tag-and-probe'' method~\cite{EWK-10-002}
at an accuracy of $2\%$.
The energy scales for electrons and muons ($\Pe$\,ES and $\Pgm$\,ES) are calibrated
using $\PJgy \to \ell\ell$, $\PUpsilon \to \ell\ell$, and $\cPZ/\Pggx \to \ell\ell$ events (with $\ell$ referring to $\Pe$ and $\Pgm$),
and have an uncertainty of $1\%$.
The $\Pe$\,ES and $\Pgm$\,ES uncertainties affect the acceptance in the $\cPZ/\Pggx \to \Pgt\Pgt$ signal
in the $\taue\tauh$, $\taum\tauh$, $\taue\taum$, and $\taum\taum$ channels by less than $1\%$.
\par}

The $\MET$ response and resolution are known within uncertainties of a few percent from studies performed in $\cPZ/\Pggx \to \Pgm\Pgm$, $\cPZ/\Pggx \to \Pe\Pe$, and $\gamma$+jets events~\cite{JME-16-004}.
The impact of these uncertainties on the acceptance in the $\cPZ/\Pggx \to \Pgt\Pgt$ signal is small, amounting to less than $1\%$.
In the $\taue\tauh$ and $\taum\tauh$ channels, the impact arises from the $\mT < 40\GeV$ selection criterion.
In the $\taue\taum$ and $\taum\taum$ channels, the impact is due to the $\Pzetamiss - 0.85 \, \Pzetavis > -20\GeV$ requirement
and the use of $\MET$ and $P_{\zeta}$ as input variables in the BDT that separates the $\cPZ/\Pggx \to \Pgt\Pgt$ signal from the $\cPZ/\Pggx \to \Pgm\Pgm$ background, respectively.
The effect of uncertainties related to the modelling of the $\MET$ on the distribution in $m_{\Pgt\Pgt}$ is small.

{\tolerance=4800
The uncertainty in the integrated luminosity is $2.3\%$~\cite{LUM-15-001}.
\par}

\begin{figure*}[h]
\centering
\includegraphics[width=\cmsSmallFigWidth]{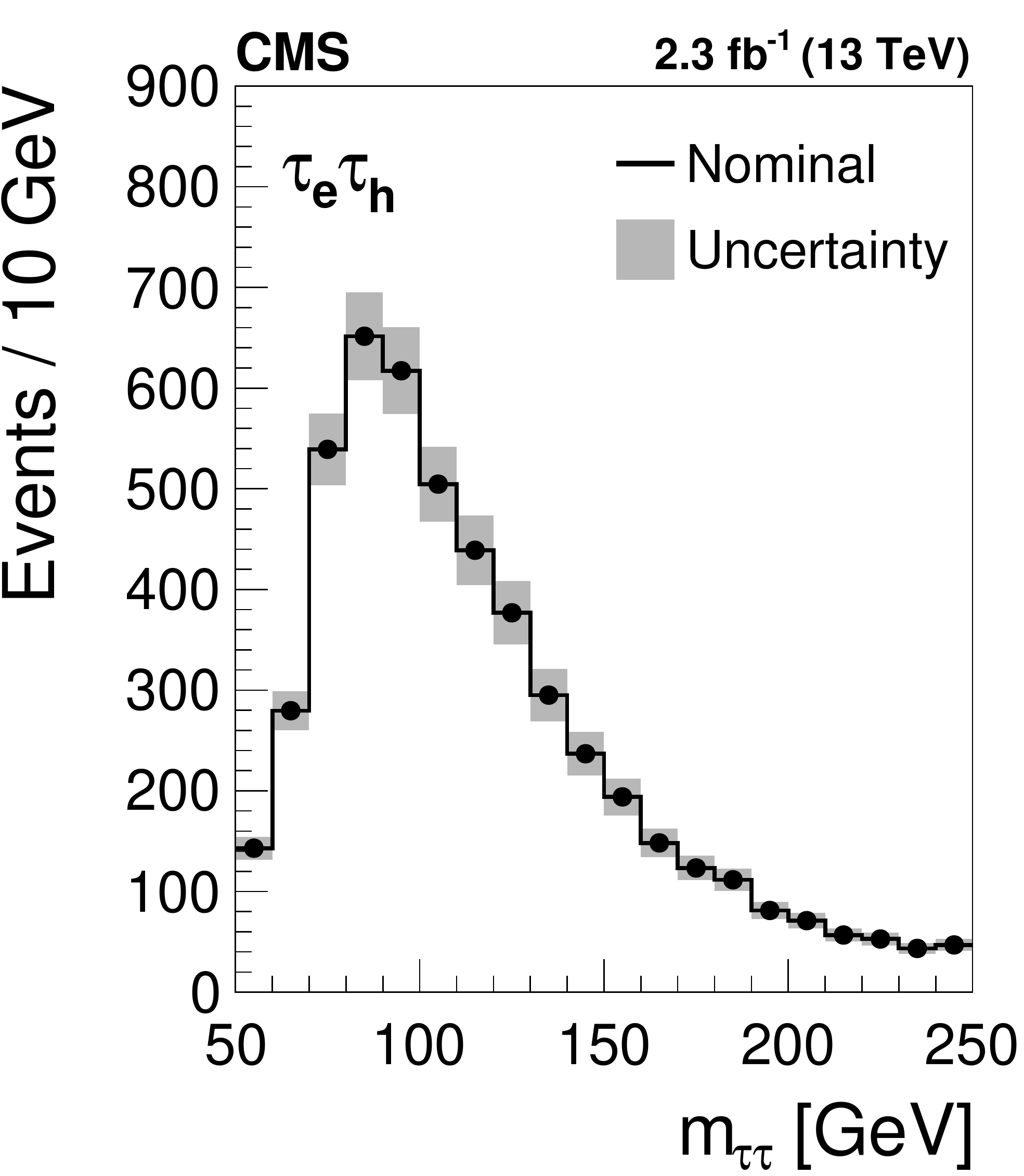} \hfil
\includegraphics[width=\cmsSmallFigWidth]{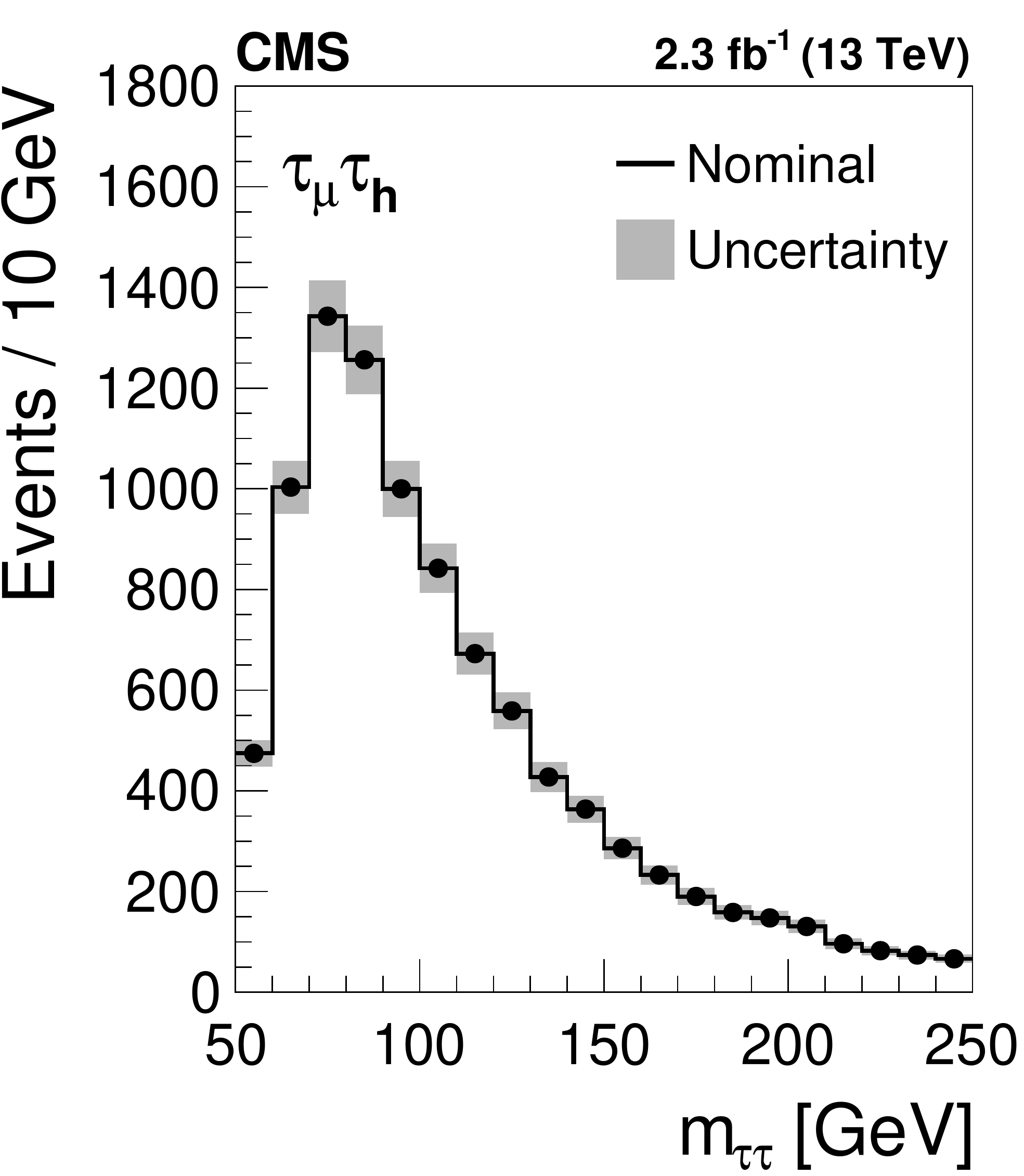} \hfil
\includegraphics[width=\cmsSmallFigWidth]{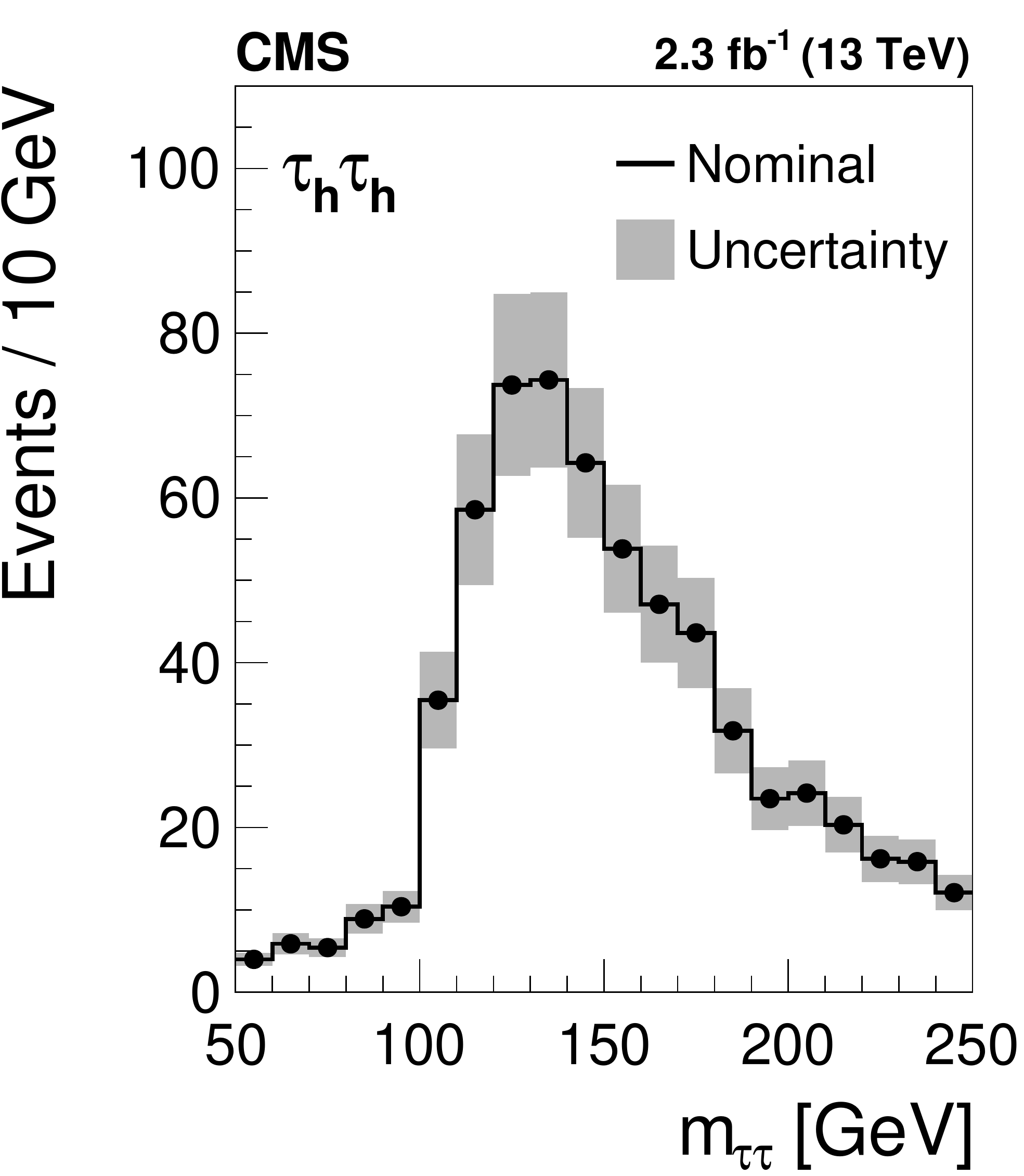}

\caption{
Distributions in $m_{\Pgt\Pgt}$ expected for the background
arising from quark or gluon jets misidentified as $\tauh$
in the (left) $\taue\tauh$, (center) $\taum\tauh$, and (right) $\tauh\tauh$ channels,
and the systematic uncertainty in the false-$\tauh$ background estimate.
The grey shaded band represents the quadratic sum of all systematic uncertainties related to the $\FF$ method:
uncertainties in the $\FF$ measured in the multijet, $\PW$+jets, and $\cPqt\cPaqt$ DR;
uncertainties in the relative fractions of multijet, $\PW$+jets, and $\cPqt\cPaqt$ backgrounds in the AR;
and uncertainties in the non-closure corrections
(described in Section~\ref{sec:backgroundEstimation_Fakes_etau(mu)tau_tautau}).
}
\label{fig:tauFakeRateUncertainties}
\end{figure*}

{\tolerance=600
The backgrounds determined from data are also subject to uncertainties
that alter the normalization and distribution (``shape'') of the $m_{\Pgt\Pgt}$ mass spectrum.
Background yields and their associated uncertainties are given in Table~\ref{tab:eventYieldsPrefit}.
The uncertainties in the backgrounds arising from the misidentification of quark and gluon jets as $\tauh$ candidates
in the $\taue\tauh$, $\taum\tauh$, and $\tauh\tauh$ channels are obtained by changing the $\FF$ values
as well as the relative fractions $R_{\textrm{p}}$ of multijet, $\PW$+jets, and $\cPqt\cPaqt$ backgrounds within their uncertainties.
The resulting uncertainties in the $m_{\Pgt\Pgt}$ distribution in the $\taue\tauh$, $\taum\tauh$, and $\tauh\tauh$ channels
are illustrated in Fig.~\ref{fig:tauFakeRateUncertainties}.
The uncertainties in the size of the false-$\tauh$ backgrounds are $8$, $6$, and $16\%$
in the $\taue\tauh$, $\taum\tauh$, and $\tauh\tauh$ channels, respectively.
In the $\taue\taum$ and $\taum\taum$ channels,
the uncertainty in the size of the multijet background is ${\approx}20\%$.
The magnitude of the $\cPqt\cPaqt$ background is known to an accuracy of $7\%$.
The uncertainty in the distribution of the $\cPqt\cPaqt$ background is estimated by changing the weights
applied to the $\cPqt\cPaqt$ MC sample, to improve the modelling of the top quark $\pT$ distribution
(described in Section~\ref{sec:datasamples_and_MonteCarloSimulation}),
between no reweighting and the reweighting applied twice.
\par}

The uncertainties in the yields of single top quark and diboson backgrounds, modelled using MC simulation,
are each ${\approx}15\%$.
Besides constituting the dominant background in the $\taum\taum$ channel,
the DY production of electron and muon pairs are relevant backgrounds in, respectively,
the decay channels $\taue\tauh$ and $\taum\tauh$,
because of the small but non-negligible rate at which electrons and muons are misidentified as $\tauh$.
The probability for electrons and muons to pass the tight-electron or tight-muon removal criteria applied, respectively, in the $\taue\tauh$ and $\taum\tauh$ channels
is measured in $\cPZ/\Pggx \to \Pe\Pe$ and in $\cPZ/\Pggx \to \Pgm\Pgm$ events.
The misidentification rates depend on $\eta$.
For electrons in the ECAL barrel and endcap regions,
the misidentifications are at respective levels of $0.2$ and $0.1\%$,
with accuracies of $13$ and $29\%$~\cite{TAU-16-002}.
The misidentification rate for muons lies
between less than one and several tenths of a percent,
and is known to within an uncertainty of $30\%$.
The contribution from $\PW$+jets background in the $\taue\taum$ and $\taum\taum$ channels is modelled using MC simulation,
and is known to an accuracy of $15\%$.
The production of SM Higgs bosons is assigned an uncertainty of $30\%$,
reflecting the present experimental uncertainty in the $\PHiggs \to \Pgt\Pgt$ rate
measured at $\sqrt{s} = 13\TeV$~\cite{HIG-16-043}.

The theoretical uncertainty in the product of signal acceptance and efficiency for the $\cPZ/\Pggx \to \Pgt\Pgt$ signal
is ${\approx}2\%$ in the $\taue\tauh$, $\taum\tauh$, $\taue\taum$, and $\taum\taum$ channels, and $6\%$ in the $\tauh\tauh$ channel.
The quoted uncertainties include the effect of missing higher-order terms in the perturbative expansion for the calculated cross section,
estimated through independent changes in the renormalization and factorization scales by factors of $2$ and $1/2$ relative to their nominal equal values~\cite{Cacciari:2003fi,Catani:2003zt},
uncertainties in the {\sc NNPDF3.0} set of PDF,
estimated following the recommendations given in Ref.~\cite{Butterworth:2015oua},
and the uncertainties in the modelling of parton showers (PS) and the underlying event (UE).
The theoretical uncertainty is larger in the $\tauh\tauh$ channel,
as the acceptance depends crucially on the modelling of the $\pT$ distribution of the $\cPZ$ boson,
which is also affected by the missing higher-order terms in the calculation.

{\tolerance=600
The systematic uncertainties are summarized in Table~\ref{tab:systematicUncertainties_summary}.
The table also quantifies the impact that each systematic uncertainty has on the measurement of the $\cPZ/\Pggx \to \Pgt\Pgt$ cross section,
defined as the percent change in the measured cross section when individual sources are changed by one standard deviation relative to their nominal values.
The impacts are computed for the values of nuisance parameters obtained in the ML fit used to extract the signal (described in Section~\ref{sec:signalExtraction}).
\par}

\begin{table*}
\centering
\topcaption{
Effect of experimental and theoretical uncertainties in the measurement of the $\cPZ/\Pggx \to \Pgt\Pgt$ cross section.
The sources of systematic uncertainty are specified in the leftmost column,
and apply to the processes given in the second column.
The relative changes in the acceptance $\mathcal{A}$ for the $\cPZ/\Pggx \to \Pgt\Pgt$ signal, and in the yield from background processes
that correspond to a one standard deviation change in a given source of uncertainty is given in the third column.
The range in this column represents the range in signal acceptance or background yield across all decay channels and background processes.
The impact that each change produces is quantified by its effect on the measured $\cPZ/\Pggx \to \Pgt\Pgt$ cross section,
given in the rightmost column.
}
\label{tab:systematicUncertainties_summary}
\begin{tabular}{l|c|c|r}
Source & Applies to & Change in $\mathcal{A}$ or yield & Impact \\
\hline
Integrated luminosity & Simulated processes & $2.3\%$ & $1.9\%$ \\
Hadronic $\Pgt$ ID and trigger & Simulated processes & $6$--$12\%$ & $1.5\%$ \\
$\tauh$\,ES & Simulated processes & $2$--$17\%$ & ${<} 0.1\%$ \\
Rate of $\Pe$ misidentified as $\tauh$ & $\cPZ/\Pggx \to \Pe\Pe$ & $13$--$29\%$ & $0.4\%$ \\
Rate of $\Pgm$ misidentified as $\tauh$ & $\cPZ/\Pggx \to \Pgm\Pgm$ & $30\%$ & $0.2\%$ \\
Electron ID and trigger & Simulated processes & $2\%$ & $1.5\%$ \\
$\Pe$\,ES & Simulated processes & ${<} 1\%$ & $0.2\%$ \\
Muon ID and trigger & Simulated processes & $2\%$ & $1.6\%$ \\
$\Pgm$\,ES & Simulated processes & ${<} 1\%$ & ${<} 0.1\%$ \\
$\MET$ response and resolution & Simulated processes & $1$--$10\%$ & $0.2\%$ \\
\hline
Norm. $\cPZ/\Pggx \to \Pe\Pe$, $\Pgm\Pgm$ & $\cPZ/\Pggx \to \Pe\Pe$, $\Pgm\Pgm$ & Unconstrained & $1.8\%$ \\
Norm. and shape of false $\tauh$ & $\taue\tauh$, $\taum\tauh$, $\tauh\tauh$ channels & $6$--$16\%$ & ${<} 0.1\%$ \\
Norm. and shape of multijet & $\taue\taum$, $\taum\taum$ channels & $20\%$ & $0.2\%$ \\
Norm. $\cPqt\cPaqt$ & $\cPqt\cPaqt$ & $7\%$ & $1.0\%$ \\
Shape $\cPqt\cPaqt$ & $\cPqt\cPaqt$ & $1$--$6\%$ & ${<} 0.1\%$ \\
Norm. SM $\PHiggs$ & SM $\PHiggs$ & $30\%$ & ${<} 0.1\%$ \\
Norm. single top quark & Single top quark& $15\%$ & ${<} 0.1\%$ \\
Norm. diboson & Diboson & $15\%$ & $0.2\%$ \\
Norm. $\PW$+jets & $\PW$+jets & $15\%$ & ${<} 0.1\%$ \\
\hline
PDF & Signal & $1\%$ & $1.0\%$ \\
Scale dependence & Signal & ${<} 6\%$ & $0.5\%$ \\
UE and PS & Signal & $1\%$ & $1.0\%$ \\
\end{tabular}

\end{table*}

{\tolerance=600
The uncertainties in the integrated luminosity, in the cross section for DY production of electron and muon pairs,
and in the electron, muon, and $\tauh$ reconstruction and identification efficiencies have greatest impact on the results.

The impact of the uncertainty on the integrated luminosity amounts to $1.9\%$.
This is smaller than the $2.3\%$ uncertainty in the integrated luminosity measurement,
because of correlations of the nuisance parameter representing the integrated luminosity with other nuisance parameters.
When the integrated luminosity changes by $2.3\%$,
the ML fit readjusts the nuisance parameters that represent the rates for background processes obtained from MC simulation,
as well as identification and trigger efficiencies for $\Pe$, $\Pgm$, and $\tauh$,
such that the measured $\cPZ/\Pggx \to \Pgt\Pgt$ cross section changes by only $1.9\%$.
The uncertainty in the integrated luminosity is not constrained in the ML fit.

The impact of the uncertainty in the production rate of $\cPZ/\Pggx \to \Pe\Pe$ and $\cPZ/\Pggx \to \Pgm\Pgm$ background processes amounts to $1.8\%$.
The impact is sizeable, because of the small statistical uncertainty in the $\cPZ/\Pggx \to \Pgm\Pgm$ background in the $\taum\taum$ channel,
which, in the absence of uncertainties in the $\cPZ/\Pggx \to \Pgm\Pgm$ production rate, would constrain the efficiency for muon reconstruction and identification, as well as the integrated luminosity.

The impact of uncertainties in the efficiencies to reconstruct and identify electrons and muons amounts to $1.5$ and $1.6\%$, respectively.
Their impact is considerable, 
because these uncertainties are not reduced greatly in the ML fit,
as they affect all channels, except the $\tauh\tauh$ channel, in a similar way.

The impact of the uncertainty in the efficiency to reconstruct and identify $\tauh$ is of similar size,
amounting to $1.5\%$, despite that the uncertainty in the $\tauh$\,ID efficiency is significantly larger than the uncertainties in the electron and muon ID efficiencies.
This is because the simultaneous fit to the $m_{\Pgt\Pgt}$ distributions in all five channels 
reduces the uncertainties in the $\tauh$\,ID efficiency and the $\tauh$\,ES significantly,
diminishing thereby the impact that these uncertainties have on the $\cPZ/\Pggx \to \Pgt\Pgt$ cross section.
When the $\cPZ/\Pggx \to \Pgt\Pgt$ cross section is measured in the individual $\taue\tauh$, $\taum\tauh$, and $\tauh\tauh$ channels,
the impact of the uncertainty on the $\tauh$\,ID efficiency increases to $6$, $6$, and $10\%$, respectively.

The uncertainty in $\tauh$\,ES becomes relevant for the $\tauh\tauh$ channel 
when the $\cPZ/\Pggx \to \Pgt\Pgt$ cross section is measured in this channel alone, and amounts to $9\%$.
In the $\taue\tauh$ and $\taum\tauh$ channels, the impact of the $\tauh$\,ES uncertainty amounts to less than $1\%$,
even when the $\cPZ/\Pggx \to \Pgt\Pgt$ cross section is measured just in these channels.
\par}

\section{Signal extraction}
\label{sec:signalExtraction}

The cross section $\sigma(\Pp\Pp \rightarrow \cPZ/\Pggx\text{+X}) \,  \mathcal{B}(\cPZ/\Pggx \to \Pgt\Pgt)$ for DY production of $\Pgt$ pairs
is obtained through a simultaneous ML fit to the observed $m_{\Pgt\Pgt}$ distributions in the five decay channels:
$\taue\tauh$, $\taum\tauh$, $\tauh\tauh$, $\taue\taum$, and $\taum\taum$.
The likelihood function $\mathcal{L}\left(\text{data} \, \vert \, \xi, \Theta \right)$
depends on the value of the cross section, denoted by the symbol $\xi$, which defines the parameter of interest (POI) in the fit,
and it also depends on the values of nuisance parameters $\theta_{k}$
that represent the systematic uncertainties discussed in Section~\ref{sec:systematicUncertainties}:
\begin{linenomath}
\begin{equation}
\mathcal{L}\left(\text{data} \, \vert \, \xi, \Theta \right)
= \prod_{i} \, \mathcal{P}\left(n_{i} \vert \xi, \Theta\right) \, \prod_{k} \, \rho\left(\tilde{\theta}_{k} \vert \theta_{k}\right) .
\label{eq:likelihoodFunction}
\end{equation}
\end{linenomath}
The index $i$ refers to individual bins of the $m_{\Pgt\Pgt}$ distribution in each of the five final states.
The set of all nuisance parameters $\theta_{k}$ is denoted by the symbol $\Theta$.
Correlations among decay channels as well as between the $\cPZ/\Pggx \to \Pgt\Pgt$ signal and background processes
are taken into account through relationships among
channels, processes, and nuisance parameters in the ML fit.
The probability to observe $n_{i}$ events in a given bin $i$, when $\nu_{i}(\xi, \Theta)$ events are expected in that bin
is given by the Poisson distribution:
\begin{linenomath}
\begin{equation}
\mathcal{P}\left(n_{i} \vert \xi, \Theta \right)
= \frac{\left(\nu_{i}(\xi, \Theta)\right)^{n_{i}}}{n_{i}!} \, \exp \left( -\nu_{i}(\xi, \Theta) \right)  .
\label{eq:Poisson}
\end{equation}
\end{linenomath}
The number of events expected in each bin corresponds to the sum of the number of signal ($\nu_{i}^{\textrm{S}}$) and background ($\nu_{i}^{\textrm{B}}$) events:
$\nu_{i}(\xi, \Theta) = \nu_{i}^{\textrm{S}}(\xi, \Theta) + \nu_{i}^{\textrm{B}}(\Theta)$.
The estimate in the number of background events is obtained as described in Section~\ref{sec:backgroundEstimation}.
The number of signal events is proportional to $\xi$, with the
coefficient of proportionality depending on the signal acceptance and on the signal selection efficiency, with both obtained from MC simulation.

The function $\rho\left(\tilde{\theta}_{k} \vert \theta_{k}\right)$ represents the probability to
observe a value $\tilde{\theta}_{k}$ in an auxiliary measurement of the nuisance parameter,
given that the true value is $\theta_{k}$.
The nuisance parameters are treated via the frequentist paradigm,
as described in Refs.~\cite{ATL-PHYS-PUB-2011-011,HIG-11-032}.
Systematic uncertainties that affect only the normalization, but not the distribution in $m_{\Pgt\Pgt}$,
are represented by the Gamma function if they are statistical in origin,
e.g. corresponding to the number of events observed in a control region,
and otherwise by log-normal probability density functions.
Systematic uncertainties that affect the distribution in $m_{\Pgt\Pgt}$
are incorporated into the ML fit via the technique detailed in Ref.~\cite{Conway:2011in},
and represented by Gaussian probability density functions.
Nuisance parameters representing systematic uncertainties of the latter type
can also affect the normalization of the $\cPZ/\Pggx \to \Pgt\Pgt$ signal or of its backgrounds.
The nuisance parameters corresponding to the cross sections for DY production of electron and muon pairs are left unconstrained in the fit.

\begin{figure}[ht!]
\centering
\includegraphics[width=\cmsFigWidth]{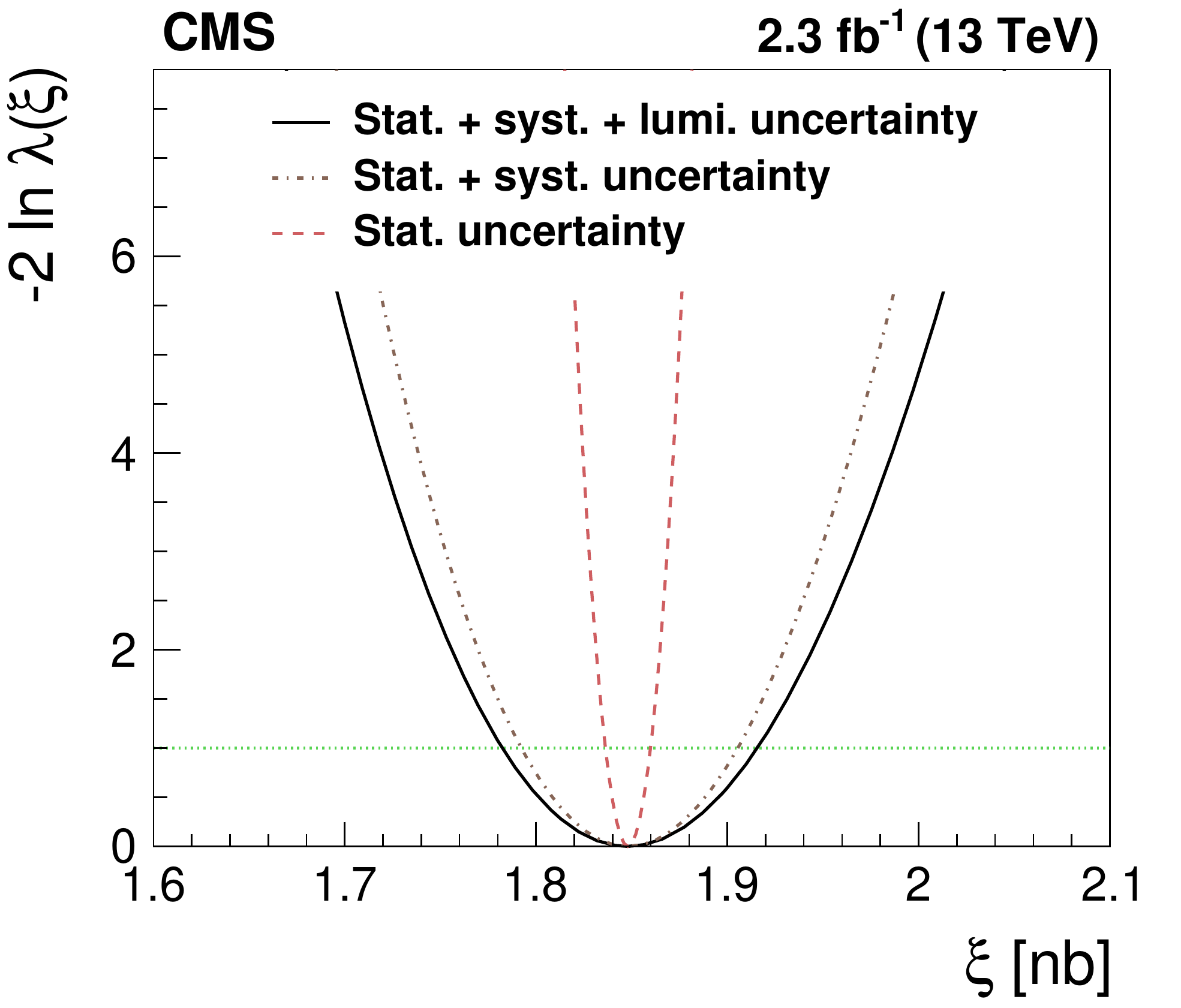}

\caption{
Dependence of $-2 \ln \lambda\left(\xi\right)$ on the cross section $\xi$ for DY production of $\Pgt$ pairs.
The PLR is computed for the simultaneous ML fit to the observed $m_{\Pgt\Pgt}$ distributions in the
$\taue\tauh$, $\taum\tauh$, $\tauh\tauh$, $\taue\taum$, and $\taum\taum$ channels.
The dashed, dash-dotted, and solid curves
correspond to situations when just the statistical uncertainties are used in the fit,
when the uncertainty in integrated luminosity is also included,
and when all uncertainties are included in the fit.
The values of nuisance parameters, corresponding to uncertainties that are ignored,
are fixed at the values that yield the best fit to the data.
The horizontal line represents the value of $-2 \ln \lambda\left(\xi\right)$ that is used to determine the $68\%$ CI on $\xi$.
}
\label{fig:likelihoodParabola}
\end{figure}

The best fit value $\hat{\xi}$ of the POI is the value that maximizes the likelihood $\mathcal{L}\left(\text{data} \, \vert \, \xi, \Theta \right)$
in Eq.~(\ref{eq:likelihoodFunction}).
A $68\%$ confidence interval (CI) on the POI is obtained using the profile likelihood ratio (PLR)~\cite{ATL-PHYS-PUB-2011-011,HIG-11-032,HIG-14-009}:
\begin{linenomath}
\begin{equation}
\lambda\left(\xi\right)
= \frac{\mathcal{L}\left(\text{data} \, \vert \, \xi, \hat{\Theta}_{\xi} \right)}{\mathcal{L}\left(\text{data} \, \vert \, \hat{\xi}, \hat{\Theta} \right)}.
\label{eq:PLR}
\end{equation}
\end{linenomath}
The symbol $\hat{\Theta}_{\xi}$ denotes the values of nuisance parameters that maximize the likelihood
for a given value of $\xi$.
The combination of $\hat{\xi}$ and $\hat{\Theta}$ correspond to the values of $\xi$ and $\Theta$ for which the likelihood function reaches its maximum.
The $68\%$ CI is defined by the values of $\xi$ for which $-2 \ln \lambda\left(\xi\right)$ increases by one unit relative to its minimum.
To quantify the effects from individual statistical uncertainties,
the uncertainty in the integrated luminosity,
and other systematic uncertainties,
we ignore some single source of uncertainties at a time, and recompute the $68\%$ CI.
The nuisance parameters $\theta_{k}$ corresponding to uncertainties that are ignored are fixed at the values $\hat{\theta}_{k}$ that yield the best fit to the data.
The square root of the quadratic difference between the CI,
computed for all sources of uncertainties in the fit,
and for the case that some given source is ignored,
reflects the estimate of the uncertainty in the POI resulting from a single source.
The procedure is illustrated in Fig.~\ref{fig:likelihoodParabola} for the combined fit of all five final states.
Correlations among different sources of uncertainty are estimated through this procedure.

The cross section for DY production of $\Pgt$ pairs
is quoted within the mass window $60 < m_{\Pgt\Pgt}^{\textrm{true}} < 120\GeV$.
The contribution from $\cPZ/\Pggx \to \Pgt\Pgt$ events that pass the selection criteria described in Section~\ref{sec:eventSelection},
but have a mass outside of this window is at the level of a few percent in the
$\taue\tauh$, $\taum\tauh$, $\taue\taum$, and $\taum\taum$ channels.
In the $\tauh\tauh$ channel,
this contribution from outside of the mass window is ${\approx}40\%$,
the reason for this being so large is the high
$\pT$ threshold on the $\tauh$ candidates required in the trigger.
The $\cPZ/\Pggx \to \Pgt\Pgt$ events that have two $\tauh$ with $\pT > 40\GeV$
contain either a $\cPZ$ boson of high $\pT$ or a $\Pgt$ lepton pair above the mass of the $\cPZ$ boson.
Only a small fraction of signal events pass either of these two conditions,
which leads to the smallest
event yield from the $\cPZ/\Pggx \to \Pgt\Pgt$ signal in the $\tauh\tauh$ channel (as shown in Table~\ref{tab:eventYieldsPostfit}),
and to the largest fraction of signal events containing a $\Pgt$ lepton pair
of mass outside of the $60 < m_{\Pgt\Pgt}^{\textrm{true}} < 120\GeV$ window.

The PLR depends on the $\tauh$\,ID efficiency and on the $\tauh$\,ES
through its dependence on the corresponding two nuisance parameters.
The $\tauh$\,ID efficiency and $\tauh$\,ES are determined by promoting these nuisance parameters to the role of POI.
The cross section for DY production of $\Pgt$ pairs, the $\tauh$\,ID efficiency, and the $\tauh$\,ES are left unconstrained in the fit,
and the PLR is minimized as a function of all three parameters.

\section{Results}
\label{sec:results}

\begin{figure*}[ht!]
\centering
\includegraphics[width=\cmsFigWidth]{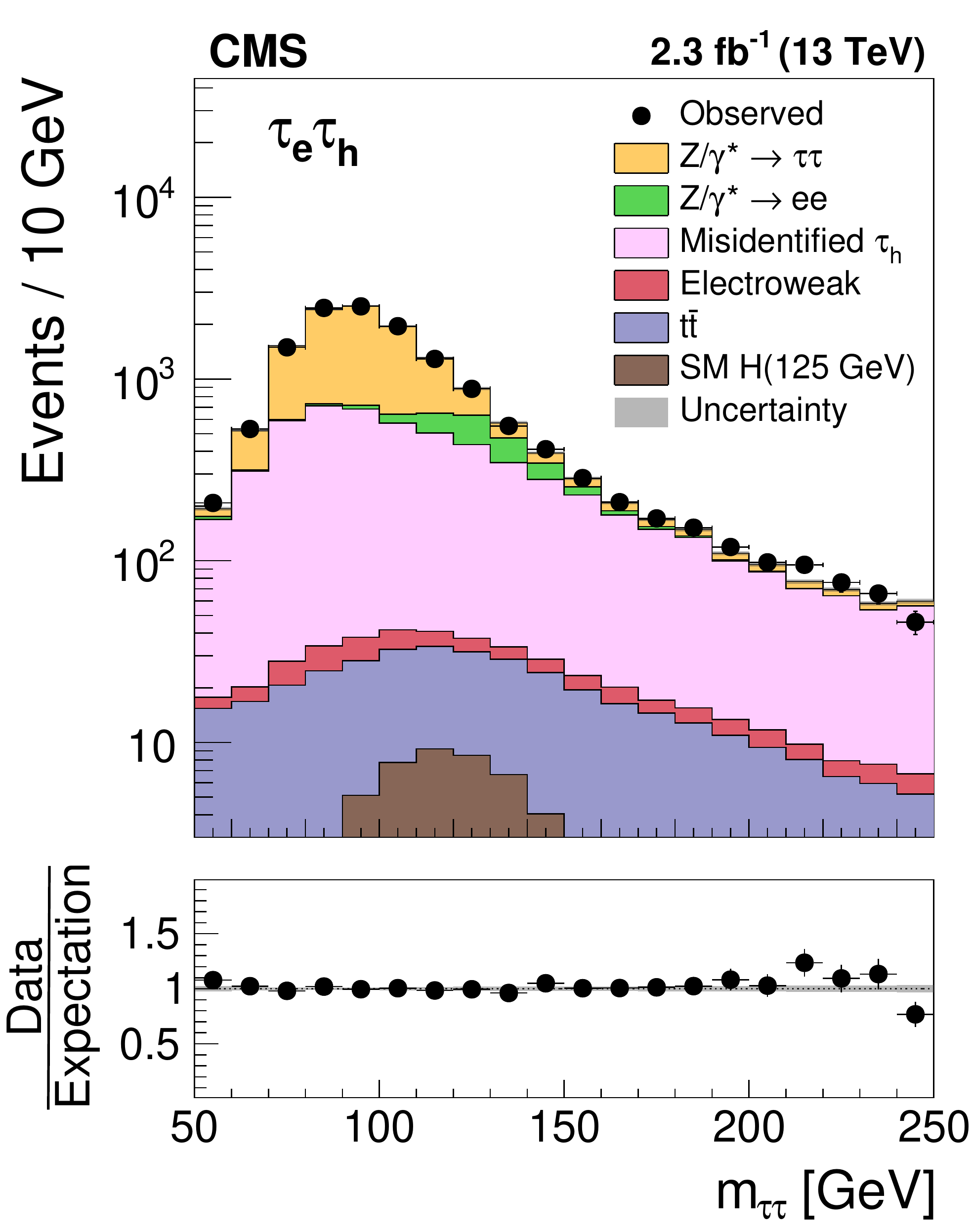} \hfil
\includegraphics[width=\cmsFigWidth]{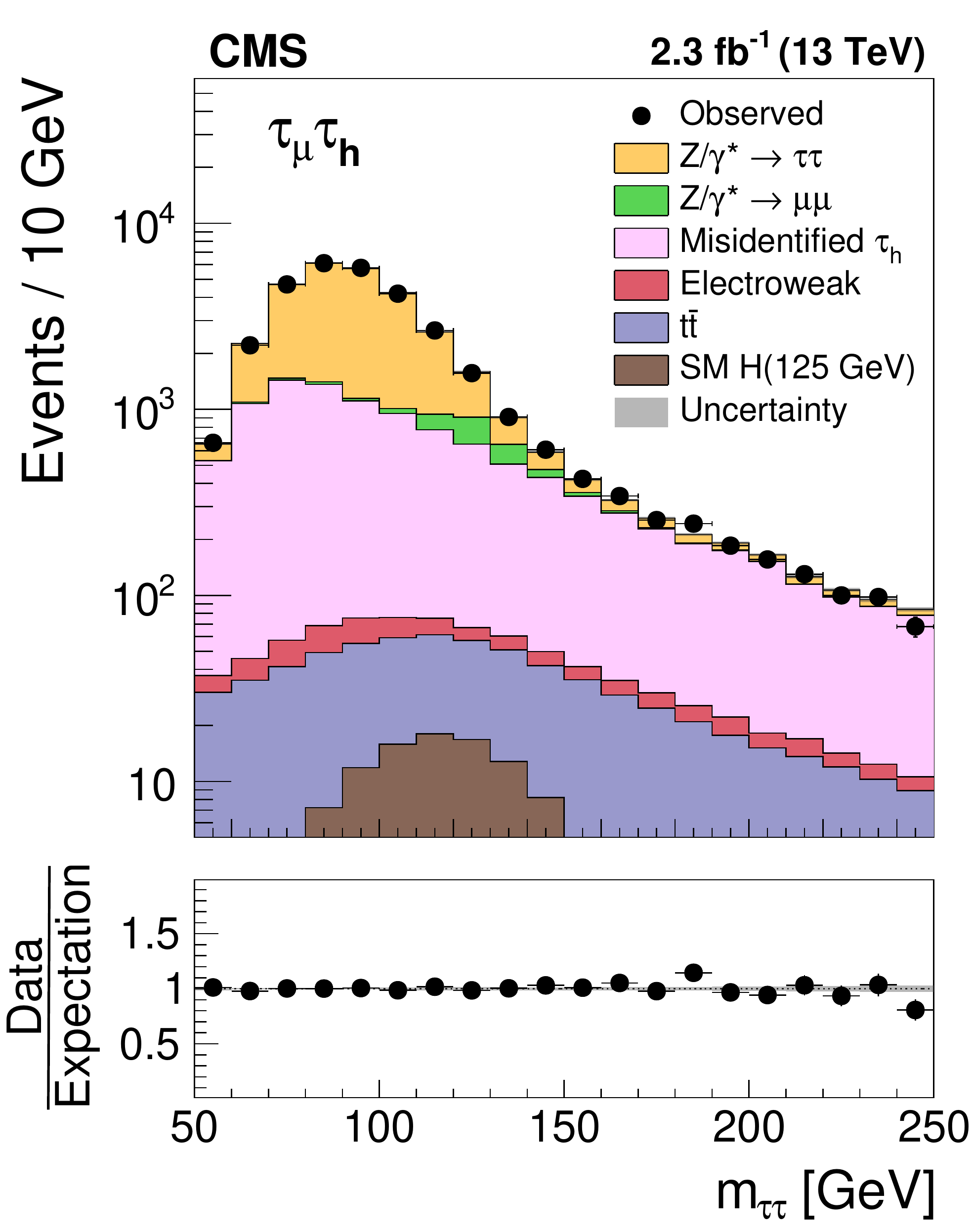} \\
\includegraphics[width=\cmsFigWidth]{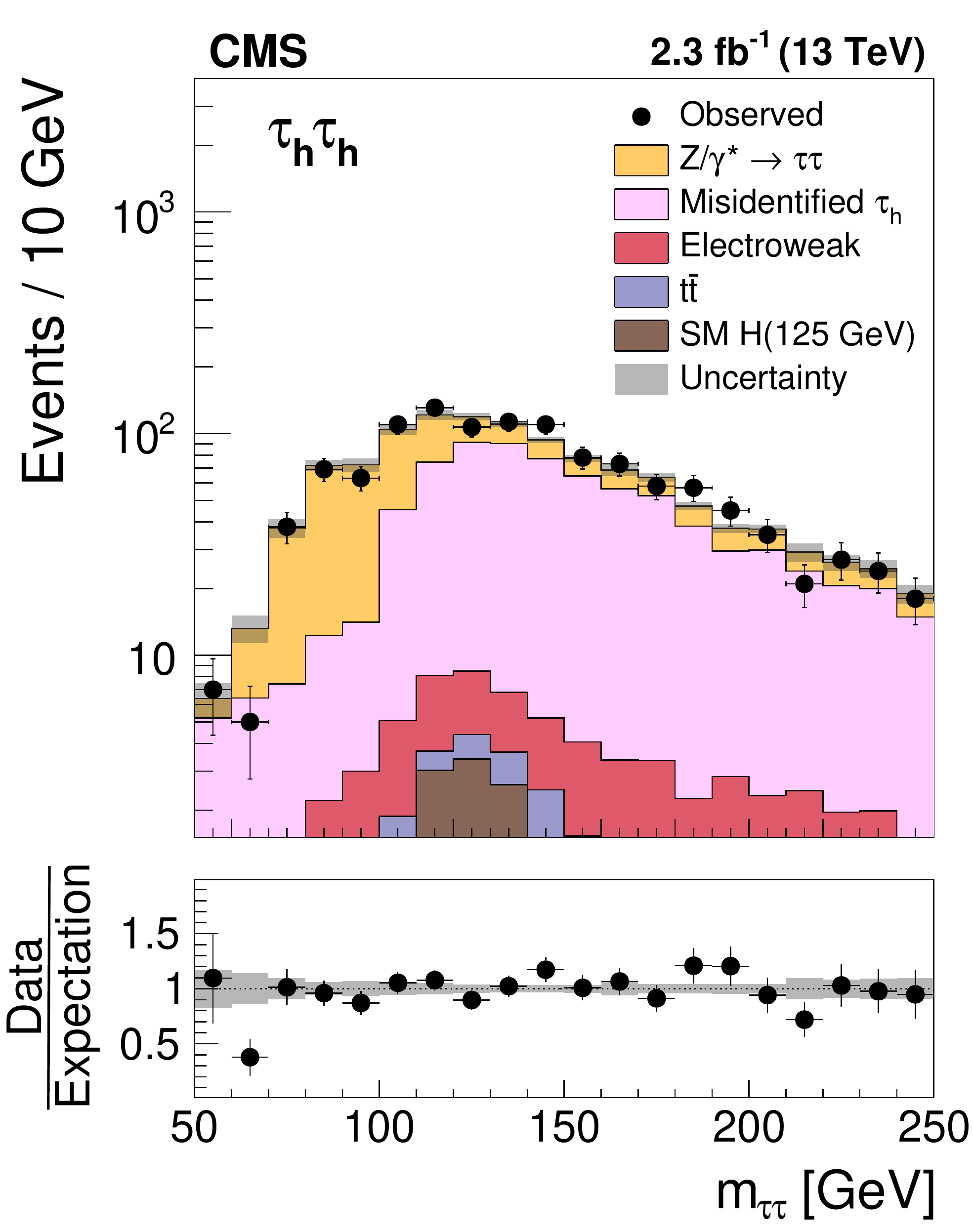}

\caption{
Distributions in $m_{\Pgt\Pgt}$
for events selected in the (upper left) $\taue\tauh$, (upper right) $\taum\tauh$, and (lower) $\tauh\tauh$ channels.
Signal and background contributions are shown for values of nuisance parameters obtained in the ML fit to the data.
}
\label{fig:postfitPlots_mTauTau1}
\end{figure*}

\begin{figure*}[ht!]
\centering
\includegraphics[width=\cmsFigWidth]{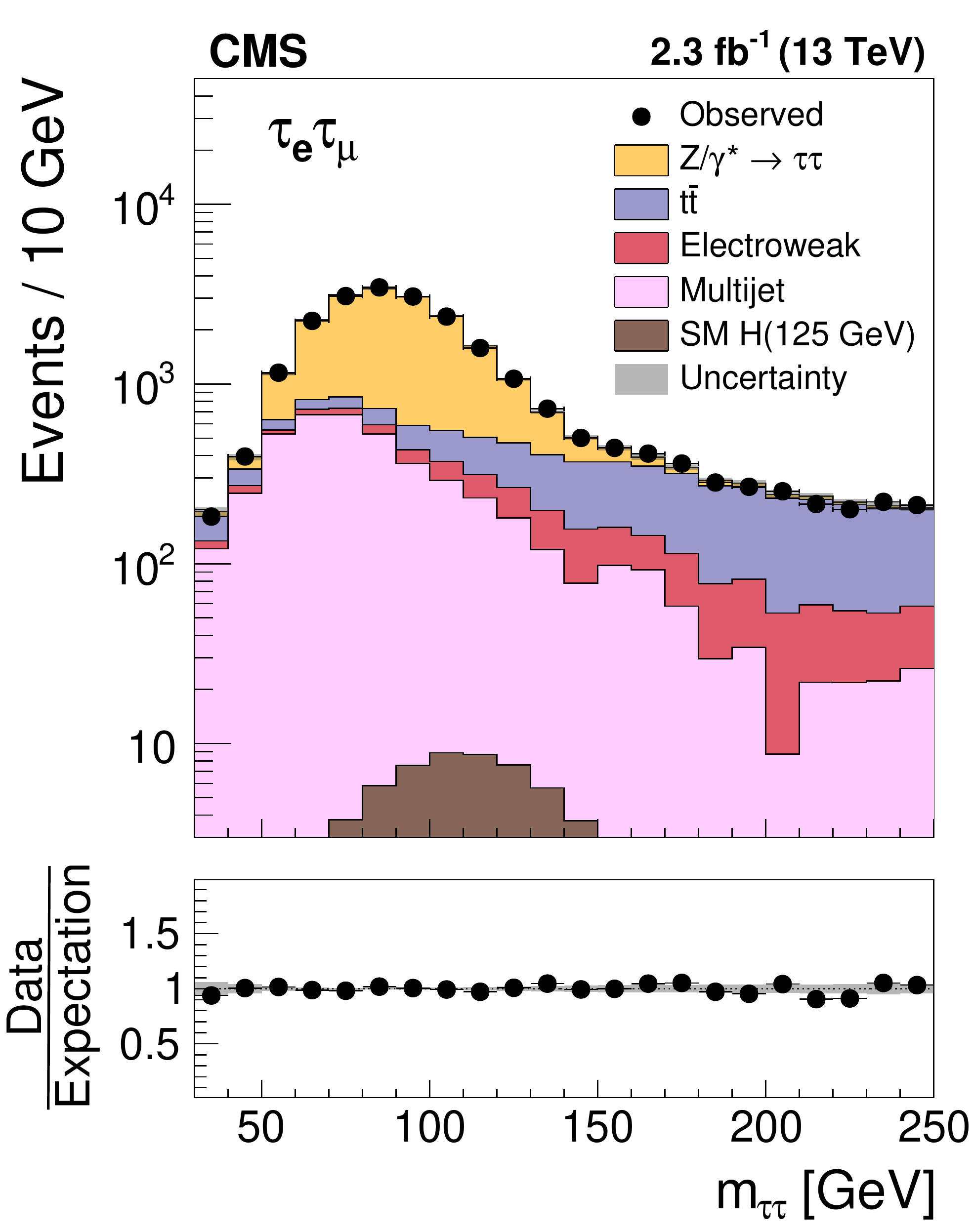} \hfil
\includegraphics[width=\cmsFigWidth]{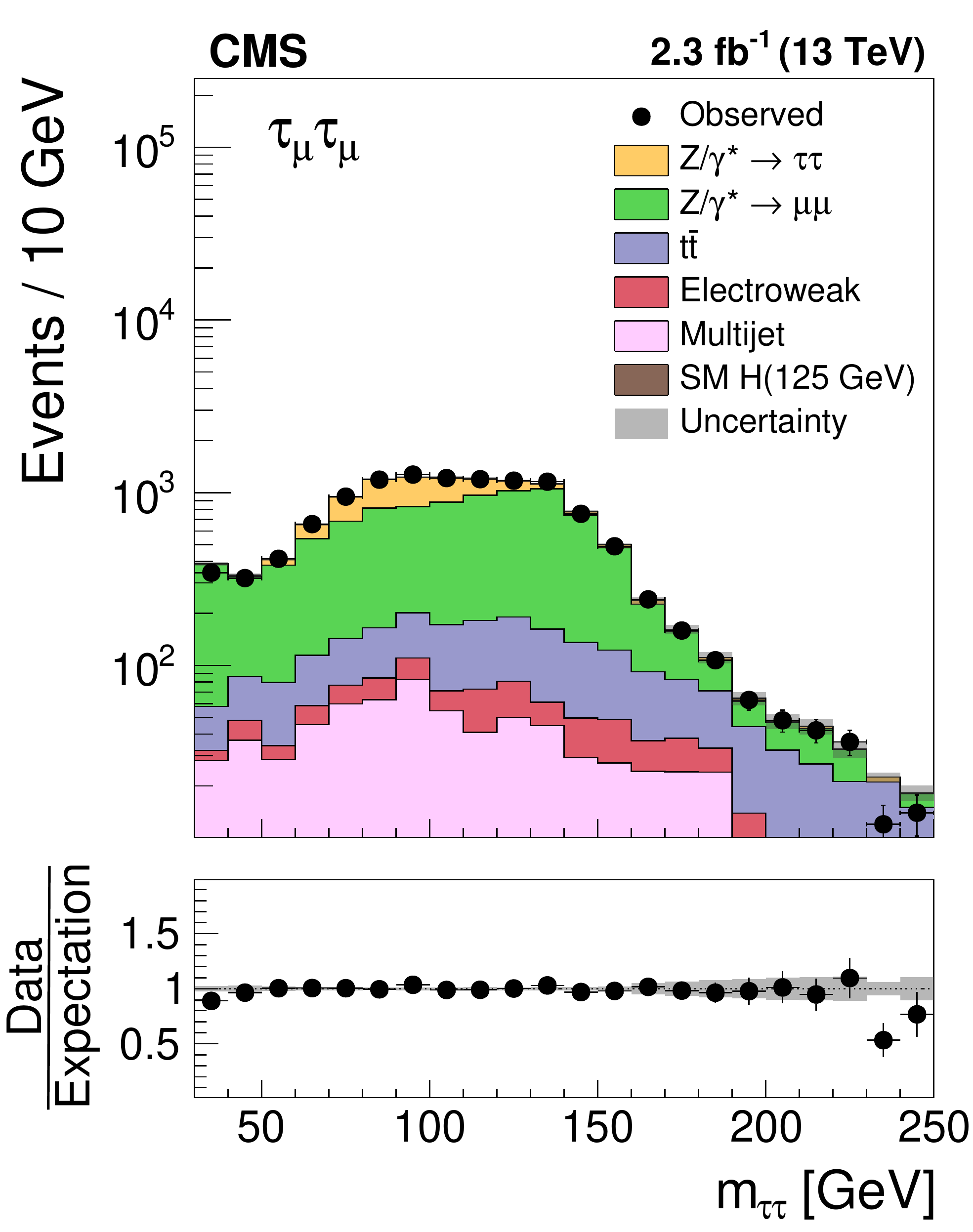}

\caption{
Distributions in $m_{\Pgt\Pgt}$
for events selected in the (left) $\taue\taum$ and (right) $\taum\taum$ channels.
Signal and background contributions are shown for the values of nuisance parameters obtained in the ML fit to the data.
}
\label{fig:postfitPlots_mTauTau2}
\end{figure*}

The yields expected in $\cPZ/\Pggx \to \Pgt\Pgt$ signal and in background contributions from
the ML fit to the $m_{\Pgt\Pgt}$ distributions in the different decay channels
are given in Table~\ref{tab:eventYieldsPostfit}.
The cross sections are displayed in Table~\ref{tab:xSectionResultsIndividual},
and the distributions in $m_{\Pgt\Pgt}$ for the selected events are shown in Figs.~\ref{fig:postfitPlots_mTauTau1} and~\ref{fig:postfitPlots_mTauTau2}.

\begin{table*}[ht!]
\topcaption{
Yields expected in $\cPZ/\Pggx \to \Pgt\Pgt$ signal events and backgrounds
in the $\taue\tauh$, $\taum\tauh$, $\tauh\tauh$, $\taue\taum$, and $\taum\taum$ channels,
obtained from the ML fit described in Section~\ref{sec:signalExtraction}.
The uncertainties are rounded to two significant digits,
except when they are $< 10$, in which case they are rounded to one significant digit,
and the event yields are rounded to match the precision in the uncertainties.
The analysed data corresponds to an integrated luminosity of $2.3~\mathrm{fb}^{-1}$.
}
\label{tab:eventYieldsPostfit}
\centering
\begin{tabular}{l|r@{$ \,\,\pm\,\, $}lr@{$ \,\,\pm\,\, $}lr@{$ \,\,\pm\,\, $}l}
Process & \multicolumn{2}{c}{$\taue\tauh$} & \multicolumn{2}{c}{$\taum\tauh$} & \multicolumn{2}{c}{$\tauh\tauh$} \\
\hline
$\cPZ/\Pggx \to \Pgt\Pgt$                        & $7\,160$ & $130$ & $20\,020$ & $220$ & $415$ & $32$ \\
\hline
Jets misidentified as $\tauh$                    & $5\,690$ & $160$ & $10\,550$ & $220$ & $770$ & $49$ \\
$\cPqt\cPaqt$                                    & $354$ & $26$ & $639$ & $47$ & $17$ & $2$ \\
$\cPZ/\Pggx \to \Pe\Pe$, $\Pgm\Pgm$ ($\Pe$ or $\Pgm$ misidentified as $\tauh$) & $718$ & $96$ & $840$ & $130$ & \multicolumn{2}{c}{---} \\
Electroweak                                      & $93$ & $13$ & $183$ & $28$ & $40$ & $6$ \\
SM $\PHiggs$                                     & $49$ & $11$ & $103$ & $23$ & $13$ & $3$ \\
\hline
Total expected background                        & $6\,900$ & $130$ & $12\,310$ & $180$ & $841$ & $46$ \\
\hline
Total SM expectation                             & $14\,060$ & $120\phantom{\,00}$ & $32\,340$ & $180\phantom{\,00}$ & $1\,255$ & $40\phantom{\,00}$ \\
\hline
Observed data                                    & \multicolumn{2}{c}{$14\,063$} & \multicolumn{2}{c}{$32\,350$} & \multicolumn{2}{c}{$1\,255$} \\
\end{tabular}

\vspace*{0.3 cm}

\begin{tabular}{l|r@{$ \,\,\pm\,\, $}lr@{$ \,\,\pm\,\, $}l}
Process & \multicolumn{2}{c}{$\taue\taum$} & \multicolumn{2}{c}{$\taum\taum$} \\
\hline
$\cPZ/\Pggx \to \Pgt\Pgt$                        & $13\,600$ & $220$ & $2\,067$ & $34$ \\
\hline
Multijet                                         & $4\,620$ & $240$ & $710$ & $110$ \\
$\cPZ/\Pggx \to \Pgm\Pgm$                        & \multicolumn{2}{c}{---} & $8\,010$ & $170$ \\
$\cPqt\cPaqt$                                    & $3\,500$ & $140$ & $1\,239$ & $79$ \\
Electroweak                                      & $1\,146$ & $98$ & $293$ & $30$ \\
SM $\PHiggs$                                     & $57$ & $12$ & $18$ & $4$ \\
\hline
Total expected background                        & $9\,300$ & $210$ & $10\,270$ & $120$ \\
\hline
Total SM expectation                             & $22\,930$ & $130\phantom{\,00}$ & $12\,340$ & $120\phantom{\,00}$ \\
\hline
Observed data                                    & \multicolumn{2}{c}{$22\,930$} & \multicolumn{2}{c}{$12\,327$} \\
\end{tabular}

\end{table*}

\begin{table}
\centering
\topcaption{
Cross section $\sigma(\Pp\Pp \to \cPZ/\Pggx\text{+X}) \, \mathcal{B}(\cPZ/\Pggx \to \Pgt\Pgt)$ measured in individual final states.
}
\label{tab:xSectionResultsIndividual}
\renewcommand{\arraystretch}{1.1}
\begin{tabular}{c|c@{$\, \,\pm\,\, $}r@{$\stat\,\pm\,\, $}r@{$\syst \,\pm\,\, $}c}
Channel & \multicolumn{4}{c}{$\sigma(\Pp\Pp \to \cPZ/\Pggx\text{+X}) \, \mathcal{B}(\cPZ/\Pggx \to \Pgt\Pgt)$~[pb]} \\
\hline
$\taue\tauh$ & $1799$ & $29$ & $120$ & $34\lum$ \\
$\taum\tauh$ & $1784$ & $17$ & $117$ & $34\lum$ \\
$\tauh\tauh$ & $1477$ & $137$ & $270$ & $30\lum$ \\
$\taue\taum$ & $1851$ & $19$ & $58$ & $34\lum$ \\
$\taum\taum$ & $1967$ & $121$ & $92$ & $37\lum$ \\
\end{tabular}

\end{table}

\begin{figure*}[ht!]
\centering
\includegraphics[width=\cmsFigWidth]{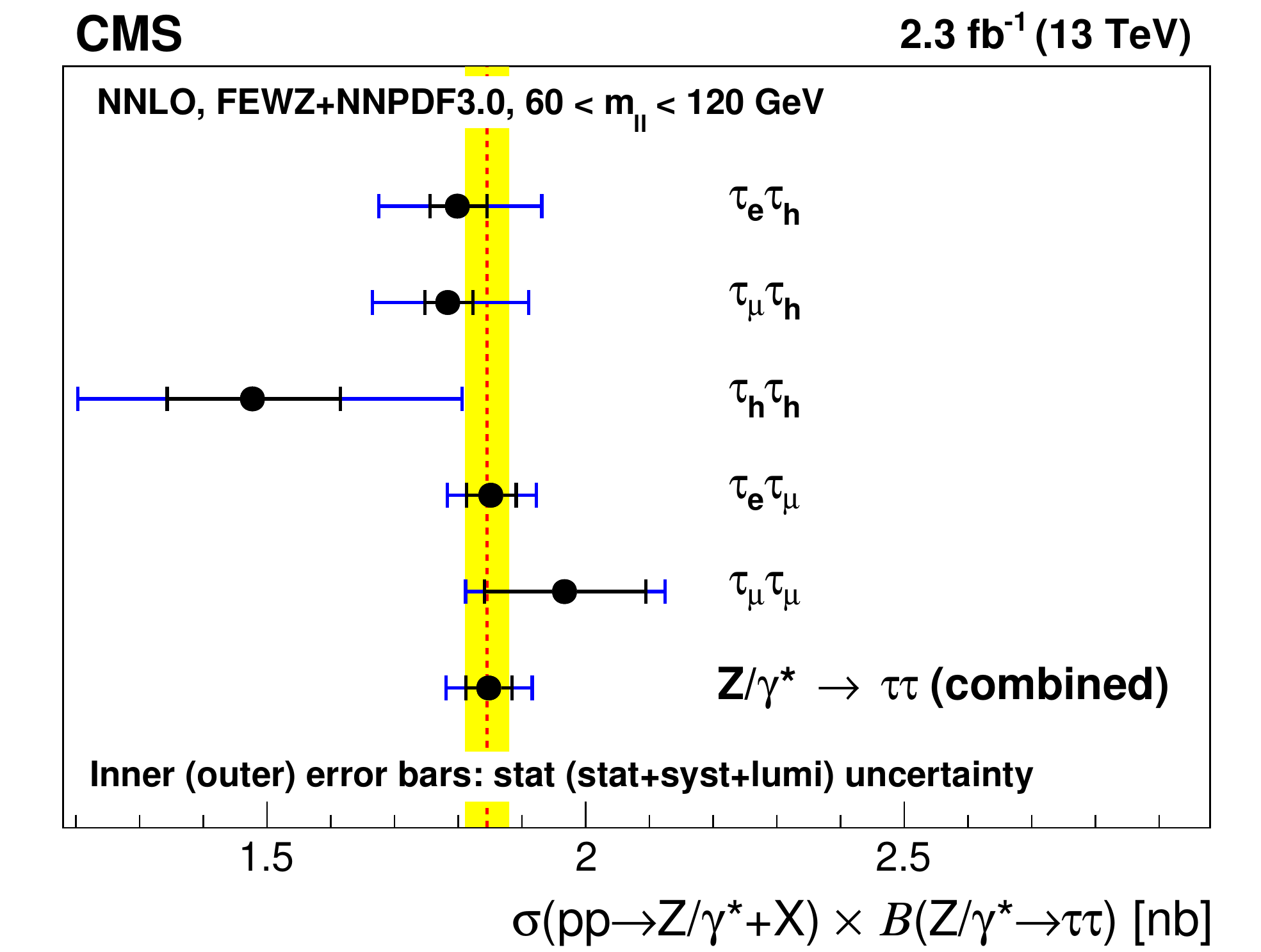}

\caption{
The inclusive cross section $\sigma(\Pp\Pp \to \cPZ/\Pggx\text{+X}) \,  \mathcal{B}(\cPZ/\Pggx \to \Pgt\Pgt)$ measured in individual channels,
and in the combination of all final states,
compared to the theoretical prediction~\cite{FEWZ3}.
}
\label{fig:xSectionResult}
\end{figure*}

{\tolerance=10000
The total uncertainty in the cross section is decomposed into statistical contributions, uncertainty in the integrated luminosity of the data,
and other systematic uncertainties, as described in Section~\ref{sec:signalExtraction}.
The measured values are compatible with each other.
The largest deviation, amounting to a little more than one standard deviation, is observed in the $\tauh\tauh$ channel.
A deviation of this magnitude is expected.
We proceed to a simultaneous fit of the $m_{\Pgt\Pgt}$ distributions in the five final states.
The value of the cross section obtained from the combined fit is:
\begin{multline}
\sigma(\Pp\Pp \to \cPZ/\Pggx\text{+X}) \,  \mathcal{B}(\cPZ/\Pggx \to \Pgt\Pgt) = \\
1848 \pm 12\stat \pm 57\syst \pm 35\lum\unit{pb}.
\label{eq:xSectionResultCombined}
\end{multline}
The result is compatible with the prediction of $1845^{+12}_{-6}\text{~(scale)} \pm 33\text{~(PDF)}$~pb,
computed at NNLO accuracy~\cite{FEWZ3} using the {NNPDF3.0} PDF.
The results are illustrated in Fig.~\ref{fig:xSectionResult}.
The inner and outer error bars represent, respectively, the statistical uncertainties,
and the quadratic sum of the uncertainties in the statistical, systematic, and integrated-luminosity components.
The uncertainty in $\sigma(\Pp\Pp \to \cPZ/\Pggx\text{+X}) \,  \mathcal{B}(\cPZ/\Pggx \to \Pgt\Pgt)$
arising from the uncertainty in the integrated luminosity
is smaller than the uncertainty in the integrated luminosity, for the reasons discussed in Section~\ref{sec:systematicUncertainties}.
\par}

As a side note, the values of the nuisance parameters that correspond to the cross sections in the $\cPZ/\Pggx \to \Pe\Pe$ and $\cPZ/\Pggx \to \Pgm\Pgm$ backgrounds,
obtained from the simultaneous fit to the $m_{\Pgt\Pgt}$ distributions in the five final states in data,
are also compatible with the expected values.

\begin{figure*}[ht!]
\centering
\includegraphics[width=\cmsFigWidth]{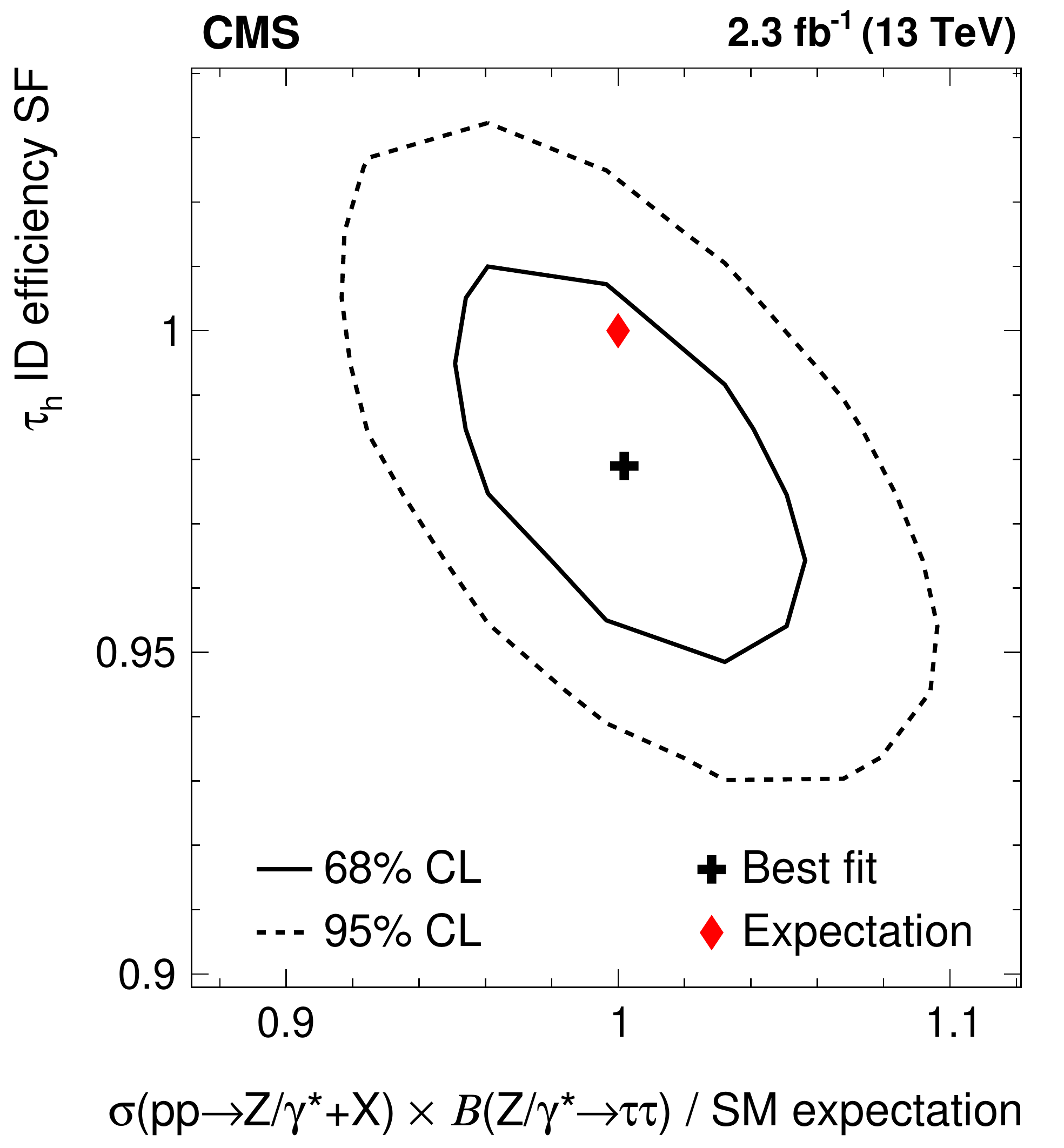} \hfil
\includegraphics[width=\cmsFigWidth]{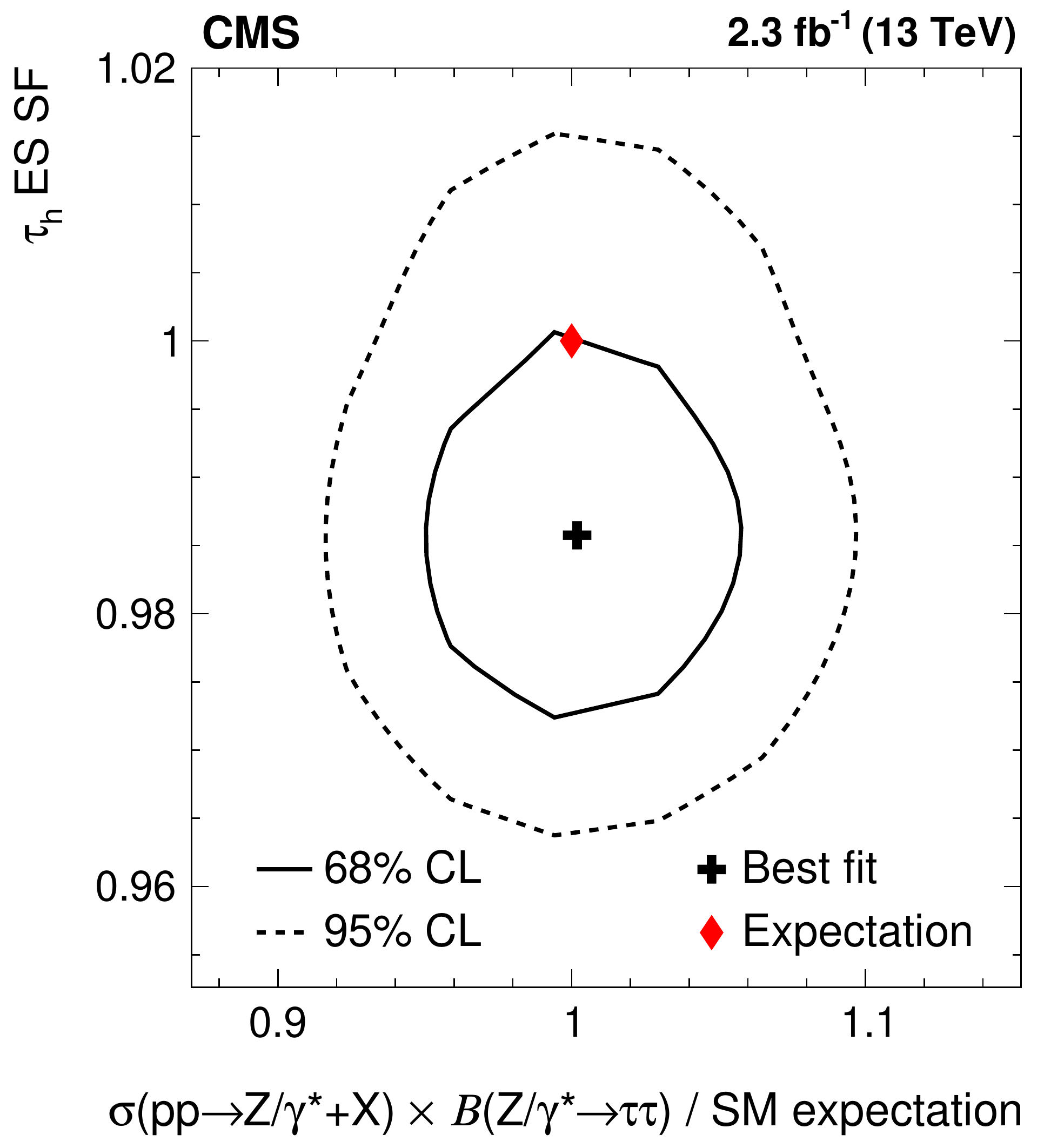} \\
\includegraphics[width=\cmsFigWidth]{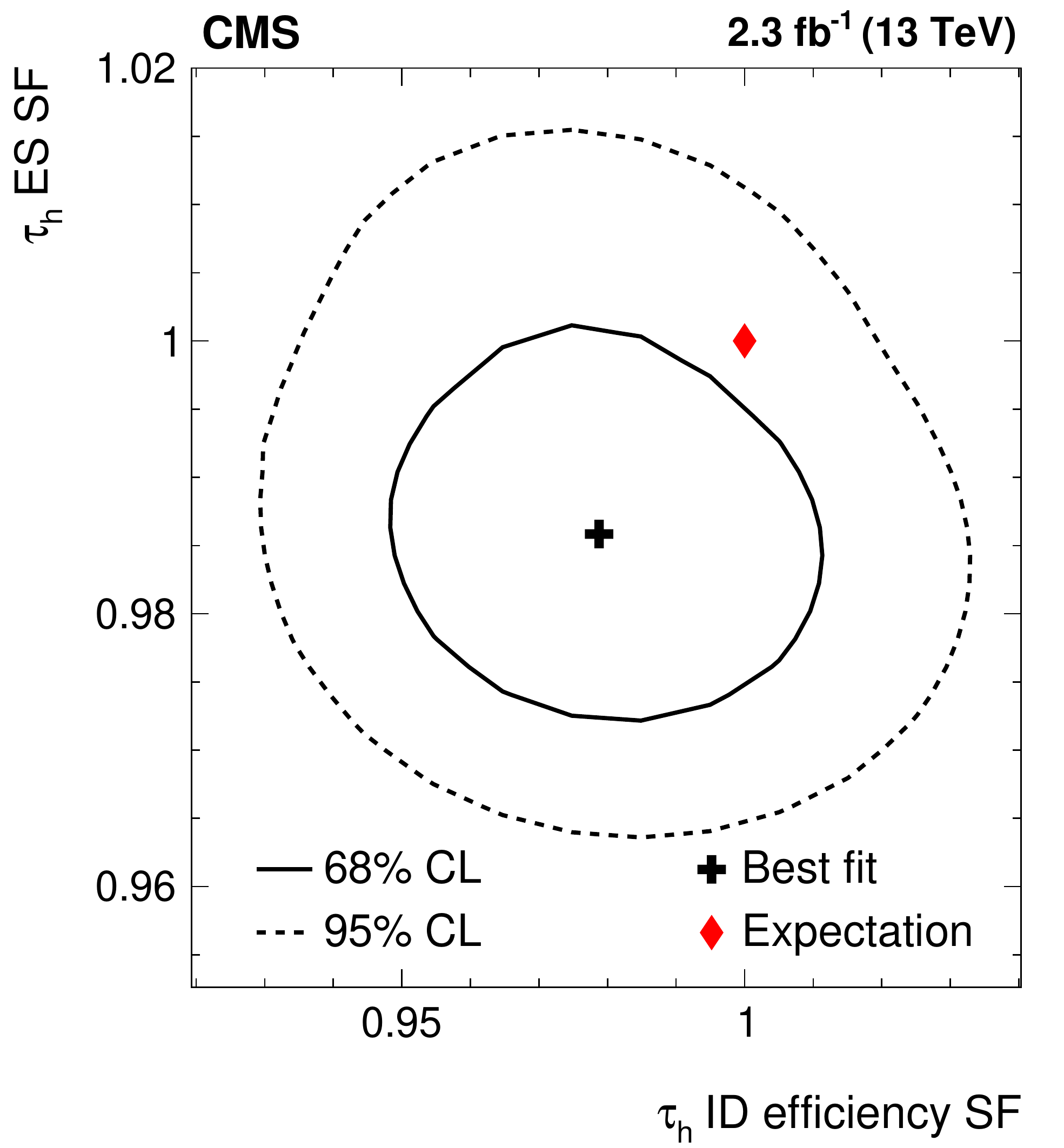}

\caption{
Likelihood contours for the joint parameter estimation of
(upper left) $\sigma(\Pp\Pp \to \cPZ/\Pggx\text{+X}) \,  \mathcal{B}(\cPZ/\Pggx \to \Pgt\Pgt)$ and the $\tauh$\,ID efficiency,
(upper right) $\sigma(\Pp\Pp \to \cPZ/\Pggx\text{+X}) \,  \mathcal{B}(\cPZ/\Pggx \to \Pgt\Pgt)$ and $\tauh$\,ES,
and (lower) the $\tauh$\,ES and the $\tauh$\,ID efficiency,
at $68$ and $95\%$ confidence level (CL).
The values of the $\tauh$\,ID efficiency and of $\tauh$\,ES are quoted in terms of scale factors (SF) relative to their standard model, MC expectation.
}
\label{fig:likelihoodFunctionProjections}
\end{figure*}

{\tolerance=600
Two-dimensional projections of $-2 \ln \lambda\left(\xi\right)$, obtained when the $\tauh$\,ID efficiency and $\tauh$\,ES are left unconstrained in the fit,
are shown in Fig.~\ref{fig:likelihoodFunctionProjections}.
Measured values of the $\tauh$\,ID efficiency and of $\tauh$\,ES are quoted as scale factors (SF) relative to their MC expectation.
The values of $\sigma(\Pp\Pp \to \cPZ/\Pggx\text{+X}) \,  \mathcal{B}(\cPZ/\Pggx \to \Pgt\Pgt)$,
$\tauh$\,ID efficiency, and $\tauh$\,ES that minimize $-2 \ln \lambda\left(\xi\right)$,
yielding the best fit to the data,
are indicated by a cross.
Contours for which $-2 \ln \lambda\left(\xi\right)$ exceeds its minimum value by $2.30$ and $6.18$ units,
corresponding to coverage probabilities of $68$ and $95\%$ in the two-dimensional parameter plane, are also shown.
The $68\%$ CIs for the $\tauh$\,ID efficiency and $\tauh$\,ES are obtained
as the values of the respective parameter for which $-2 \ln \lambda\left(\xi\right)$ increases by one unit relative to its minimum.
The measured SF for the $\tauh$\,ID efficiency and for $\tauh$\,ES amount to $0.979 \pm 0.022$ and $0.986 \pm 0.009$, respectively.
Both SF are compatible with unity, indicating that the measured values of the $\tauh$\,ID efficiency and of the $\tauh$\,ES are in agreement with the MC expectation.
The expected $\tauh$\,ID efficiency in the LHC data is documented in Ref.~\cite{TAU-16-002}.
\par}

\section{Summary}
\label{sec:summary}

The cross section for inclusive Drell--Yan production of $\Pgt$ pairs has been measured
using $\Pp\Pp$ collisions recorded by the CMS experiment at $\sqrt{s} = 13\TeV$ at the LHC.
The analysed data correspond to an integrated luminosity of $2.3~\mathrm{fb}^{-1}$.
The signal yield was determined in a global fit to the mass distributions in five $\Pgt\Pgt$ decay channels:
$\taue\tauh$, $\taum\tauh$, $\tauh\tauh$, $\taue\taum$, and $\taum\taum$.
The measured cross section times branching fraction
$\sigma(\Pp\Pp \to \cPZ/\Pggx\text{+X}) \,  \mathcal{B}(\cPZ/\Pggx \to \Pgt\Pgt)
= 1848 \pm 12\stat \pm 57\syst \pm 35\lum\unit{pb}$
is in agreement with the standard model expectation, computed at next-to-next-to-leading order accuracy in perturbation theory.
As a byproduct of the global fit,
the efficiency for reconstructing and identifying the decays of $\Pgt$ leptons to hadrons ($\Pgt \to \mbox{hadrons} + \Pnut$),
as well as the $\tauh$ energy scale, have been determined.
The results from data agree with Monte Carlo simulation
within the uncertainties of the measurement,
amounting to $2.2\%$ relative uncertainty in the $\tauh$ identification efficiency, and $0.9\%$ in the energy scale.

\ifthenelse{\boolean{cms@external}}{}{\clearpage}

\begin{acknowledgments}

\hyphenation{Bundes-ministerium Forschungs-gemeinschaft Forschungs-zentren Rachada-pisek}

We congratulate our colleagues in the CERN accelerator departments for the excellent performance of the LHC and thank the technical and administrative staffs at CERN and at other CMS institutes for their contributions to the success of the CMS effort. In addition, we gratefully acknowledge the computing centres and personnel of the Worldwide LHC Computing Grid for delivering so effectively the computing infrastructure essential to our analyses. Finally, we acknowledge the enduring support for the construction and operation of the LHC and the CMS detector provided by the following funding agencies: the Austrian Federal Ministry of Science, Research and Economy and the Austrian Science Fund; the Belgian Fonds de la Recherche Scientifique, and Fonds voor Wetenschappelijk Onderzoek; the Brazilian Funding Agencies (CNPq, CAPES, FAPERJ, and FAPESP); the Bulgarian Ministry of Education and Science; CERN; the Chinese Academy of Sciences, Ministry of Science and Technology, and National Natural Science Foundation of China; the Colombian Funding Agency (COLCIENCIAS); the Croatian Ministry of Science, Education and Sport, and the Croatian Science Foundation; the Research Promotion Foundation, Cyprus; the Secretariat for Higher Education, Science, Technology and Innovation, Ecuador; the Ministry of Education and Research, Estonian Research Council via IUT23-4 and IUT23-6 and European Regional Development Fund, Estonia; the Academy of Finland, Finnish Ministry of Education and Culture, and Helsinki Institute of Physics; the Institut National de Physique Nucl\'eaire et de Physique des Particules~/~CNRS, and Commissariat \`a l'\'Energie Atomique et aux \'Energies Alternatives~/~CEA, France; the Bundesministerium f\"ur Bildung und Forschung, Deutsche Forschungsgemeinschaft, and Helmholtz-Gemeinschaft Deutscher Forschungszentren, Germany; the General Secretariat for Research and Technology, Greece; the National Scientific Research Foundation, and National Innovation Office, Hungary; the Department of Atomic Energy and the Department of Science and Technology, India; the Institute for Studies in Theoretical Physics and Mathematics, Iran; the Science Foundation, Ireland; the Istituto Nazionale di Fisica Nucleare, Italy; the Ministry of Science, ICT and Future Planning, and National Research Foundation (NRF), Republic of Korea; the Lithuanian Academy of Sciences; the Ministry of Education, and University of Malaya (Malaysia); the Mexican Funding Agencies (BUAP, CINVESTAV, CONACYT, LNS, SEP, and UASLP-FAI); the Ministry of Business, Innovation and Employment, New Zealand; the Pakistan Atomic Energy Commission; the Ministry of Science and Higher Education and the National Science Centre, Poland; the Funda\c{c}\~ao para a Ci\^encia e a Tecnologia, Portugal; JINR, Dubna; the Ministry of Education and Science of the Russian Federation, the Federal Agency of Atomic Energy of the Russian Federation, Russian Academy of Sciences, the Russian Foundation for Basic Research and the Russian Competitiveness Program of NRNU ``MEPhI"; the Ministry of Education, Science and Technological Development of Serbia; the Secretar\'{\i}a de Estado de Investigaci\'on, Desarrollo e Innovaci\'on, Programa Consolider-Ingenio 2010, Plan de Ciencia, Tecnolog\'{i}a e Innovaci\'on 2013-2017 del Principado de Asturias and Fondo Europeo de Desarrollo Regional, Spain; the Swiss Funding Agencies (ETH Board, ETH Zurich, PSI, SNF, UniZH, Canton Zurich, and SER); the Ministry of Science and Technology, Taipei; the Thailand Center of Excellence in Physics, the Institute for the Promotion of Teaching Science and Technology of Thailand, Special Task Force for Activating Research and the National Science and Technology Development Agency of Thailand; the Scientific and Technical Research Council of Turkey, and Turkish Atomic Energy Authority; the National Academy of Sciences of Ukraine, and State Fund for Fundamental Researches, Ukraine; the Science and Technology Facilities Council, UK; the US Department of Energy, and the US National Science Foundation.

Individuals have received support from the Marie-Curie programme and the European Research Council and Horizon 2020 Grant, contract No. 675440 (European Union); the Leventis Foundation; the A. P. Sloan Foundation; the Alexander von Humboldt Foundation; the Belgian Federal Science Policy Office; the Fonds pour la Formation \`a la Recherche dans l'Industrie et dans l'Agriculture (FRIA-Belgium); the Agentschap voor Innovatie door Wetenschap en Technologie (IWT-Belgium); the Ministry of Education, Youth and Sports (MEYS) of the Czech Republic; the Council of Scientific and Industrial Research, India; the HOMING PLUS programme of the Foundation for Polish Science, cofinanced from European Union, Regional Development Fund, the Mobility Plus programme of the Ministry of Science and Higher Education, the National Science Center (Poland), contracts Harmonia 2014/14/M/ST2/00428, Opus 2014/13/B/ST2/02543, 2014/15/B/ST2/03998, and 2015/19/B/ST2/02861, Sonata-bis 2012/07/E/ST2/01406; the National Priorities Research Program by Qatar National Research Fund; the Programa Severo Ochoa del Principado de Asturias; the Thalis and Aristeia programmes cofinanced by EU-ESF and the Greek NSRF; the Rachadapisek Sompot Fund for Postdoctoral Fellowship, Chulalongkorn University and the Chulalongkorn Academic into Its 2nd Century Project Advancement Project (Thailand); the Welch Foundation, contract C-1845; and the Weston Havens Foundation (USA).

\end{acknowledgments}

\ifthenelse{\boolean{cms@external}}{}{\newpage}

\bibliography{auto_generated}

\ifthenelse{\boolean{cms@external}}{}{\newpage}

\appendix
\section{Validation of background model in event categories}
\label{sec:appendix}

The validity of the background estimation described in Section~\ref{sec:backgroundEstimation}
is checked in event categories that are relevant for the SM $\PHiggs \to \Pgt\Pgt$ analysis as well as in searches for new physics.

Event categories based on jet multiplicity,
$\pT$ of the $\Pgt$ lepton pair, and on the multiplicity of $\Pbottom$ jets
are defined by the conditions given in Table~\ref{tab:exclEventCategories}.

\begin{table*}[ht!]
\topcaption{
Event categories used to study the modelling of backgrounds.
Similar categories have been used in previous $\PHiggs \to \Pgt\Pgt$ analyses at the LHC.
}
\label{tab:exclEventCategories}
\centering
\renewcommand{\arraystretch}{1.1}
\begin{tabular}{l|l}
Category & Selection \\
\hline
$0$-jet & No jets$^{1}$ and no $\Pbottom$ jets$^{2}$ \\
$1$-jet, low $\cPZ$ boson $\pT$ & At least one jet$^{1}$, no $\Pbottom$ jets$^{2}$, $\pT^{\,\cPZ} < 50\GeV$, \\
& excluding events selected in $2$-jet VBF category \\
$1$-jet, medium $\cPZ$ boson $\pT$ & At least one jet$^{1}$, no $\Pbottom$ jets$^{2}$, $50 < \pT^{\,\cPZ} < 100\GeV$, \\
& excluding events selected in $2$-jet VBF category \\
$1$-jet, high $\cPZ$ boson $\pT$ & At least one jet$^{1}$, no $\Pbottom$ jets$^{2}$, $\pT^{\,\cPZ} > 100\GeV$, \\
& excluding events selected in $2$-jet VBF category \\
$2$-jet VBF & At least one pair of jets$^{1}$ satisfying $m_{\textrm{jj}} > 500\GeV$ and $\Delta\eta_{\textrm{jj}} > 3.5$, \\
& no $\Pbottom$ jets$^{2}$ \\
$1$ $\Pbottom$ jet & Exactly one $\Pbottom$ jet$^{2}$  \\
$2$ $\Pbottom$ jet & Exactly two $\Pbottom$ jets$^{2}$ \\
\end{tabular}

$^{1}$ With $\pT > 30\GeV$ and $\lvert \eta \rvert < 4.7$ \\
$^{2}$ With $\pT > 20\GeV$, $\lvert \eta \rvert < 2.4$, and identified by the CSV algorithm
as originating from the hadronization of $\Pbottom$ quarks \\
\end{table*}

The transverse momentum of the $\cPZ$ boson ($\pT^{\,\cPZ}$)
is reconstructed by adding the momentum vectors from the visible $\Pgt$ decay products and the reconstructed $\vecMET$ in the transverse plane.
The observables $m_{\textrm{jj}}$ and $\Delta\eta_{\textrm{jj}}$ are used to select signal events produced through the fusion of virtual vector bosons (VBF)
in the SM $\PHiggs \to \Pgt\Pgt$ analysis, and refer, respectively, to the mass and to the separation in pseudorapidity of the two jets of highest $\pT$ in events containing two or more jets.

Background contributions arising from
$\cPZ/\Pggx \to \Pe\Pe$, $\cPZ/\Pggx \to \Pgm\Pgm$, $\PW$+jets, $\cPqt\cPaqt$, single top quark, and diboson production
to the event categories defined in Table~\ref{tab:exclEventCategories}
in the $\taue\tauh$, $\taum\tauh$, $\tauh\tauh$, and $\taue\taum$ channels
are estimated as described above.
The fractions $R_{\textrm{p}}$ of multijet, $\PW$+jets, DY, and $\cPqt\cPaqt$ backgrounds
used in Eq.~(\ref{eq:jetToTauFakeRate_weight})
are calculated separately for each of the event categories.

The contribution of $\cPZ/\Pggx \to \Pgt\Pgt$ is determined from data, using $\cPZ/\Pggx \to \Pgm\Pgm$ events.
Events passing the single-muon trigger are selected by the presence of two muons of opposite charge passing tight identification and isolation criteria.
At least one of the muons is required to have $\pT > 20\GeV$ and $\lvert \eta \rvert < 2.1$,
while the other muon is required to satisfy the conditions $\pT > 10\GeV$ and $\lvert \eta \rvert < 2.4$.
The number of $\cPZ/\Pggx \to \Pgm\Pgm$ candidate events selected in the different categories in data
is compared to the MC expectation for $\cPZ/\Pggx \to \Pgm\Pgm$ production,
and their
ratio is used as a scale factor to correct the MC expectation for the $\cPZ/\Pggx \to \Pgt\Pgt$ event yield in that category.
The expected contribution of background processes, obtained from MC simulation,
is subtracted from the data before taking the ratio.
The selection criteria applied on muon $\pT$ and $\eta$ in $\cPZ/\Pggx \to \Pgm\Pgm$,
and on $\pT$ and $\eta$ of the visible $\Pgt$ decay products in $\cPZ/\Pggx \to \Pgt\Pgt$ events
are known to cause a bias in the $\pT^{\,\cPZ}$ distribution.
The latter is correlated with the multiplicity of jets.
The bias must be corrected, as its magnitude is very different for $\cPZ/\Pggx \to \Pgm\Pgm$ and $\cPZ/\Pggx \to \Pgt\Pgt$ events.
The bias is emulated by replacing the muons reconstructed in $\cPZ/\Pggx \to \Pgm\Pgm$ candidate events with generator-level $\Pgt$ leptons.
The $\Pgt$ leptons are decayed using {\sc tauola++} $1.1.4$~\cite{TAUOLA,TAUOLApp}, and
effects of $\Pgt$ lepton polarization in the decays are modelled through weights computed with the {\sc TauSpinner}~\cite{TauSpinner} program.
A sample of $1000$ random $\Pgt$ lepton decays is generated for each $\cPZ/\Pggx \to \Pgm\Pgm$ candidate event,
and the weights computed in {\sc TauSpinner} are recorded for each decay.
The ratio of the sum of the weights for decays in which the visible products of both $\Pgt$ leptons pass selection criteria on $\pT$ and $\eta$,
to the sum of all weights computed for the $1000$ decays,
is applied as event weight to the $\cPZ/\Pggx \to \Pgm\Pgm$ candidate, which
corrects for the difference in bias of $\pT^{\,\cPZ}$ caused by selection criteria on between $\cPZ/\Pggx \to \Pgm\Pgm$ and $\cPZ/\Pggx \to \Pgt\Pgt$ events.
The procedure is validated through MC simulation.

The contributions of background processes that are modelled in the MC simulation
to the different categories are affected by uncertainties in the jet energy scale and resolution.
The energy scale of jets is measured using the $\pT$ balance of jets with $\cPZ$ bosons and photons
in $\cPZ/\Pggx \to \Pe\Pe$ and $\cPZ/\Pggx \to \Pgm\Pgm$ and $\Pgg$+jets events
and the $\pT$ balance between jets in dijet events as described in Ref.~\cite{JME-13-004}.
The uncertainty in the jet energy scale is a few percent and depends on $\pT$ and $\eta$.
The impact of jet energy scale and resolution uncertainties on the yields of background processes
is evaluated by varying the jet energy scale and resolution within their uncertainties,
redetermining the multiplicity of jets and $\Pbottom$ jets,
and reapplying the event categorization conditions given in Table~\ref{tab:exclEventCategories}.

\begin{figure*}
\centering
\includegraphics[width=\cmsFigWidth]{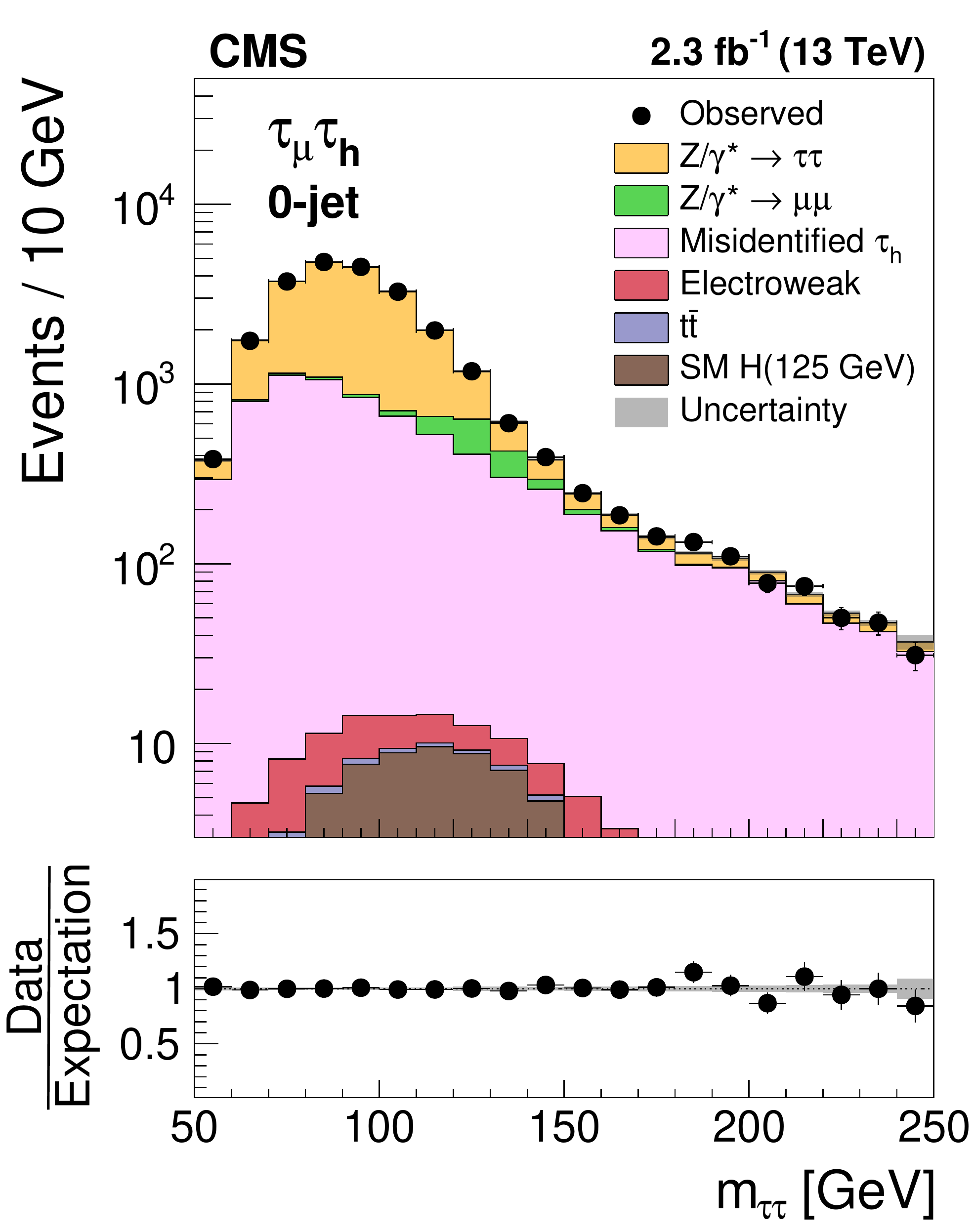} \hfil
\includegraphics[width=\cmsFigWidth]{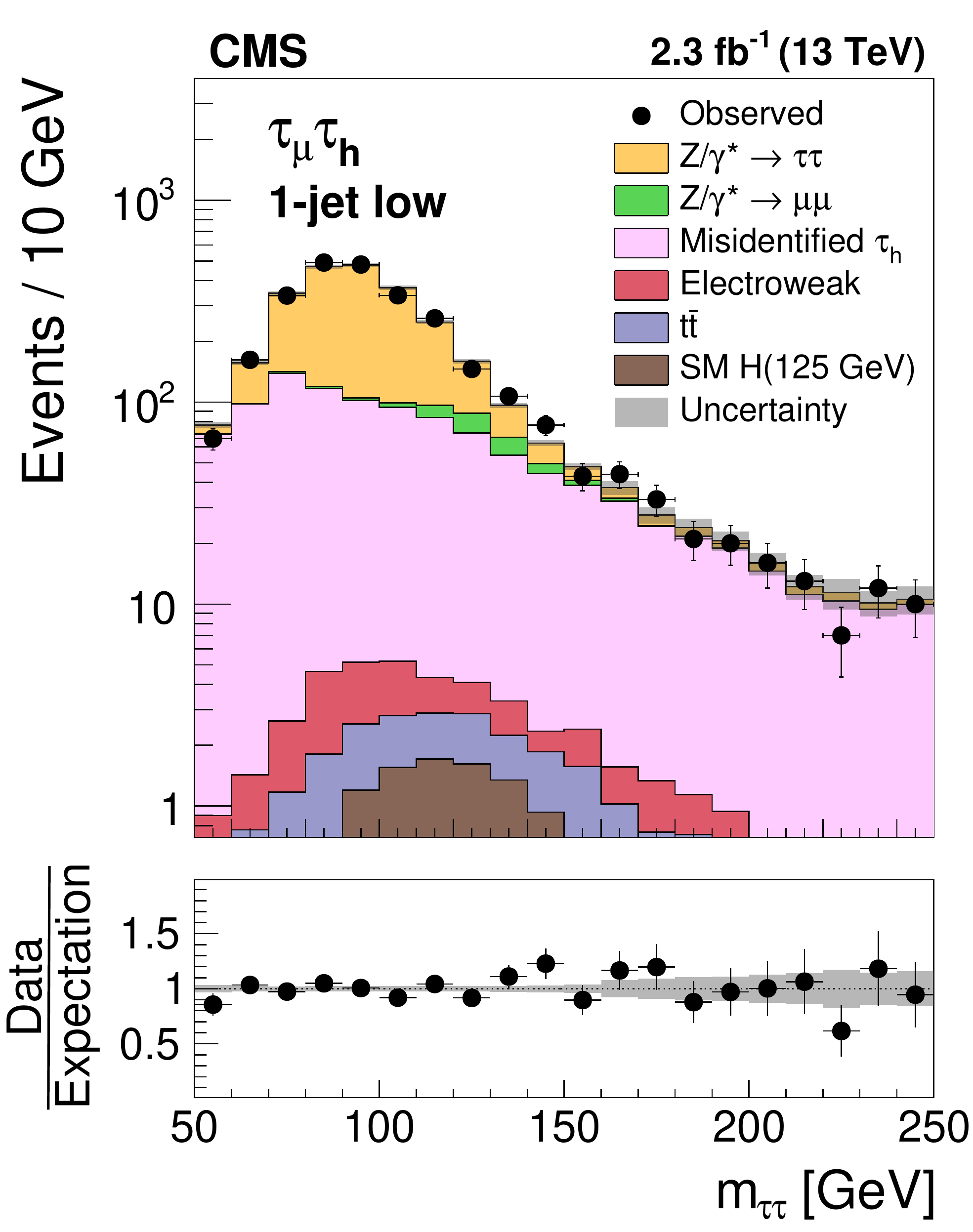} \\
\includegraphics[width=\cmsFigWidth]{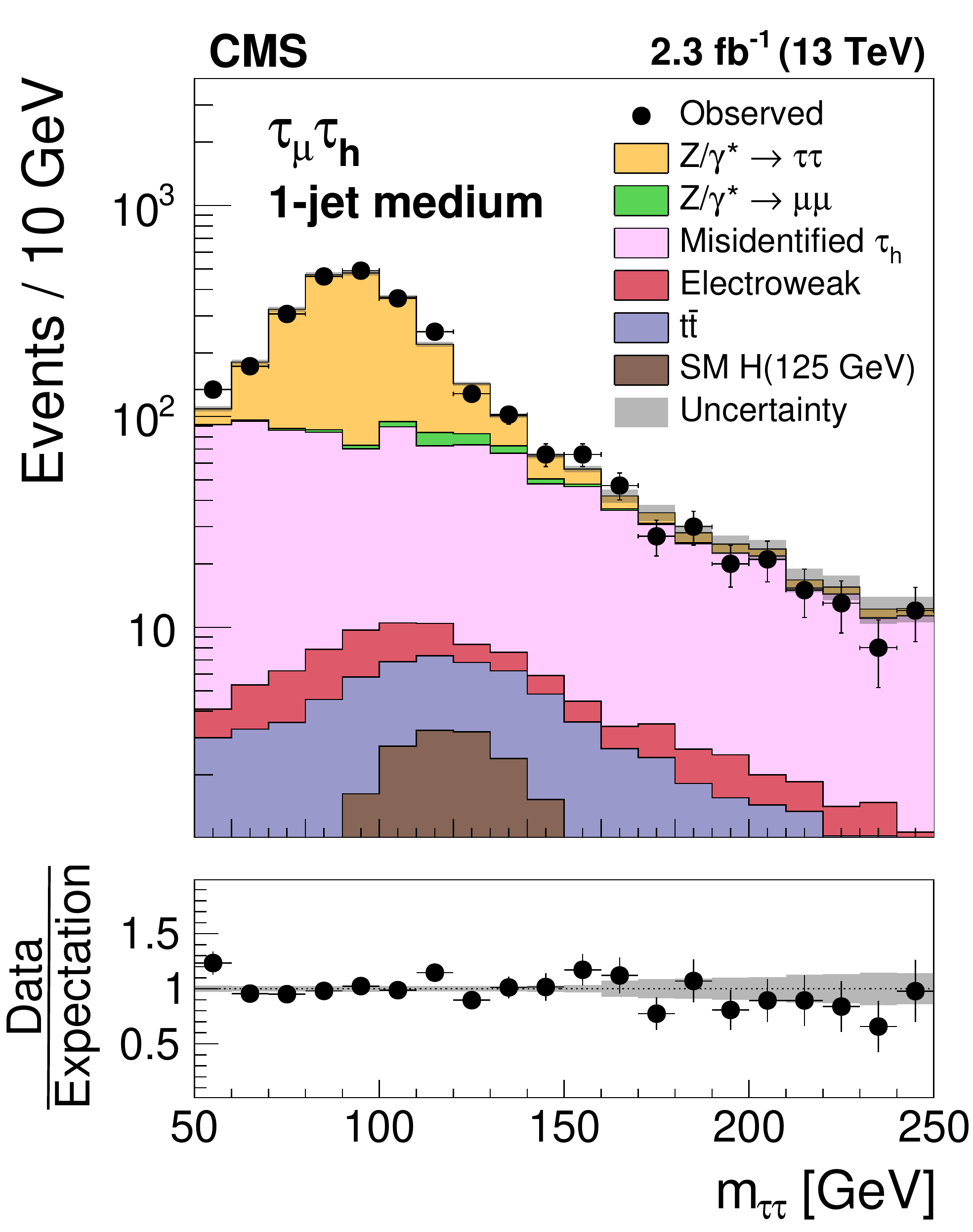} \hfil
\includegraphics[width=\cmsFigWidth]{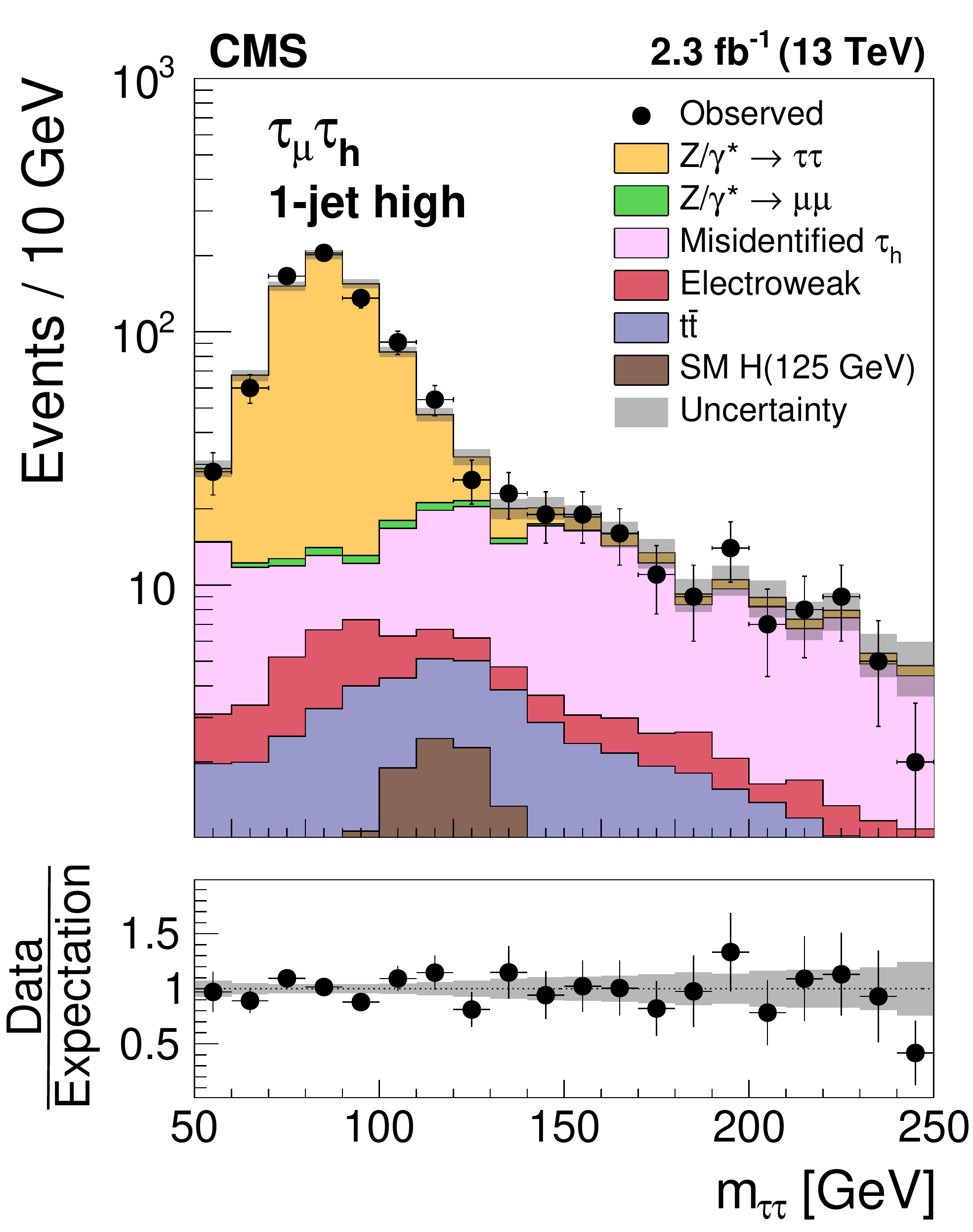}

\caption{
Distributions in $m_{\Pgt\Pgt}$
for different categories in the $\taum\tauh$ channel:
(upper left) $0$-jet,
(upper right) $1$-jet low, (lower left) medium, and (lower right) high $\cPZ$ boson $\pT$.
}
\label{fig:evtCategoryControlPlots_mutau1}
\end{figure*}

\begin{figure*}
\centering
\includegraphics[width=\cmsFigWidth]{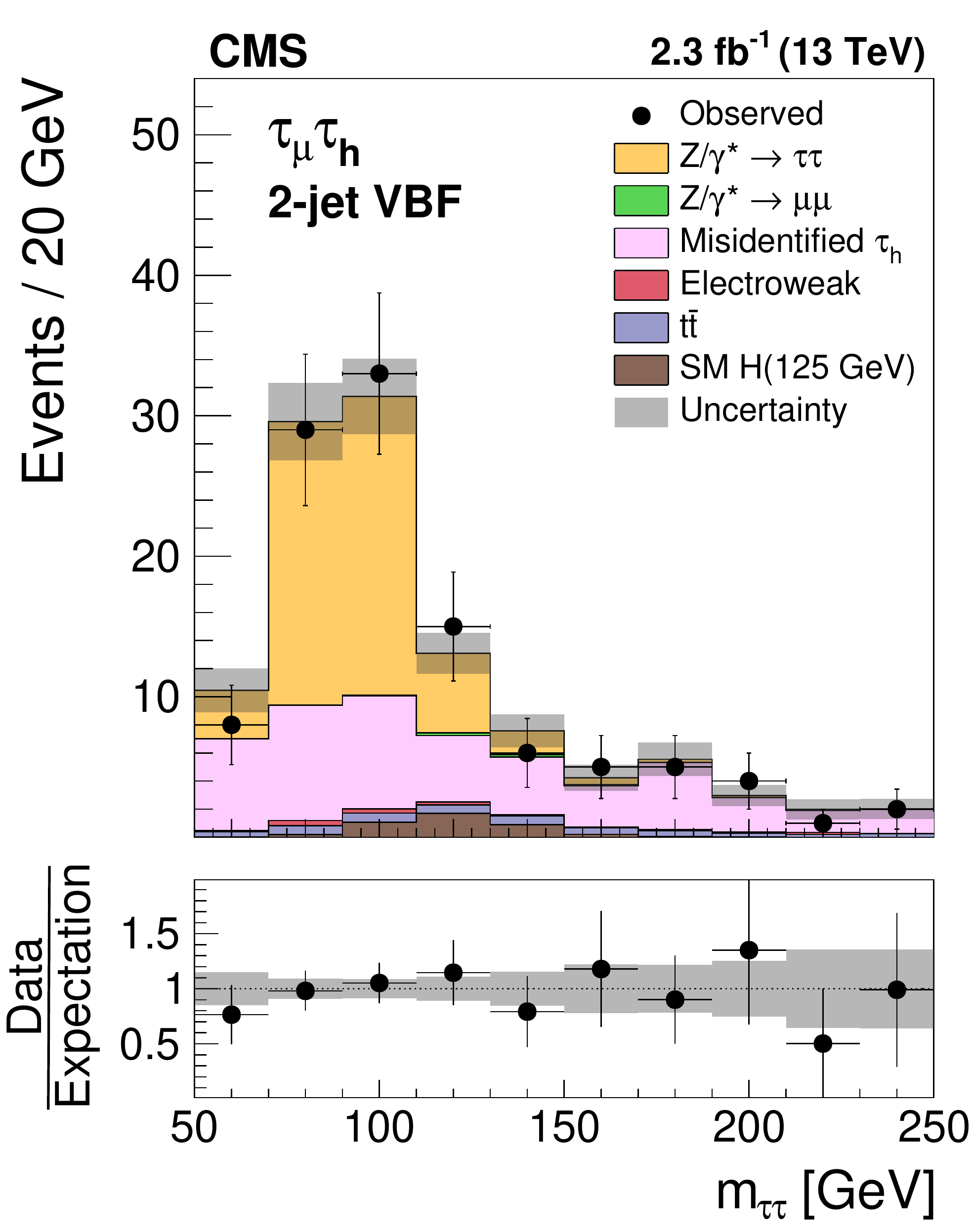} \hfil
\includegraphics[width=\cmsFigWidth]{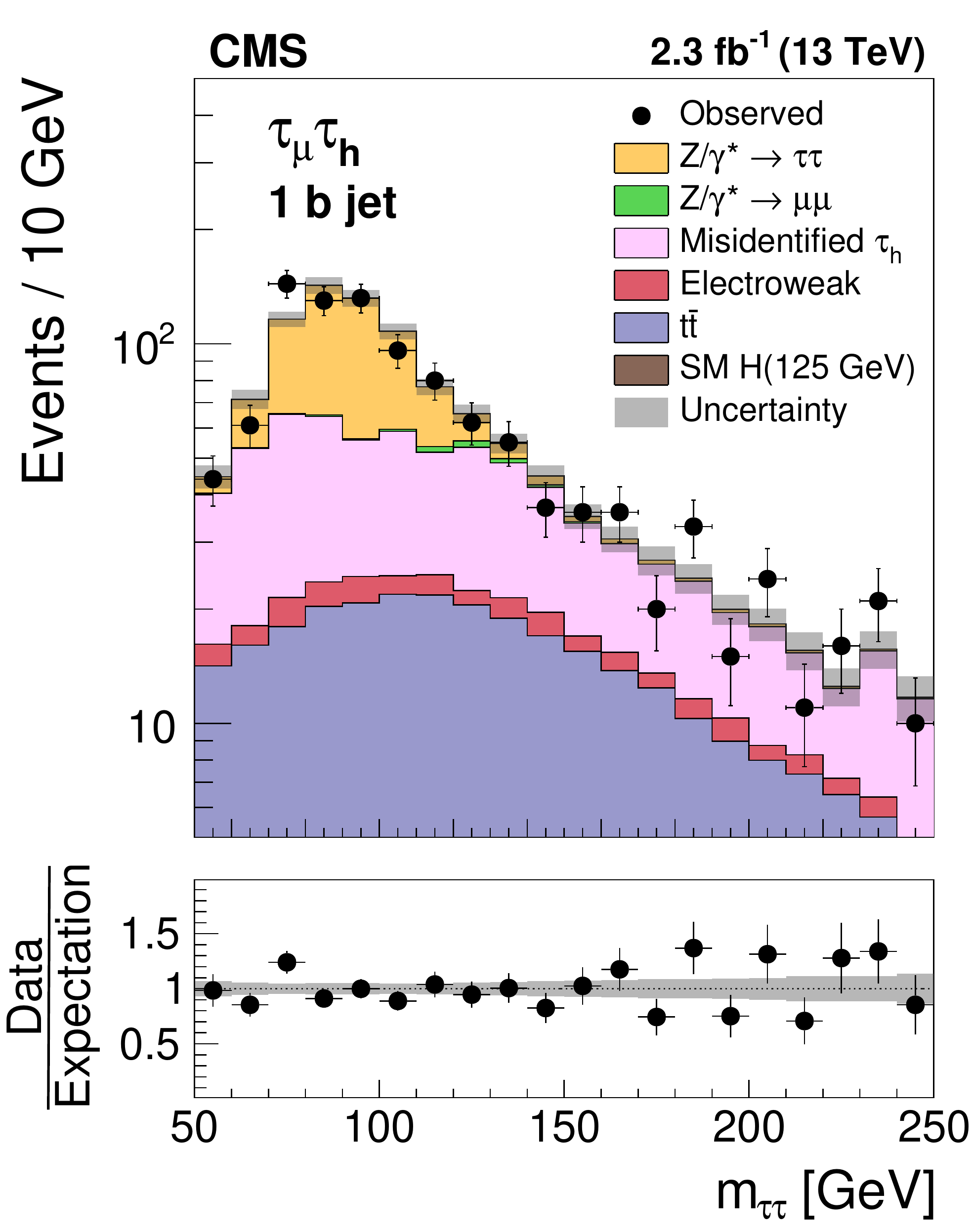} \\
\includegraphics[width=\cmsFigWidth]{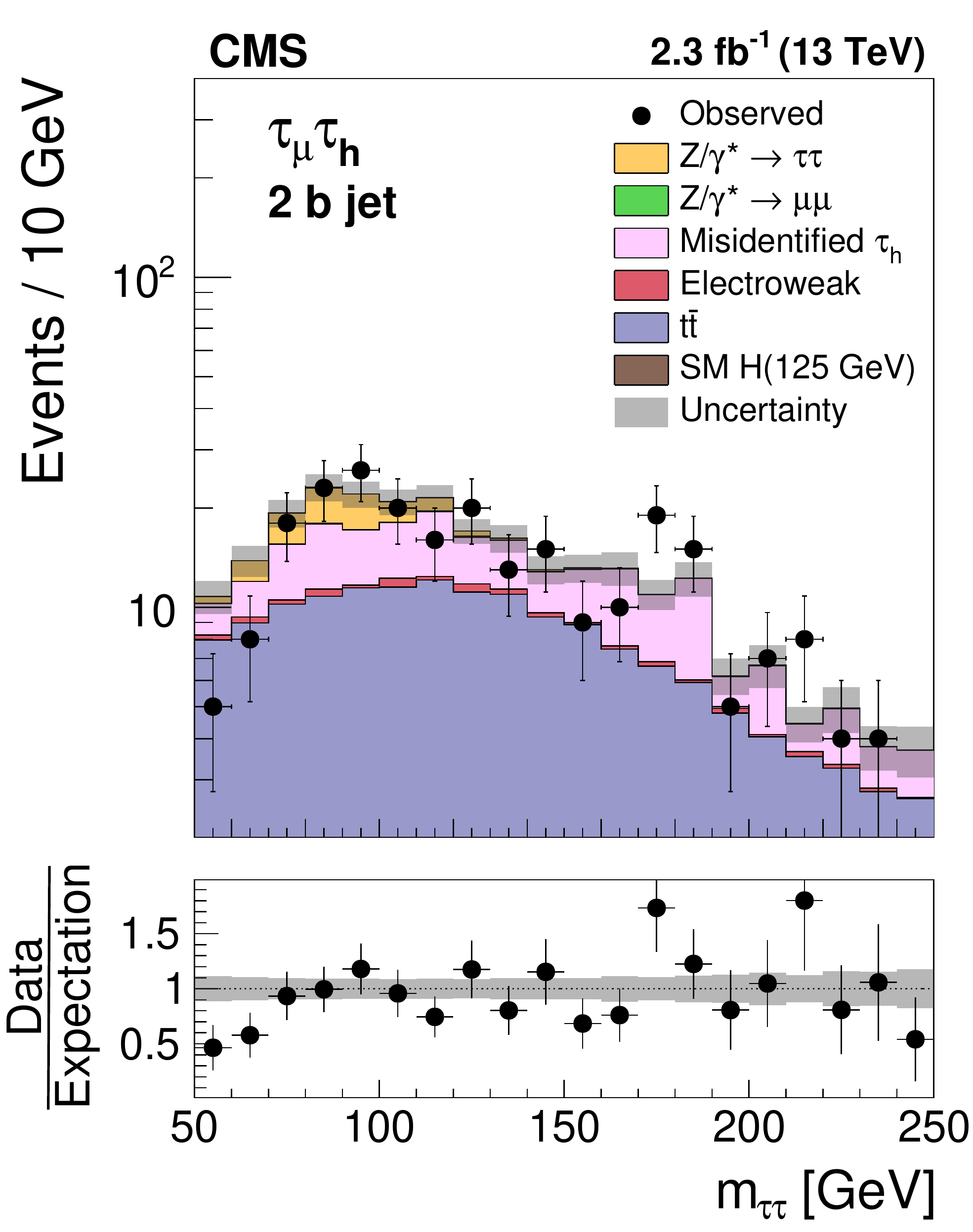}

\caption{
Distributions in $m_{\Pgt\Pgt}$
for different categories in the $\taum\tauh$ channel:
(upper) $2$-jet VBF,
(lower left) $1$ $\Pbottom$ jet, and (lower right) $2$ $\Pbottom$ jet.
}
\label{fig:evtCategoryControlPlots_mutau2}
\end{figure*}

\ifthenelse{\boolean{cms@external}}
{
Distributions in $m_{\Pgt\Pgt}$ for events selected in different event categories
are shown for the $\taum\tauh$ channel
in Figs.~\ref{fig:evtCategoryControlPlots_mutau1} and~\ref{fig:evtCategoryControlPlots_mutau2}.
The corresponding distributions for events selected in the  $\taue\tauh$, $\tauh\tauh$, $\taue\taum$, and $\taum\taum$ channels are published as \suppMaterial.
}
{
Distributions in $m_{\Pgt\Pgt}$ for events selected in different event categories
are shown for the $\taue\tauh$, $\taum\tauh$, $\tauh\tauh$, $\taue\taum$, and $\taum\taum$ channels
in Figs.~\ref{fig:evtCategoryControlPlots_etau1} to~\ref{fig:evtCategoryControlPlots_mumu2}.
}

The distributions expected for the $\cPZ/\Pggx \to \Pgt\Pgt$ signal and for backgrounds
are shown for the values of nuisance parameters obtained from the ML fit described in Section~\ref{sec:signalExtraction}.
The ML fit is performed independently for each category.
The $m_{\Pgt\Pgt}$ distributions are shown within the range $50 < m_{\Pgt\Pgt} < 250\GeV$, indicating
good agreement with background expectations over that mass range.
A similar level of agreement between the data and the background prediction is observed in the $\taue\tauh$, $\tauh\tauh$, and $\taum\taum$ channels.

The agreement confirms the reliability of the $\FF$ method to estimate the reducible backgrounds in the $\taue\tauh$, $\taum\tauh$, and $\tauh\tauh$ channels in future $\PHiggs \to \Pgt\Pgt$ analyses.
It also validates the fact that the $\cPZ/\Pggx \to \Pgt\Pgt$ contribution to event categories, 
based on jet and $\Pbottom$ jet multiplicities and on the $\pT$ of the $\Pgt$ lepton pair, 
can be modelled using $\cPZ/\Pggx \to \Pgm\Pgm$ data, 
without the so-called ``embedding'' technique~\cite{HIG-13-004,Aad:2015kxa} used previously to model the $\cPZ/\Pggx 
\to \Pgt\Pgt$ background in $\PHiggs \to \Pgt\Pgt$ analyses of ATLAS and CMS.

\ifthenelse{\boolean{cms@external}}
{}
{
\begin{figure*}
\centering
\includegraphics[width=\cmsFigWidth]{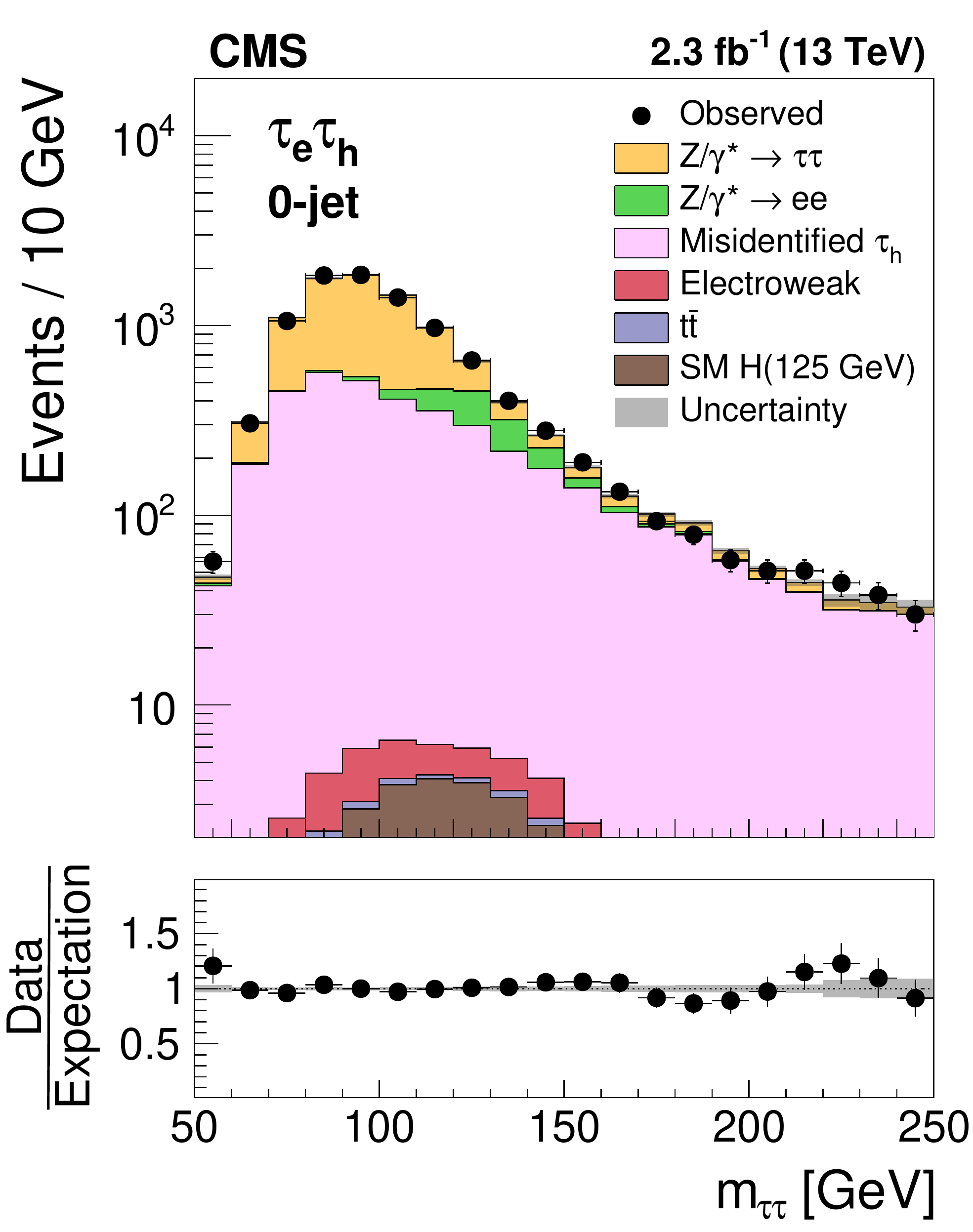} \hfil
\includegraphics[width=\cmsFigWidth]{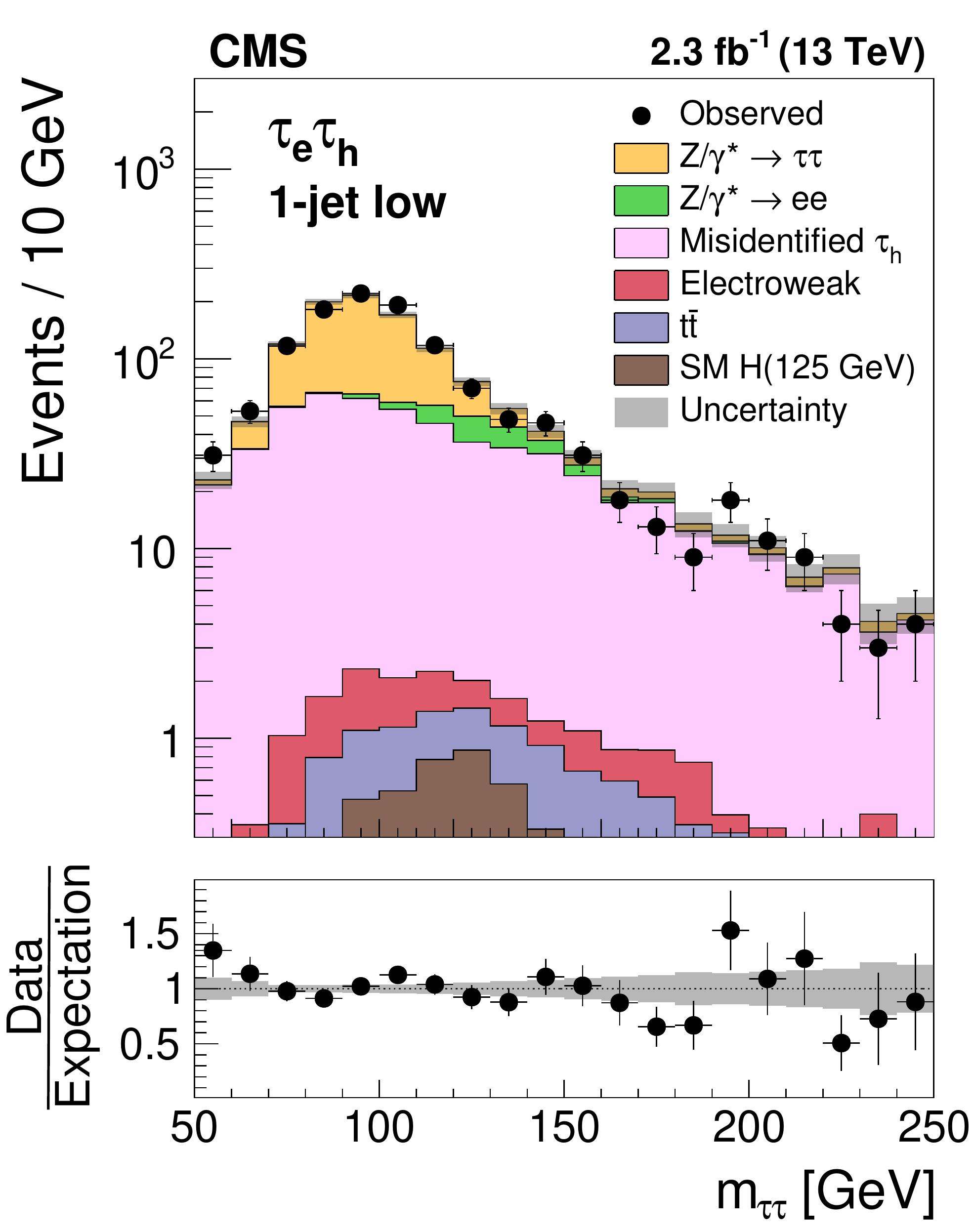} \\
\includegraphics[width=\cmsFigWidth]{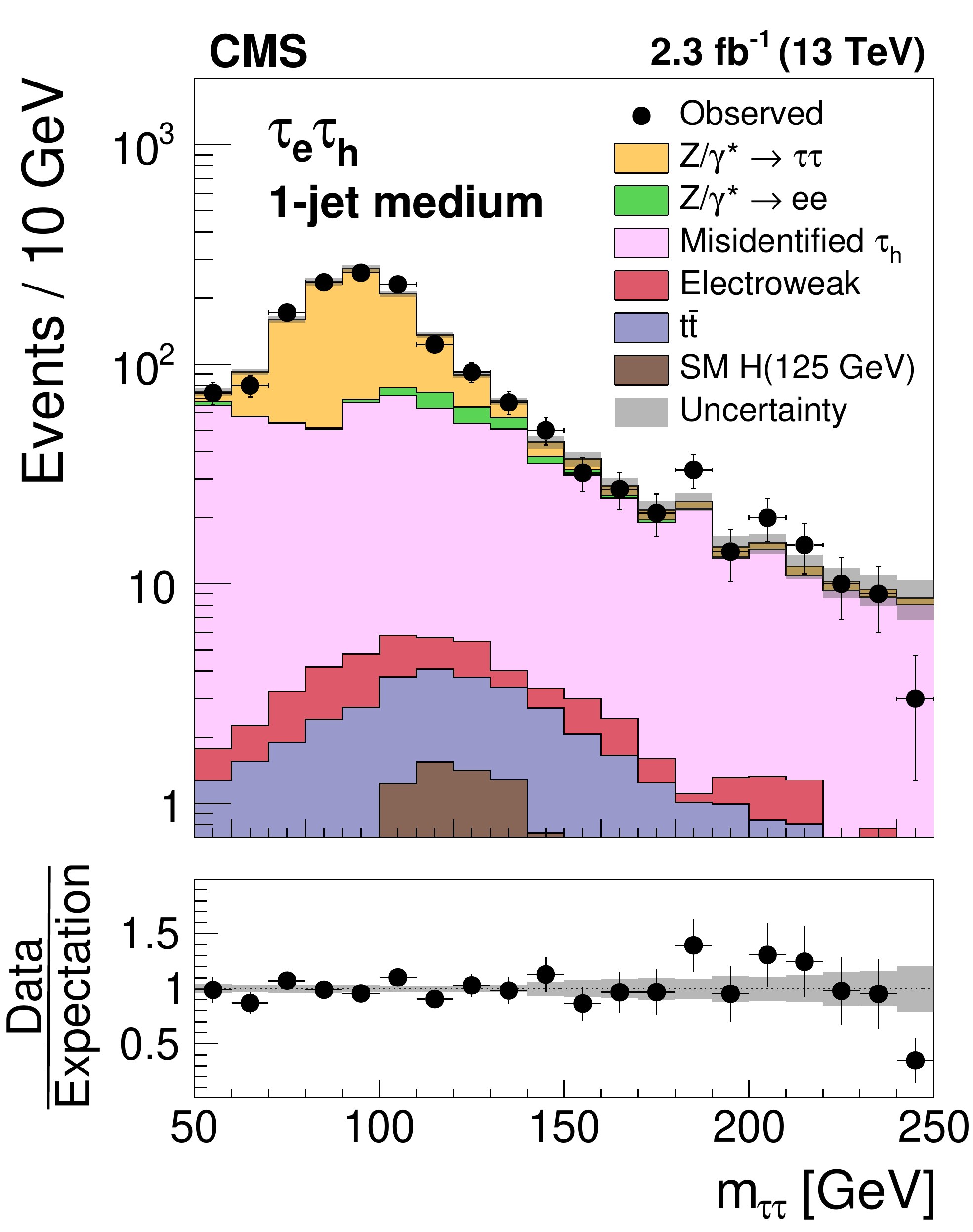} \hfil
\includegraphics[width=\cmsFigWidth]{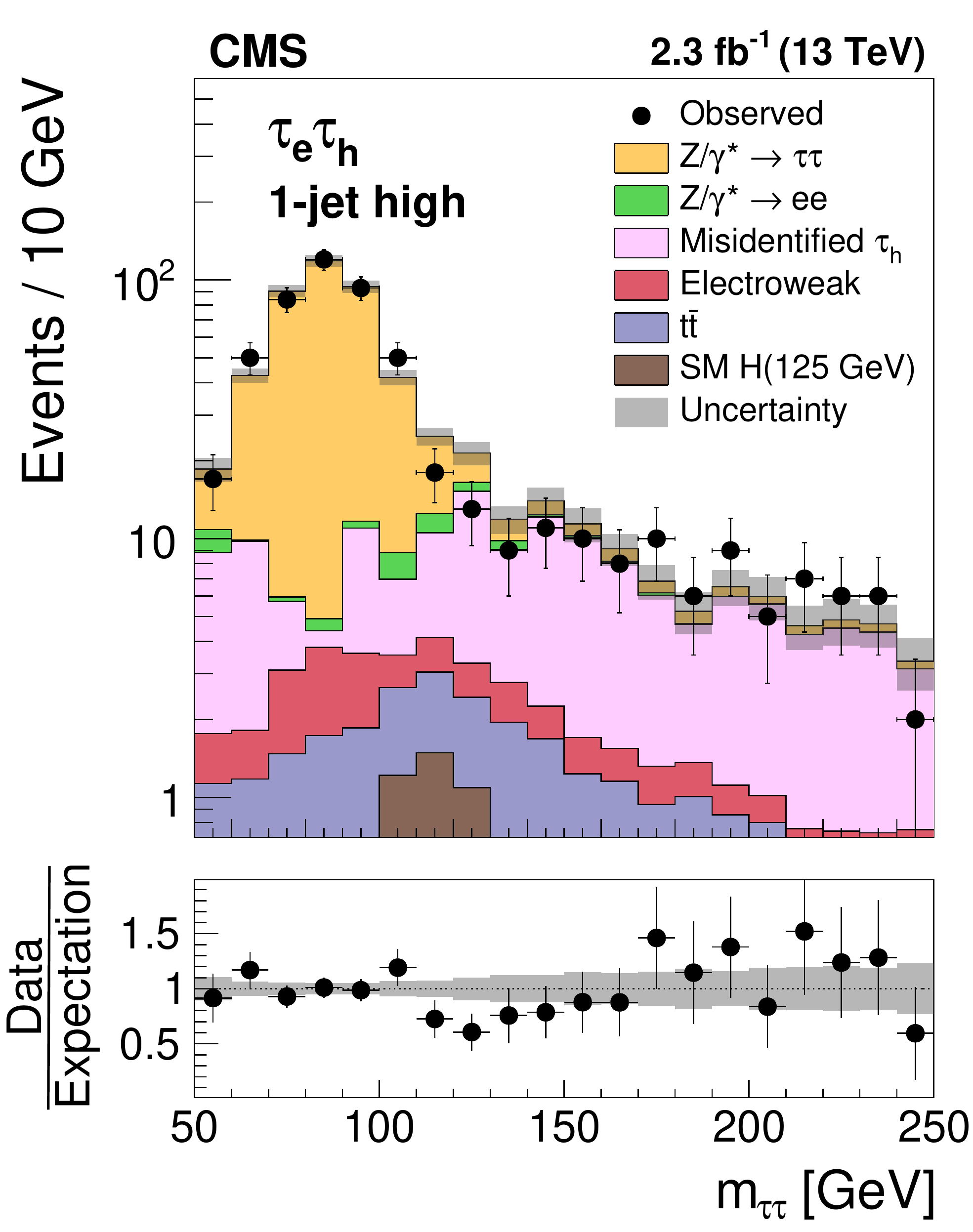}

\caption{
Distributions in $m_{\Pgt\Pgt}$
for different categories in the $\taue\tauh$ channel:
(upper left) $0$-jet,
(upper right) $1$-jet low, (lower left) medium, and (lower right) high $\cPZ$ boson $\pT$.
}
\label{fig:evtCategoryControlPlots_etau1}
\end{figure*}

\begin{figure*}
\centering
\includegraphics[width=\cmsFigWidth]{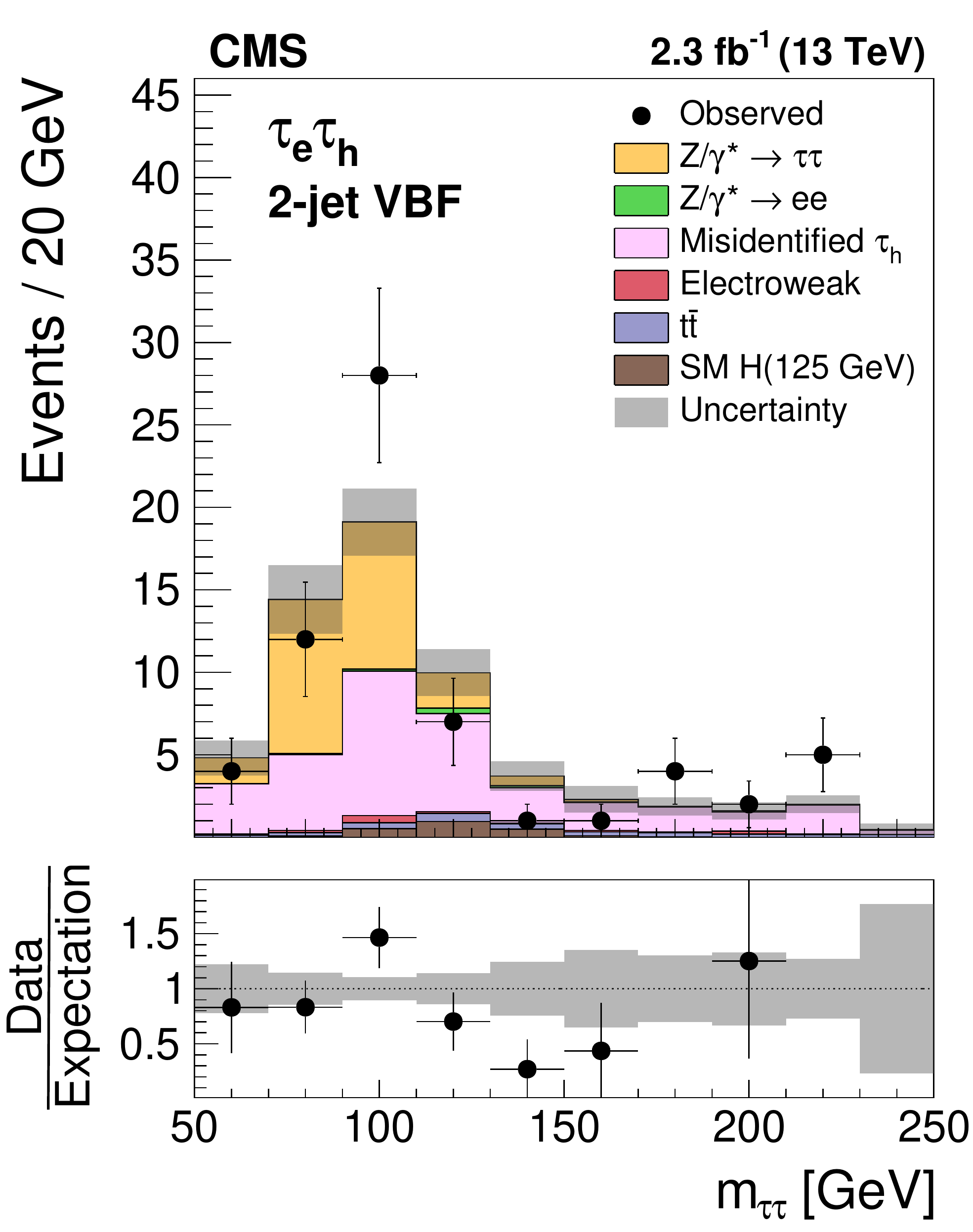} \hfil
\includegraphics[width=\cmsFigWidth]{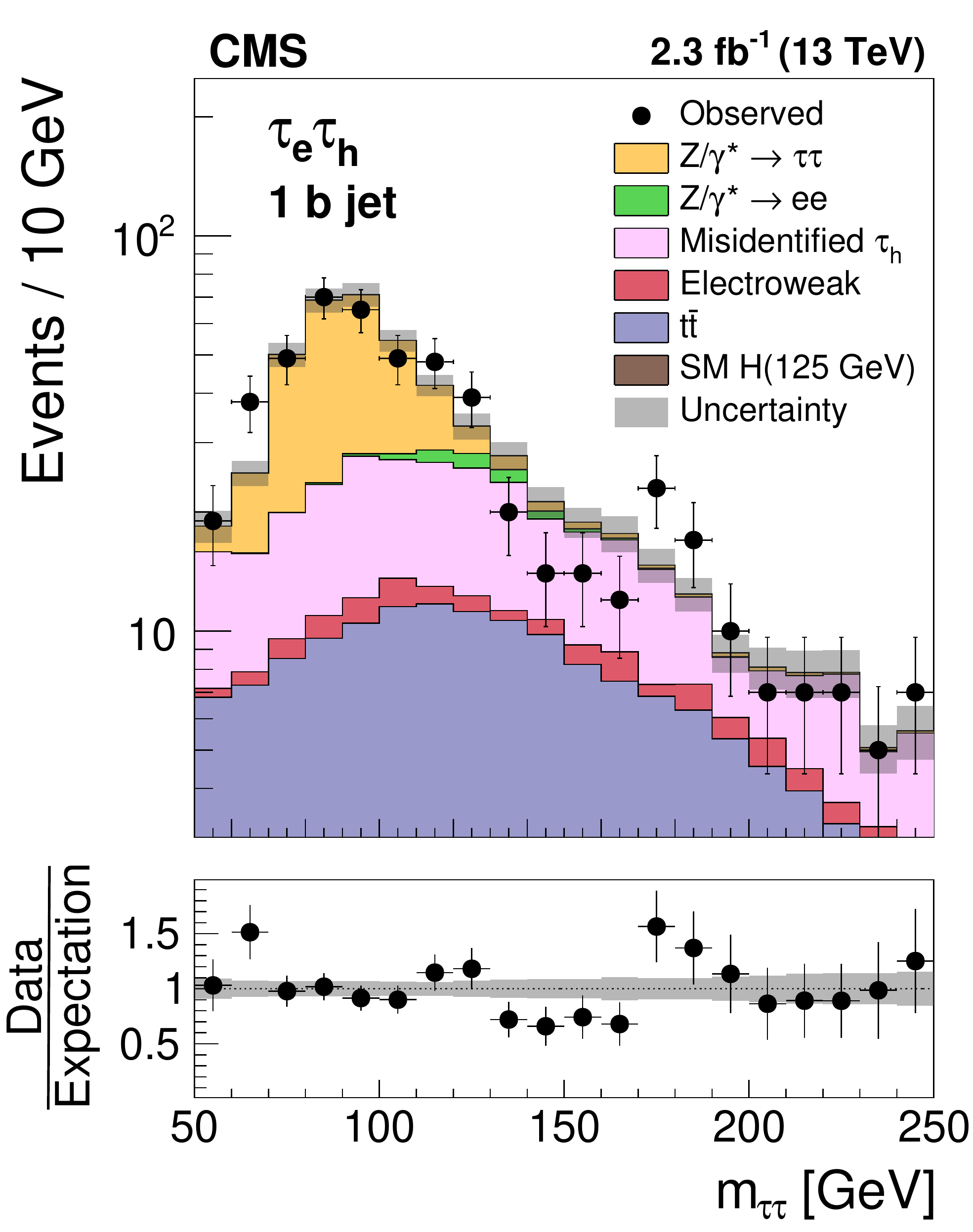} \\
\includegraphics[width=\cmsFigWidth]{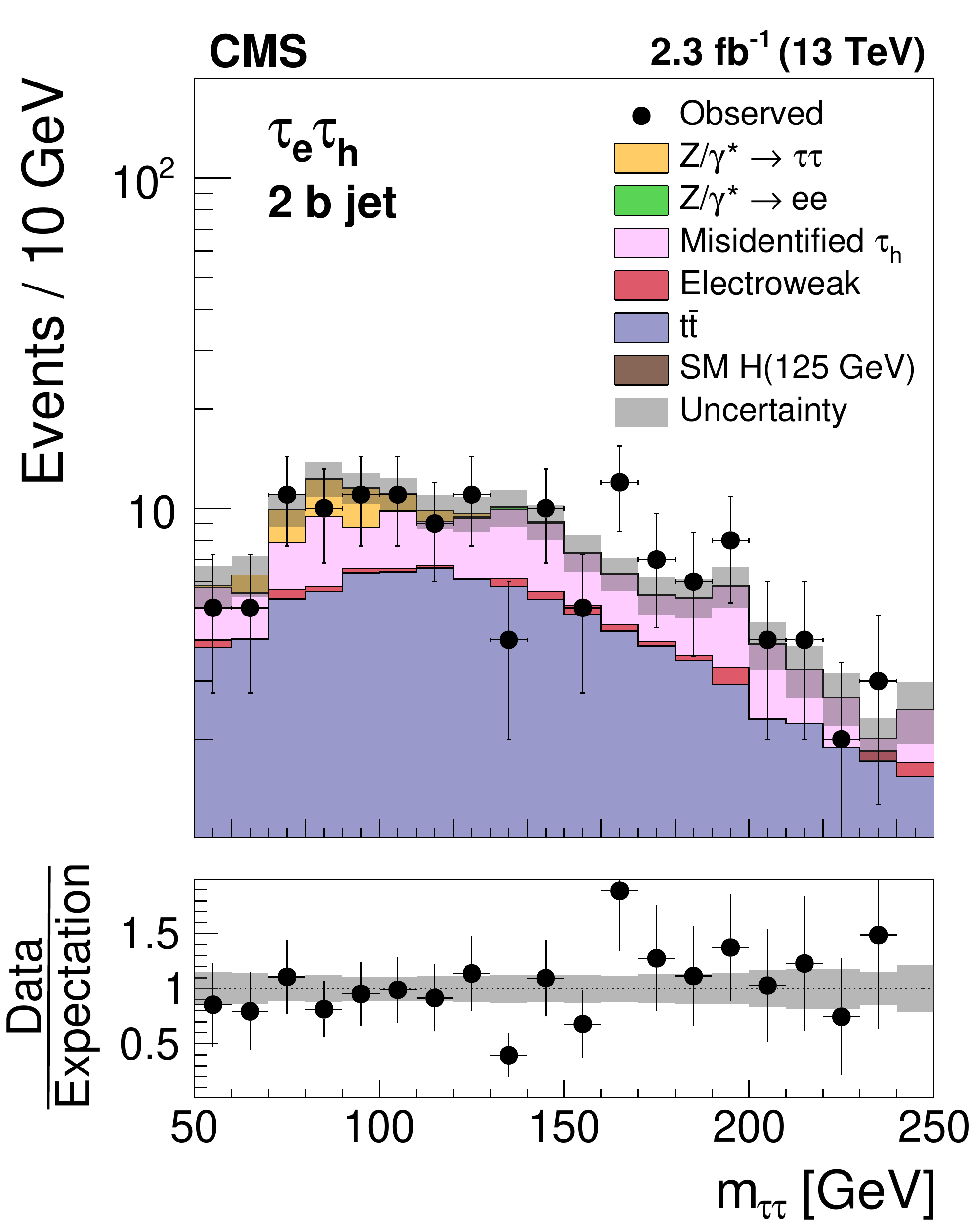}

\caption{
Distributions in $m_{\Pgt\Pgt}$
for different categories in the $\taue\tauh$ channel:
(upper) $2$-jet VBF,
(lower left) $1$ $\Pbottom$ jet, and (lower right) $2$ $\Pbottom$ jet.
}
\label{fig:evtCategoryControlPlots_etau2}
\end{figure*}

\begin{figure*}
\centering
\includegraphics[width=\cmsFigWidth]{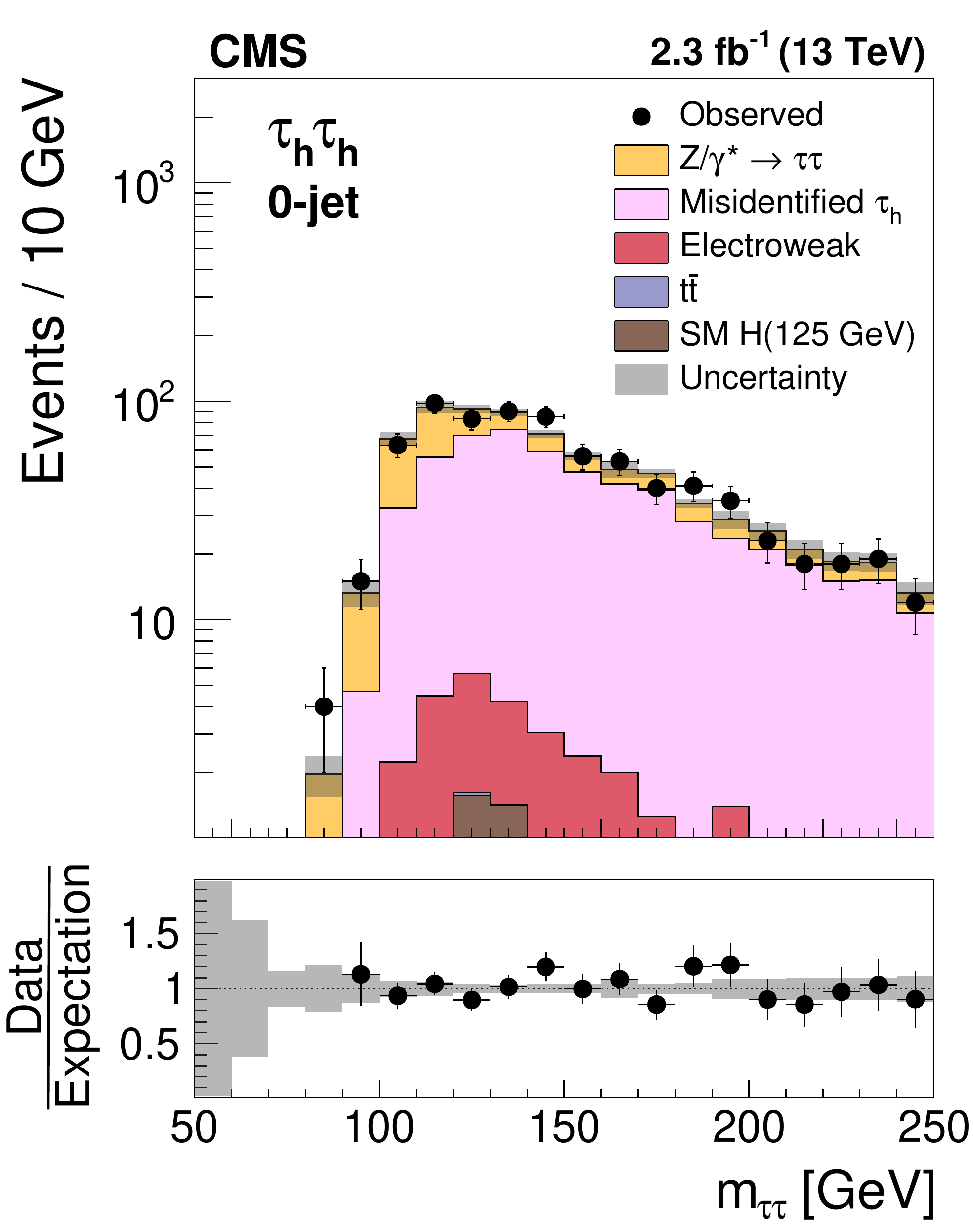} \hfil
\includegraphics[width=\cmsFigWidth]{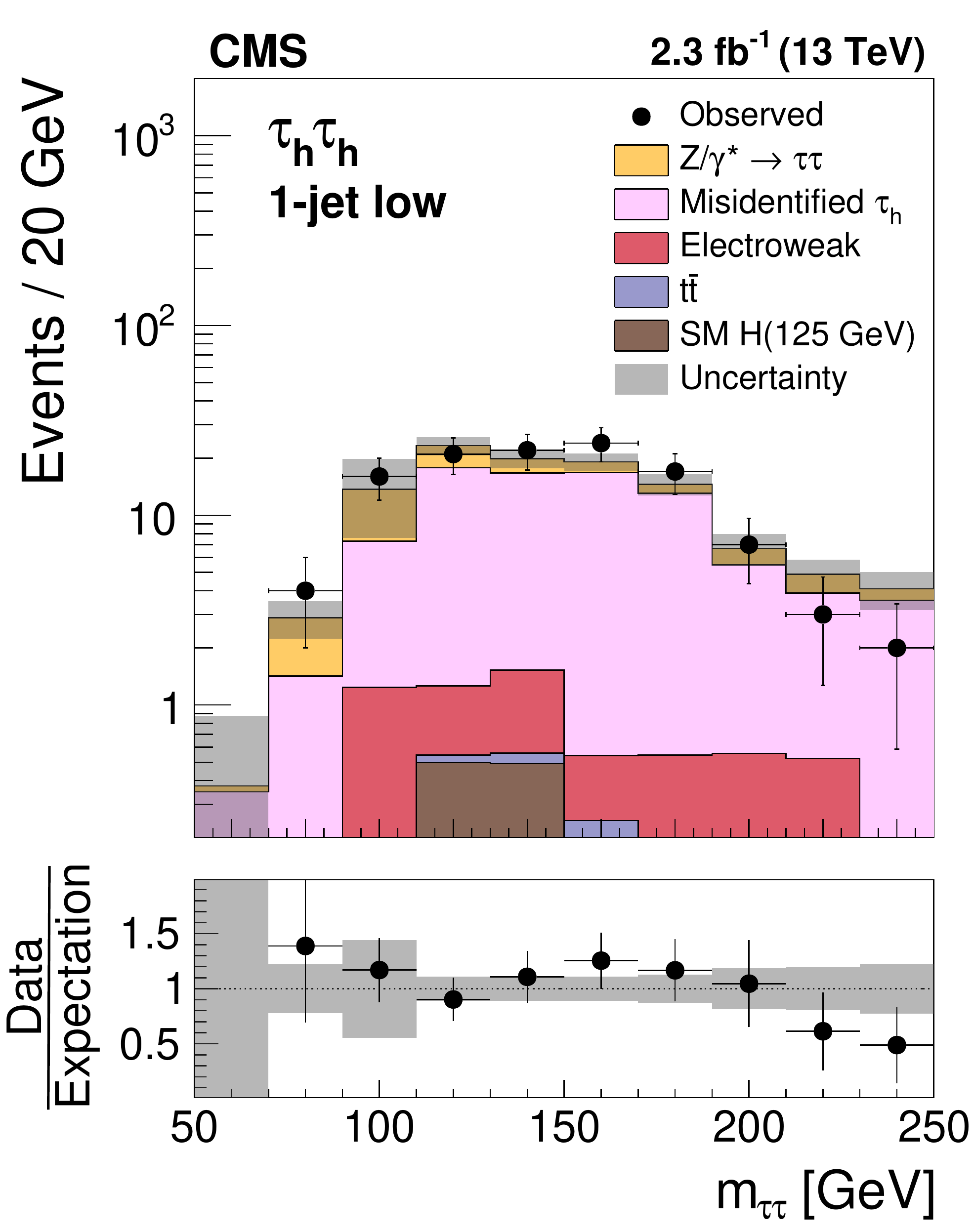} \\
\includegraphics[width=\cmsFigWidth]{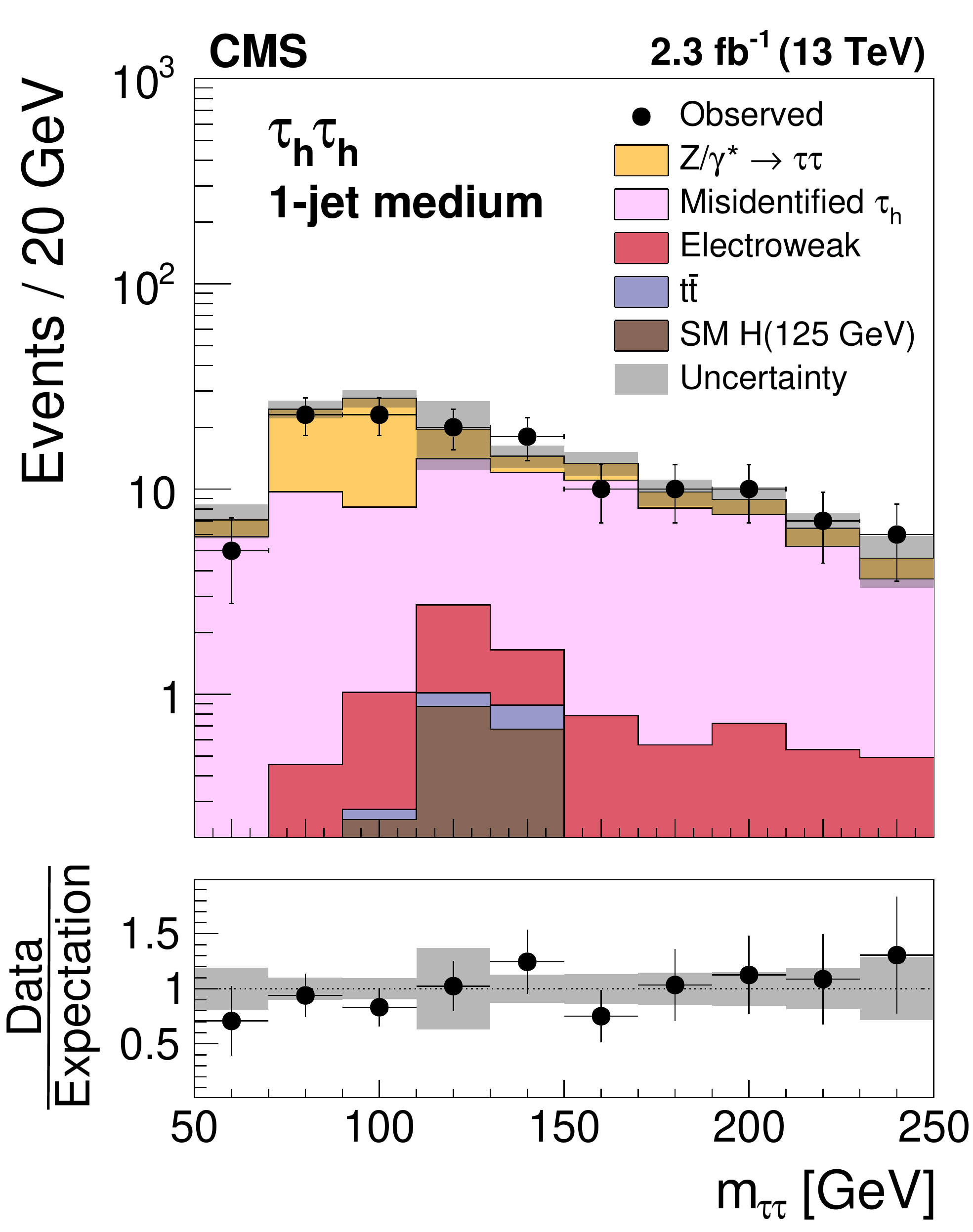} \hfil
\includegraphics[width=\cmsFigWidth]{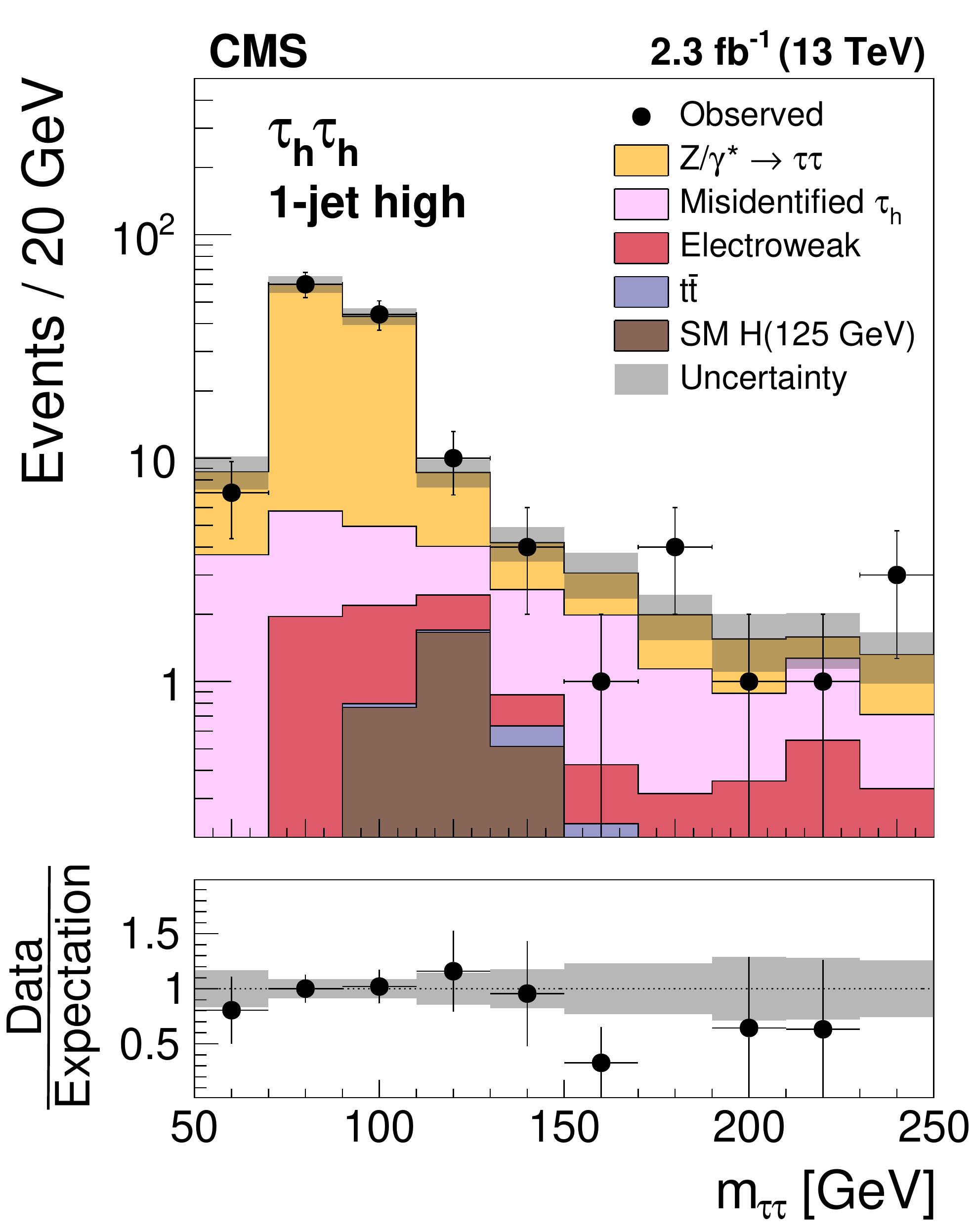}

\caption{
Distributions in $m_{\Pgt\Pgt}$
for different categories in the $\tauh\tauh$ channel:
(upper left) $0$-jet,
(upper right) $1$-jet low, (lower left) medium, and (lower right) high $\cPZ$ boson $\pT$.
}
\label{fig:evtCategoryControlPlots_tautau1}
\end{figure*}

\begin{figure*}
\centering
\includegraphics[width=\cmsFigWidth]{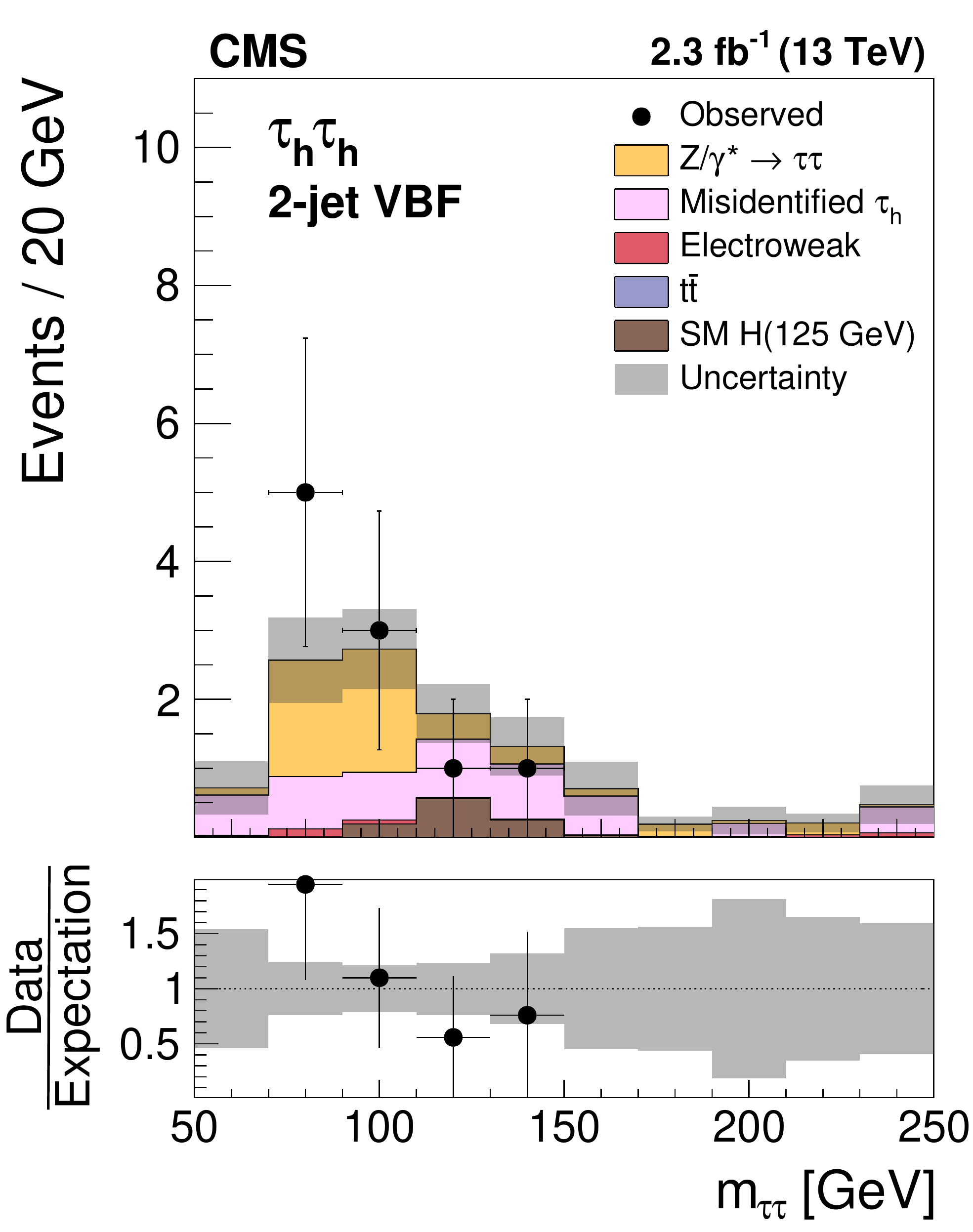} \hfil
\includegraphics[width=\cmsFigWidth]{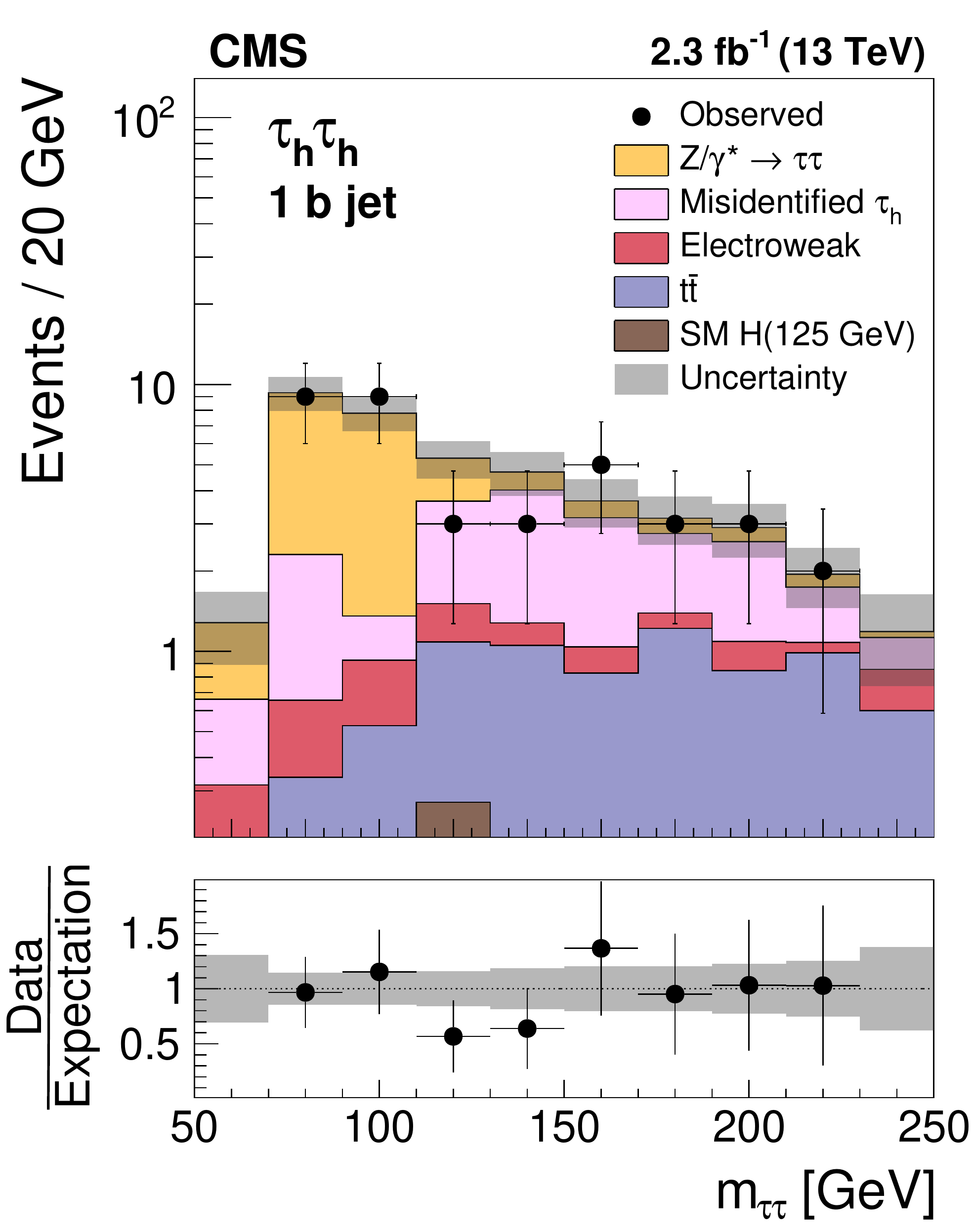} \\
\includegraphics[width=\cmsFigWidth]{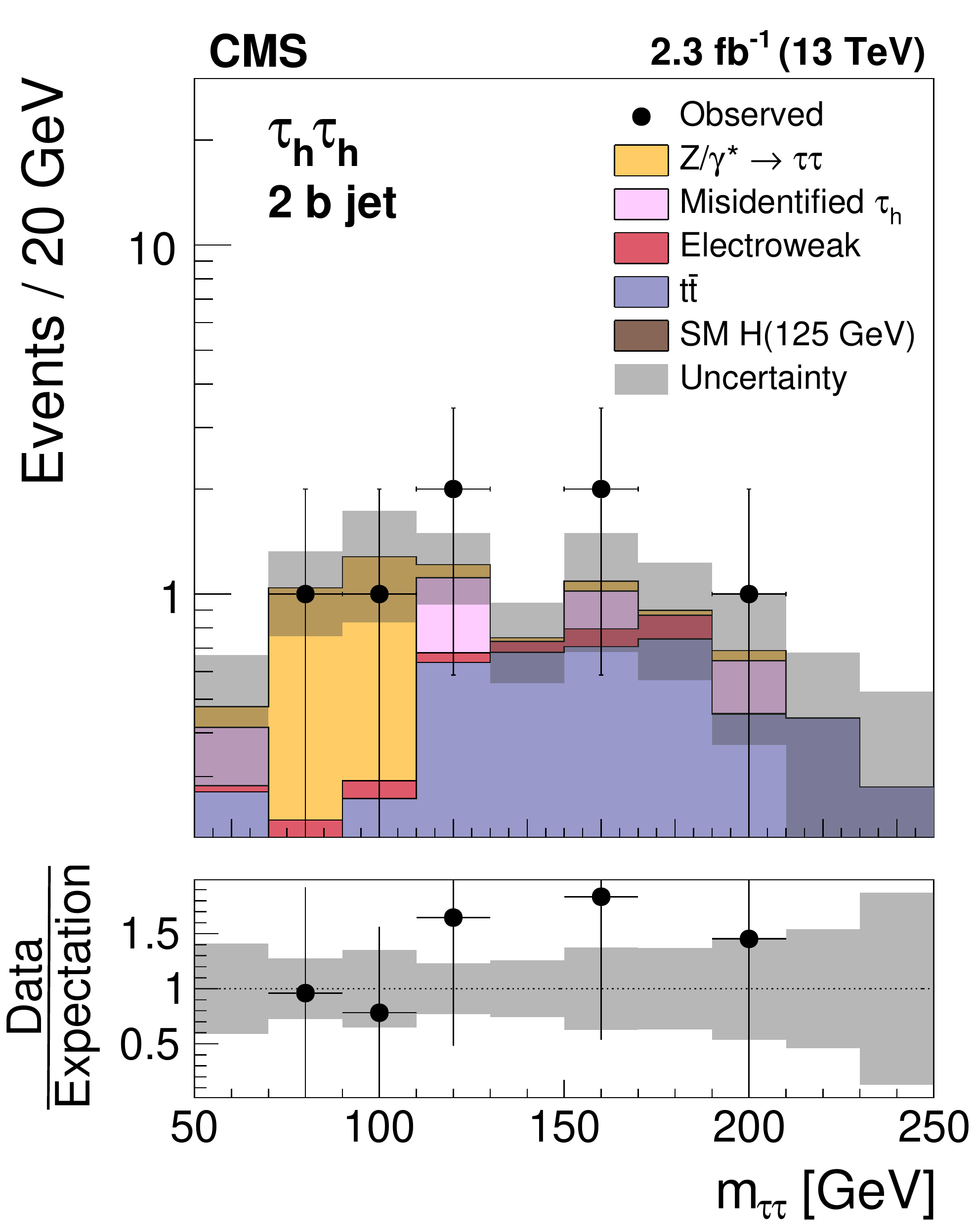}

\caption{
Distributions in $m_{\Pgt\Pgt}$
for different categories in the $\tauh\tauh$ channel:
(upper) $2$-jet VBF,
(lower left) $1$ $\Pbottom$ jet, and (lower right) $2$ $\Pbottom$ jet.
}
\label{fig:evtCategoryControlPlots_tautau2}
\end{figure*}

\begin{figure*}
\centering
\includegraphics[width=\cmsFigWidth]{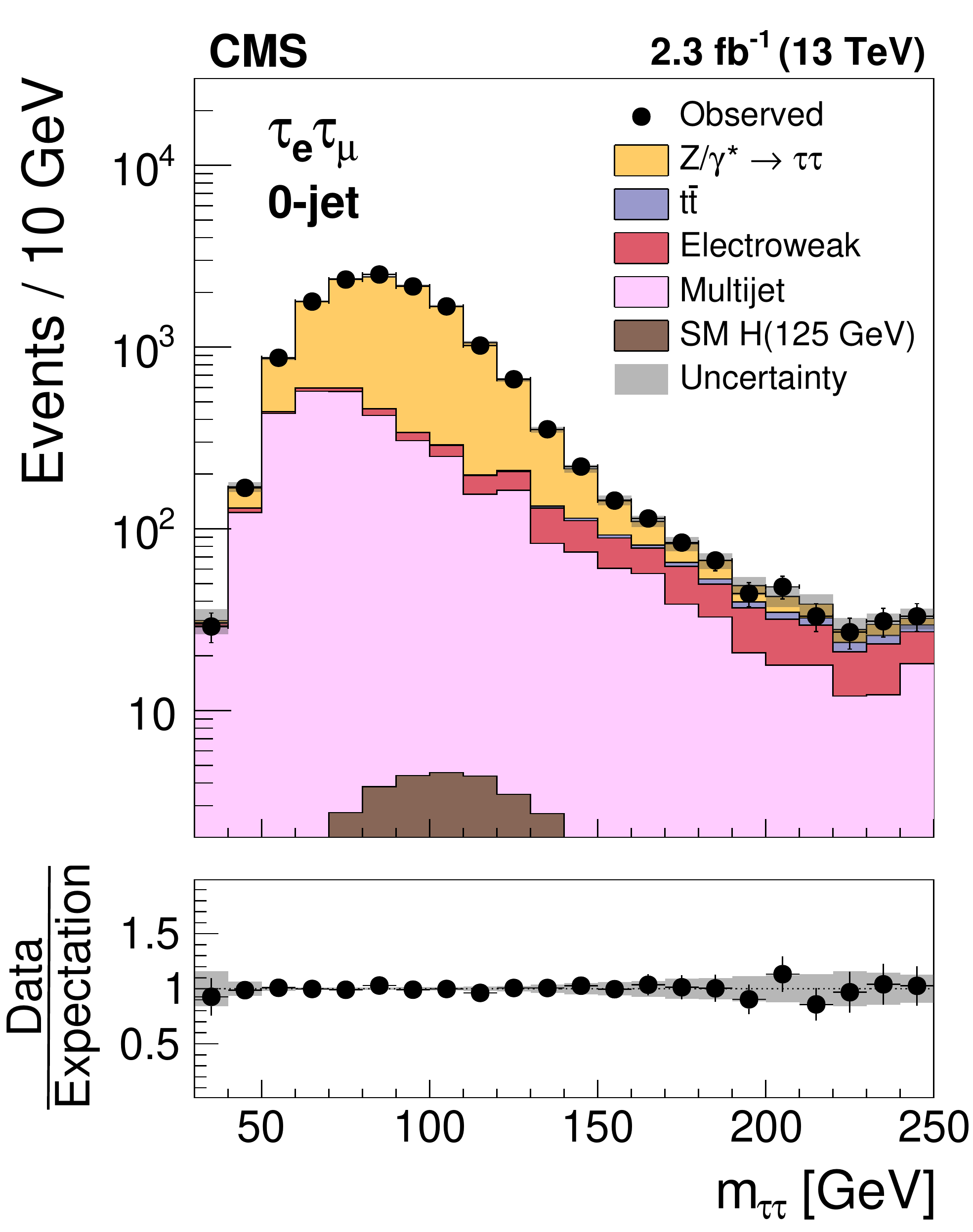} \hfil
\includegraphics[width=\cmsFigWidth]{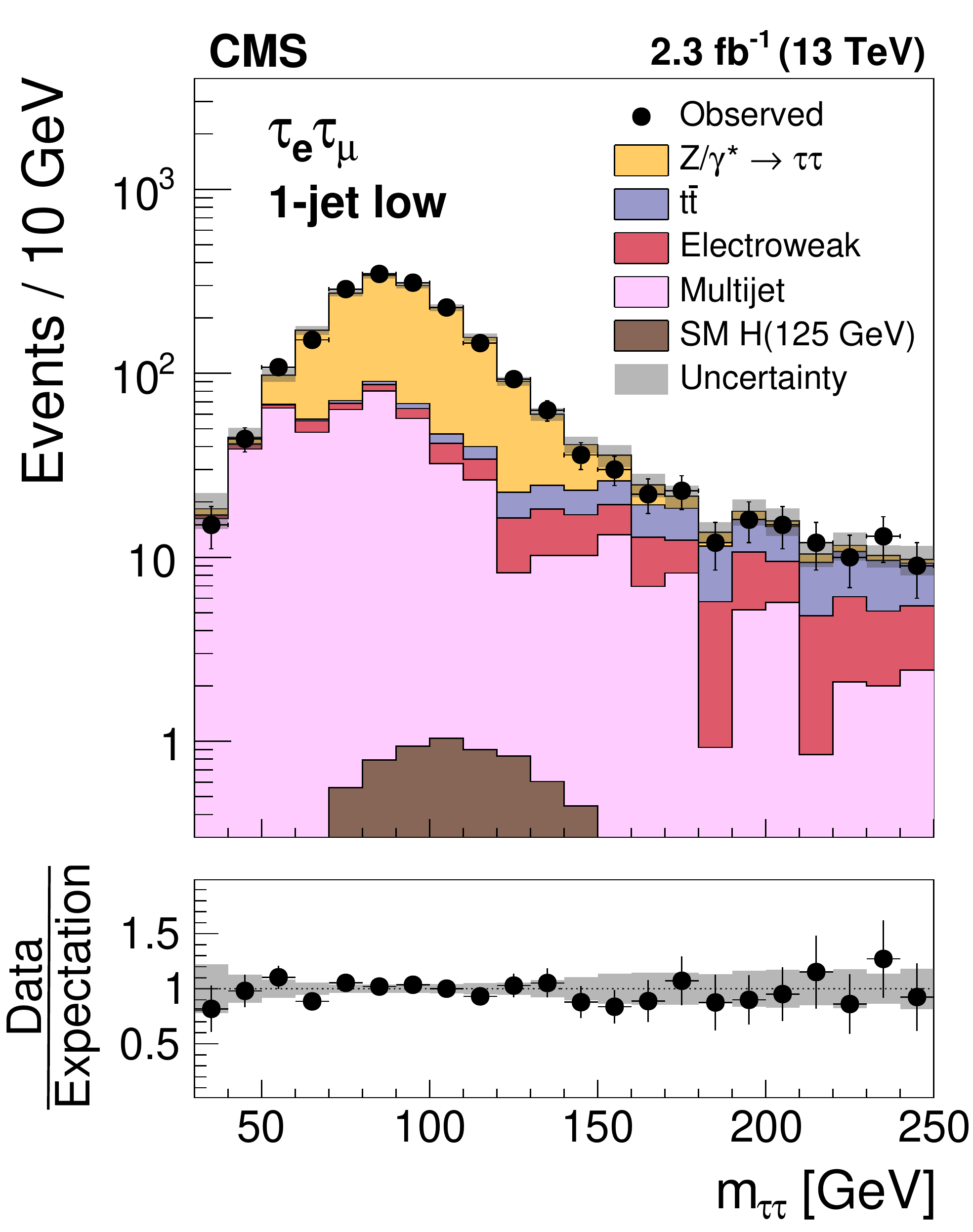} \\
\includegraphics[width=\cmsFigWidth]{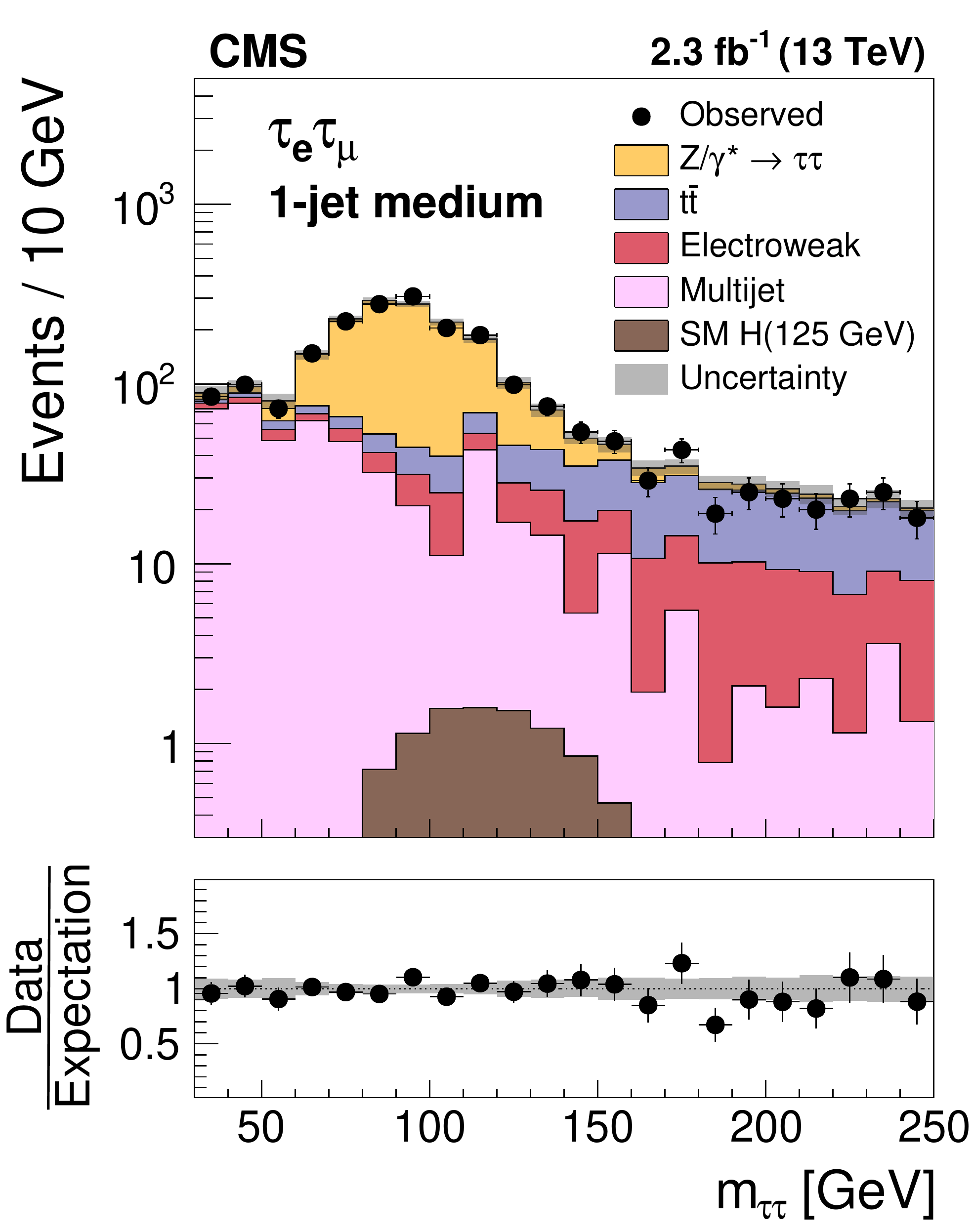} \hfil
\includegraphics[width=\cmsFigWidth]{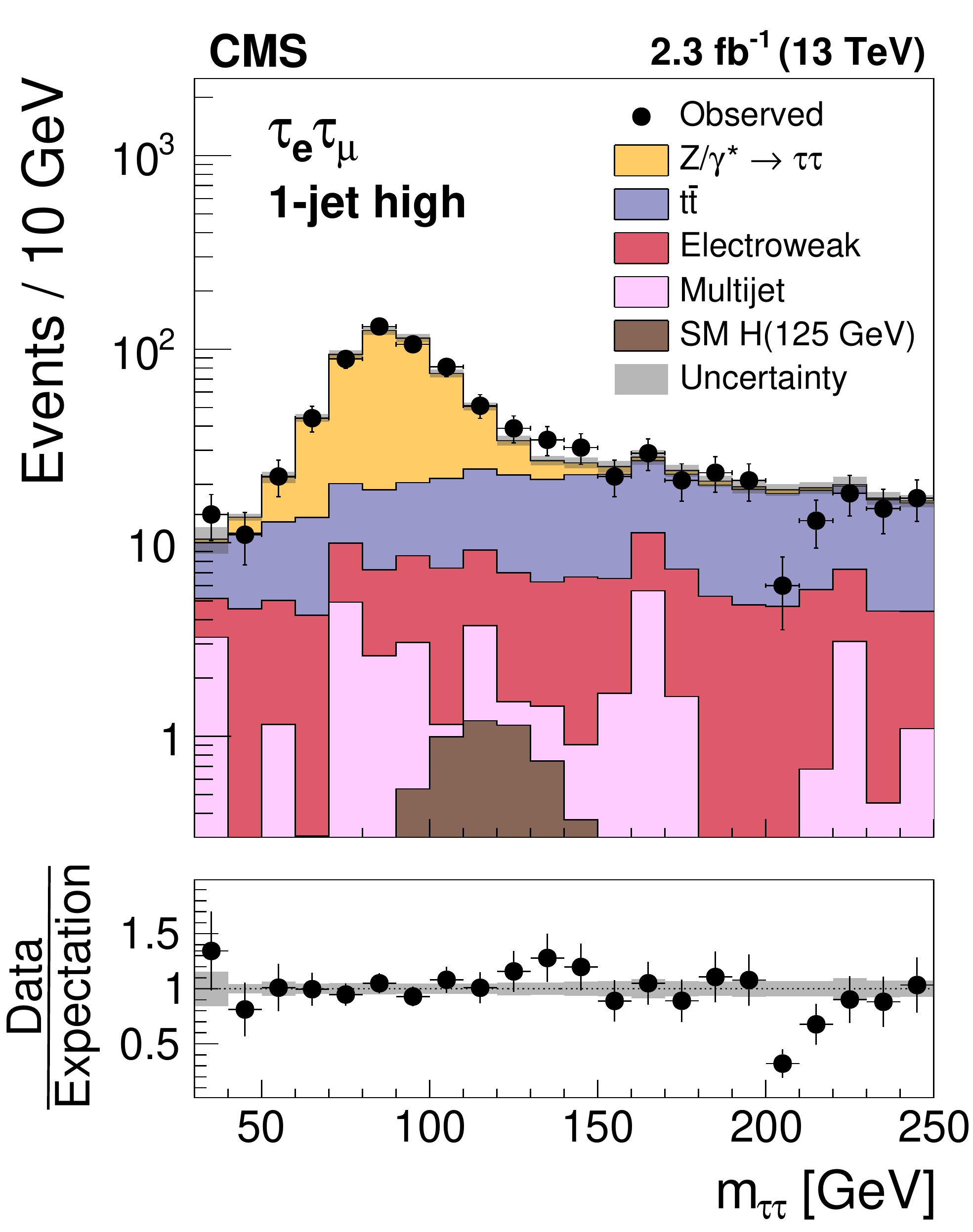}

\caption{
Distributions in $m_{\Pgt\Pgt}$
for different categories in the $\taue\taum$ channel:
(upper left) $0$-jet,
(upper right) $1$-jet low, (lower left) medium, and (lower right) high $\cPZ$ boson $\pT$.
}
\label{fig:evtCategoryControlPlots_emu1}
\end{figure*}

\begin{figure*}
\centering
\includegraphics[width=\cmsFigWidth]{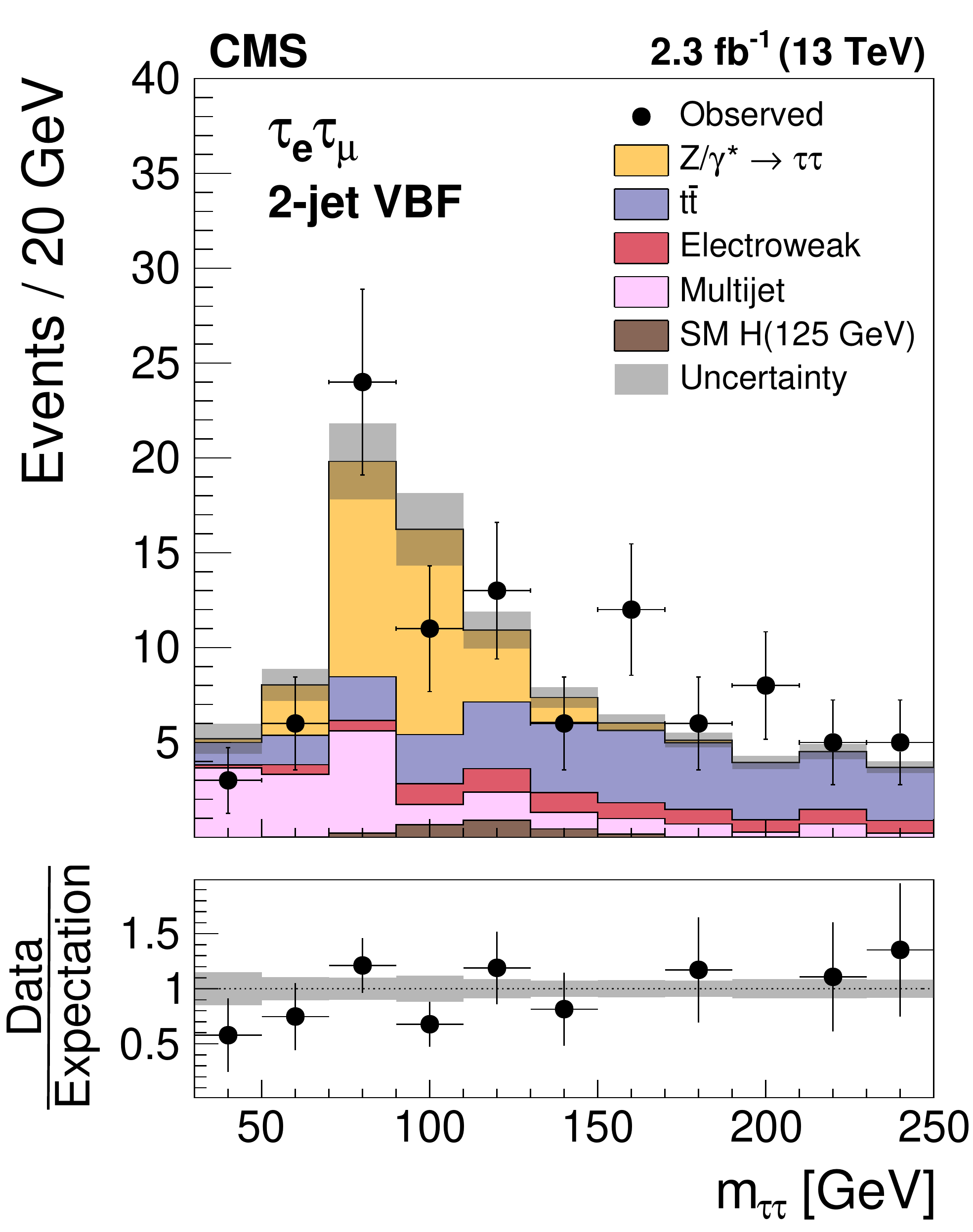} \hfil
\includegraphics[width=\cmsFigWidth]{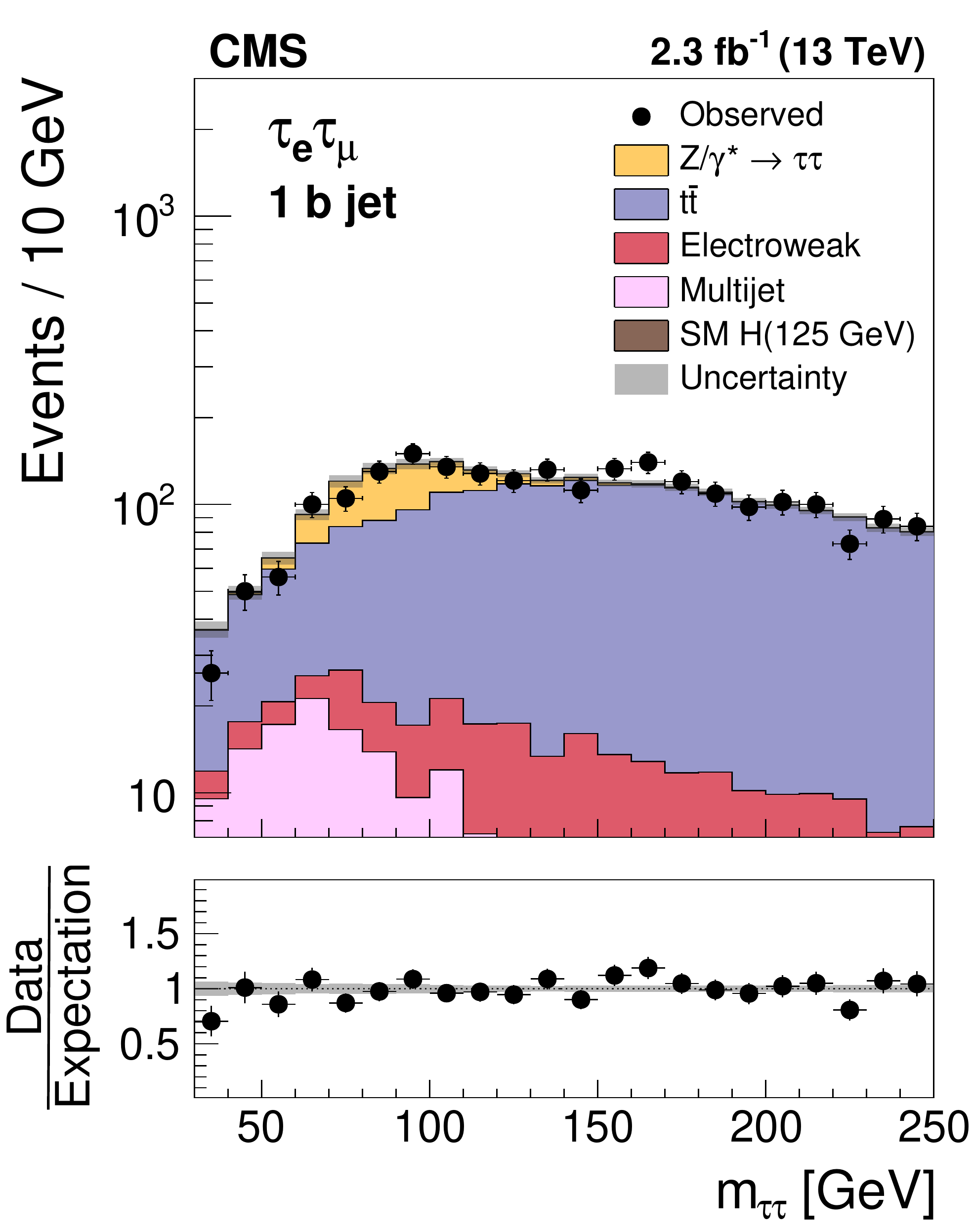} \\
\includegraphics[width=\cmsFigWidth]{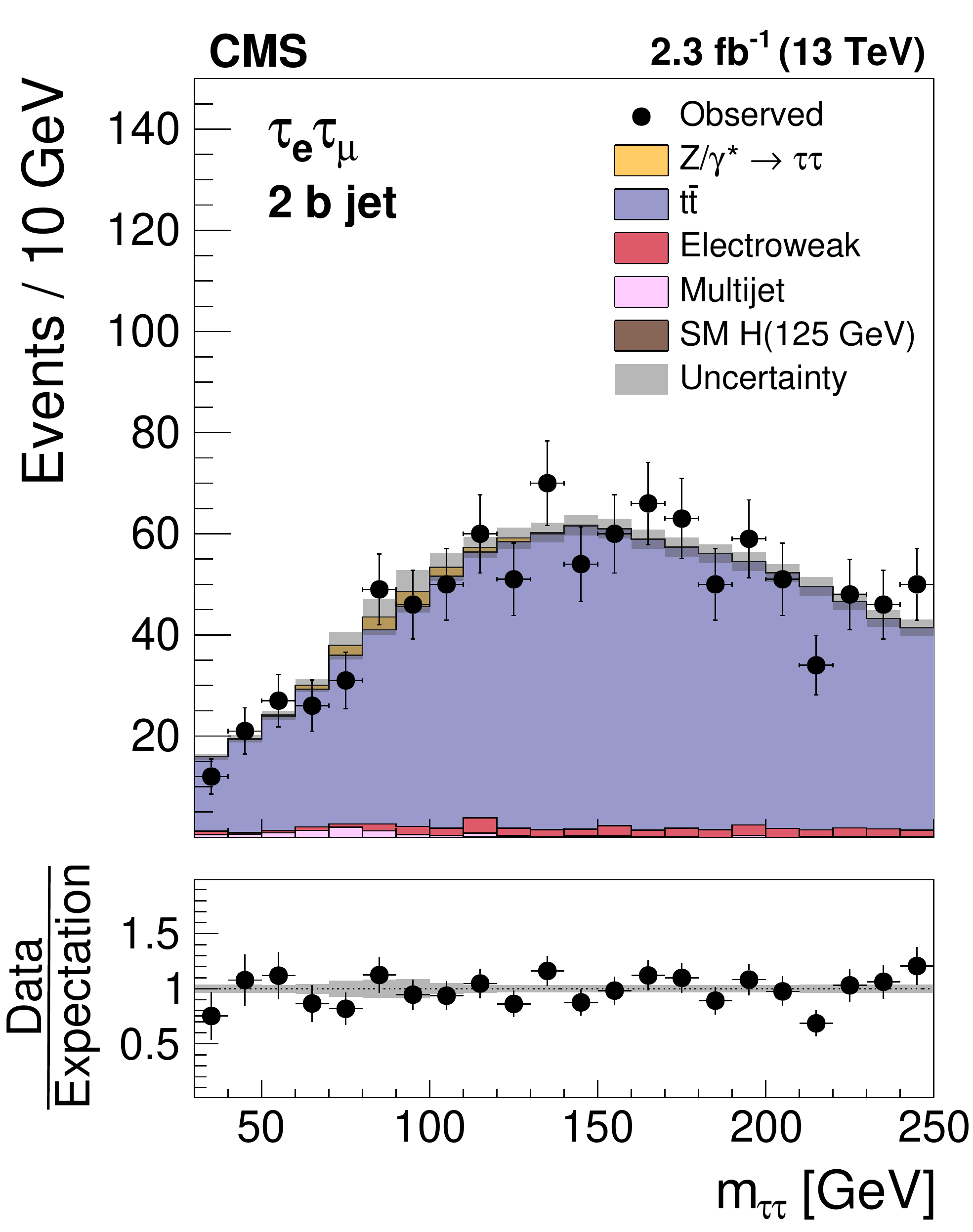}

\caption{
Distributions in $m_{\Pgt\Pgt}$
for different categories in the $\taue\taum$ channel:
(upper) $2$-jet VBF,
(lower left) $1$ $\Pbottom$ jet, and (lower right) $2$ $\Pbottom$ jet.
}
\label{fig:evtCategoryControlPlots_emu2}
\end{figure*}

\begin{figure*}
\centering
\includegraphics[width=\cmsFigWidth]{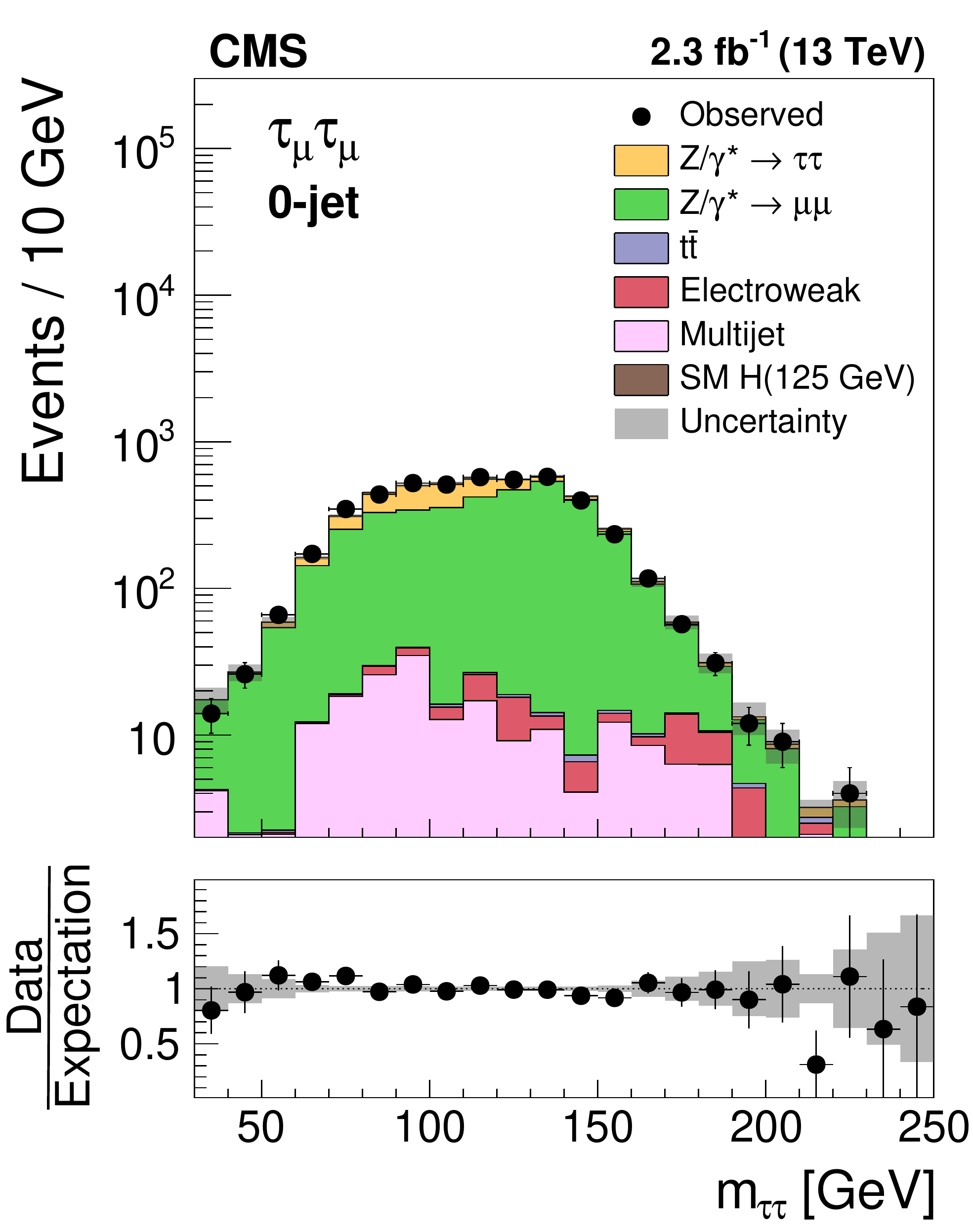} \hfil
\includegraphics[width=\cmsFigWidth]{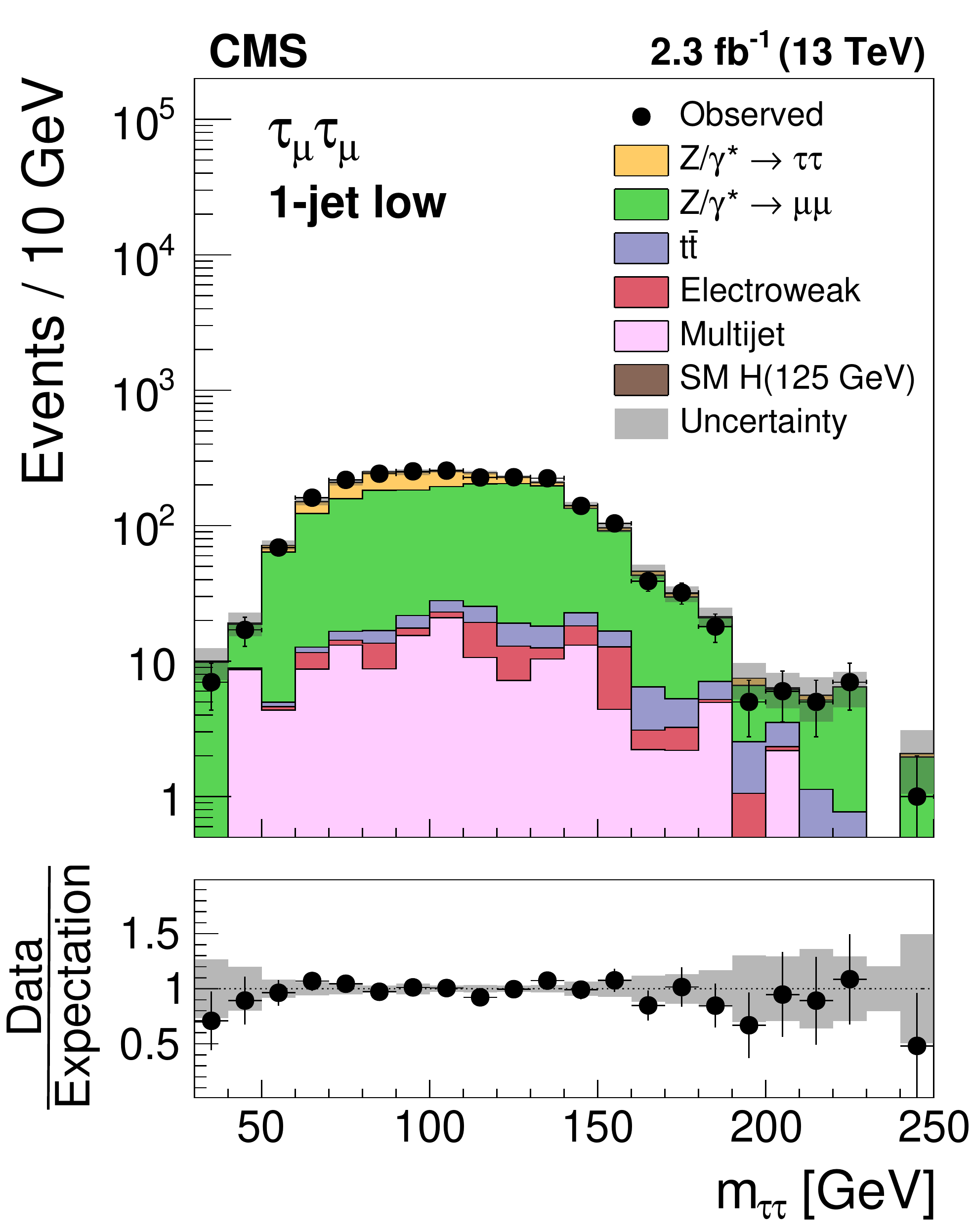} \\
\includegraphics[width=\cmsFigWidth]{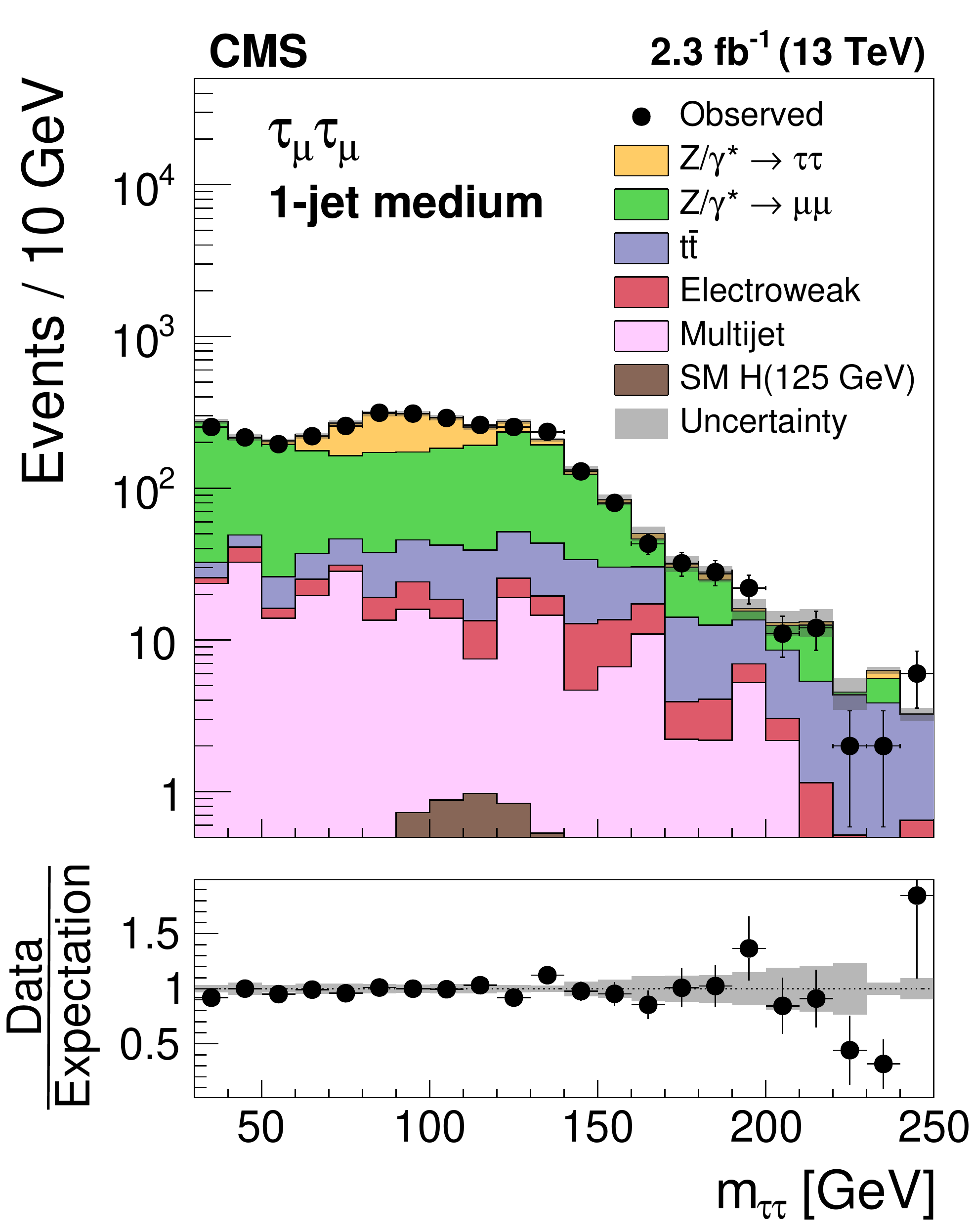} \hfil
\includegraphics[width=\cmsFigWidth]{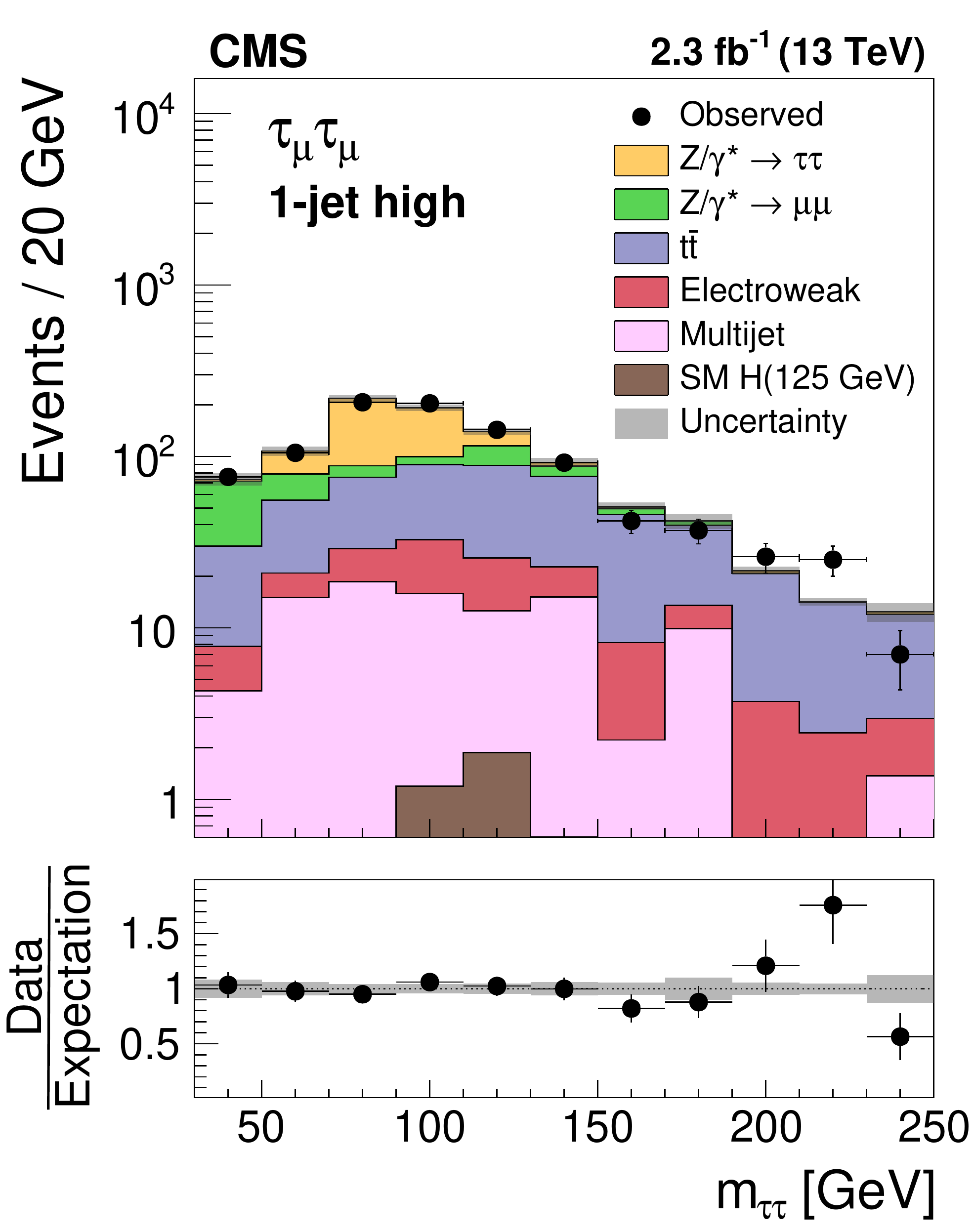}

\caption{
Distributions in $m_{\Pgt\Pgt}$
for different categories in the $\taum\taum$ channel:
(upper left) $0$-jet,
(upper right) $1$-jet low, (lower left) medium, and (lower right) high $\cPZ$ boson $\pT$.
}
\label{fig:evtCategoryControlPlots_mumu1}
\end{figure*}

\begin{figure*}
\centering
\includegraphics[width=\cmsFigWidth]{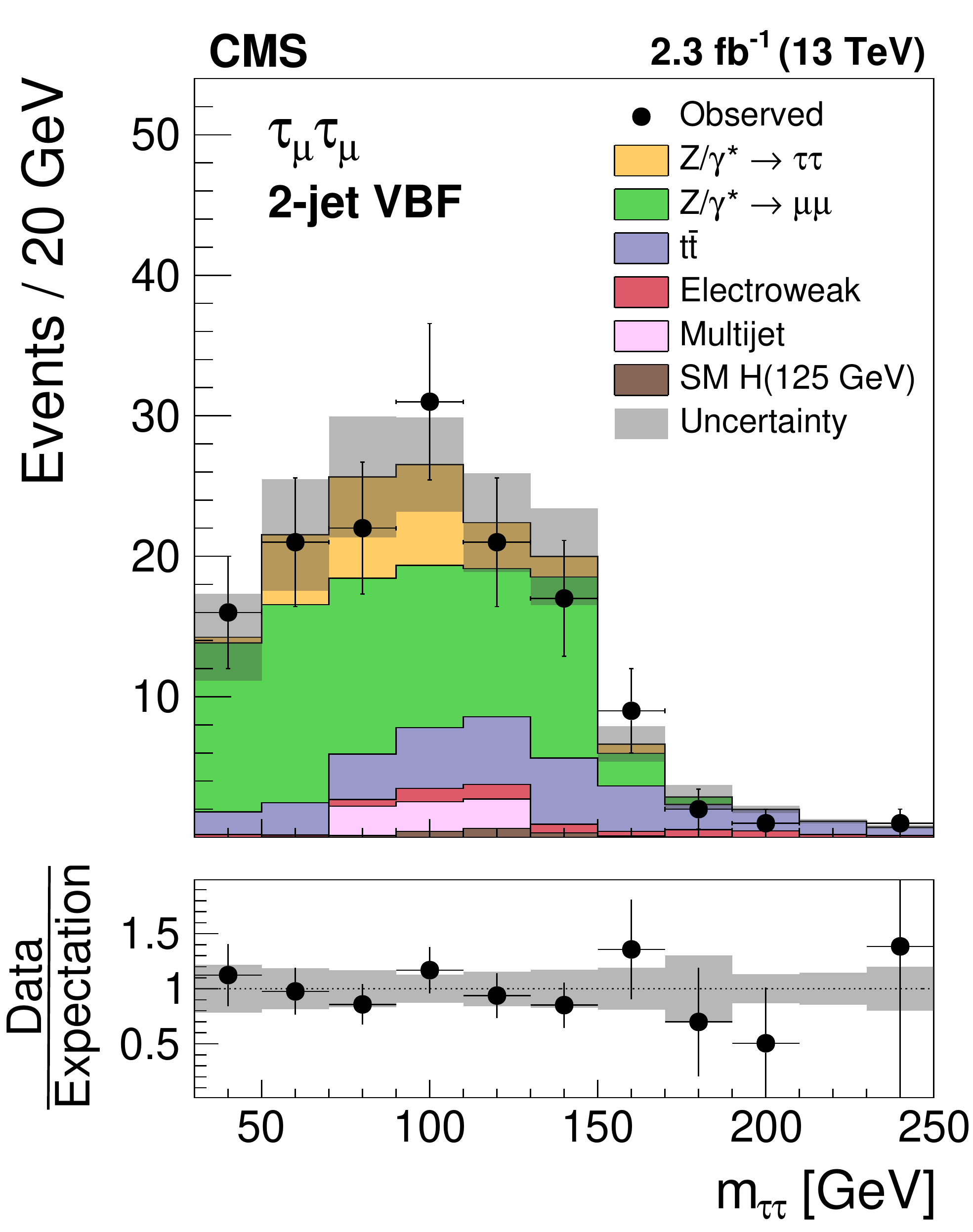} \hfil
\includegraphics[width=\cmsFigWidth]{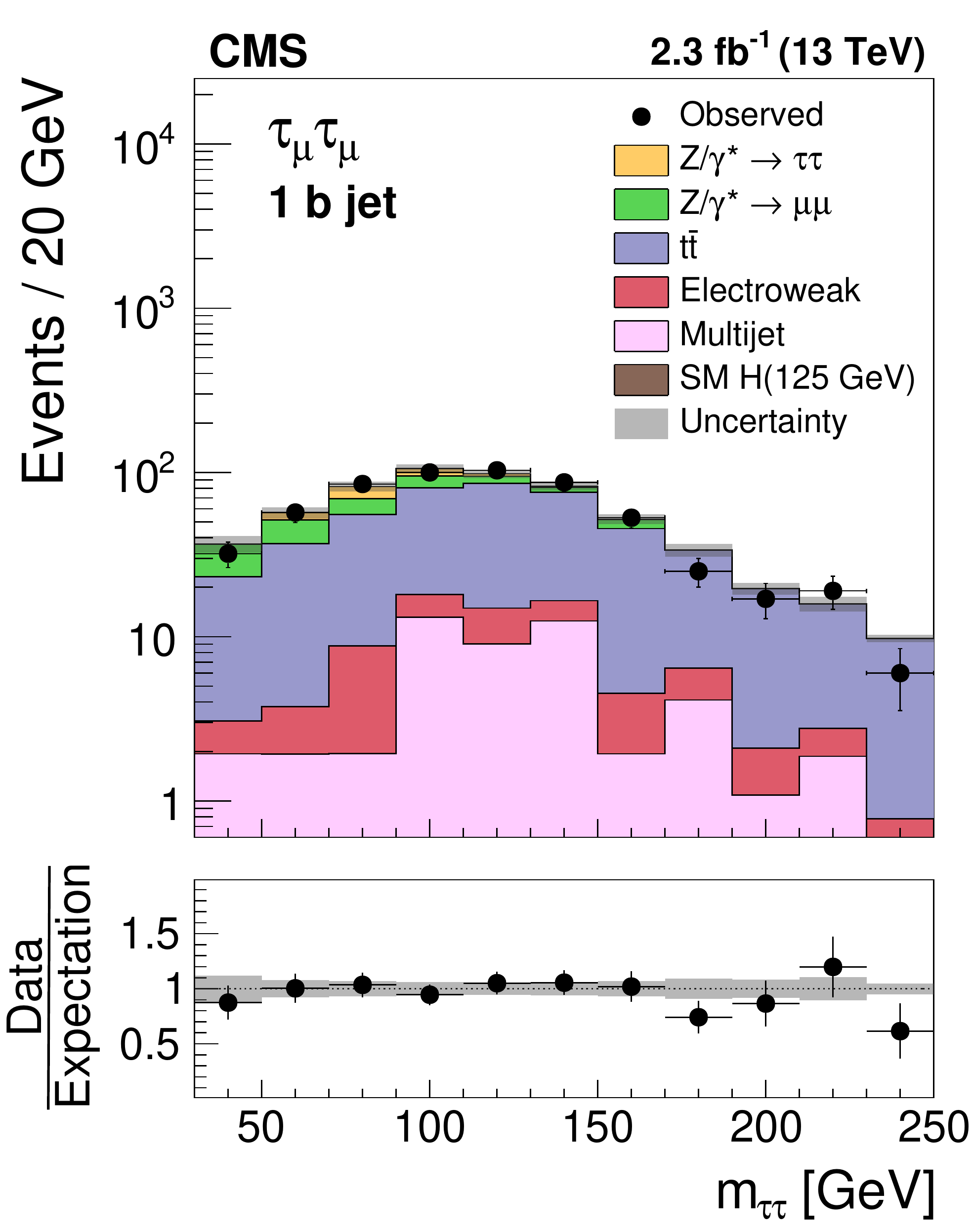} \\
\includegraphics[width=\cmsFigWidth]{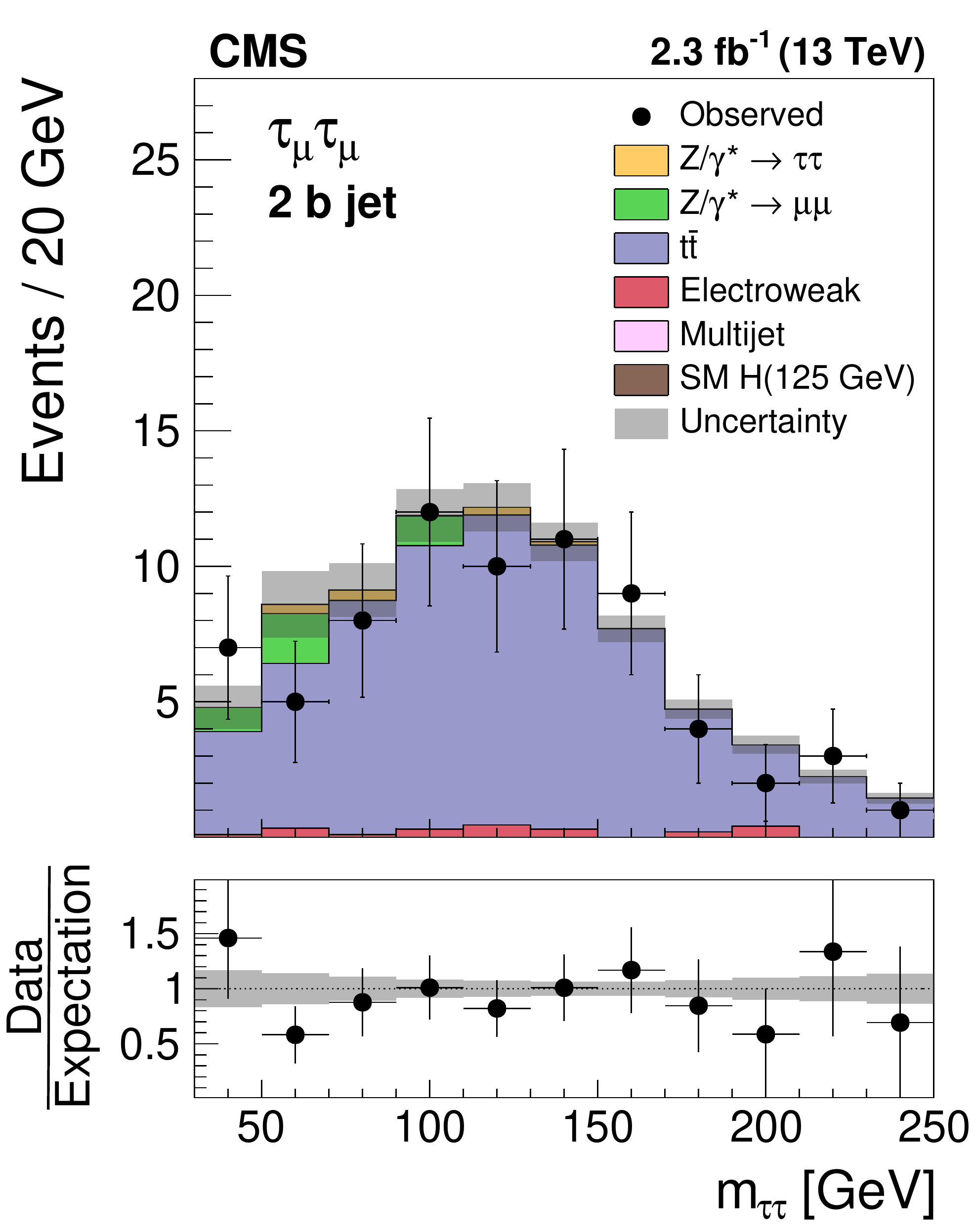}

\caption{
Distributions in $m_{\Pgt\Pgt}$
for different categories in the $\taum\taum$ channel:
(upper) $2$-jet VBF,
(lower left) $1$ $\Pbottom$ jet, and (lower right) $2$ $\Pbottom$ jet.
}
\label{fig:evtCategoryControlPlots_mumu2}
\end{figure*}
}

\cleardoublepage \section{The CMS Collaboration \label{app:collab}}\begin{sloppypar}\hyphenpenalty=5000\widowpenalty=500\clubpenalty=5000\vskip\cmsinstskip
\textbf{Yerevan~Physics~Institute,~Yerevan,~Armenia}\\*[0pt]
A.M.~Sirunyan, A.~Tumasyan
\vskip\cmsinstskip
\textbf{Institut~f\"{u}r~Hochenergiephysik,~Wien,~Austria}\\*[0pt]
W.~Adam, F.~Ambrogi, E.~Asilar, T.~Bergauer, J.~Brandstetter, E.~Brondolin, M.~Dragicevic, J.~Er\"{o}, A.~Escalante~Del~Valle, M.~Flechl, M.~Friedl, R.~Fr\"{u}hwirth\cmsAuthorMark{1}, V.M.~Ghete, J.~Grossmann, J.~Hrubec, M.~Jeitler\cmsAuthorMark{1}, A.~K\"{o}nig, N.~Krammer, I.~Kr\"{a}tschmer, D.~Liko, T.~Madlener, I.~Mikulec, E.~Pree, N.~Rad, H.~Rohringer, J.~Schieck\cmsAuthorMark{1}, R.~Sch\"{o}fbeck, M.~Spanring, D.~Spitzbart, A.~Taurok, W.~Waltenberger, J.~Wittmann, C.-E.~Wulz\cmsAuthorMark{1}, M.~Zarucki
% ...
\vskip\cmsinstskip
% ...
1:~Also at~Vienna~University~of~Technology,~Vienna,~Austria\\
% ...
\end{sloppypar}
\end{document}